\documentclass[11pt]{article}
\usepackage{amssymb,amsfonts,amsmath,graphicx,hyperref}

\newcommand{\mcm}[3]{\newcommand{#1}[#2]{{\ensuremath{#3}}}}

\usepackage{latexsym}
\usepackage{amssymb}

\mcm{\blank}{0}{(\emptybk)} \mcm{\dashbk}{0}{\mbox{---}}
\mcm{\emptybk}{0}{\:\:} \mcm{\hyph}{0}{\mbox{-}}

\mcm{\diagspace}{0}{\mbox{\hspace{2em}}}

\mcm{\cat}{1}{\mc{#1}} \mcm{\fcat}{1}{\mb{#1}}
\mcm{\mc}{1}{\mathcal{#1}} \mcm{\mr}{1}{\mathrm{#1}}
\mcm{\mi}{1}{\mathit{#1}} \mcm{\mb}{1}{\mathbf{#1}}
\mcm{\scat}{1}{\Bbb{#1}} \mcm{\twid}{1}{\widetilde{#1}}

\mcm{\elt}{0}{\in} \mcm{\sub}{0}{\,\subseteq\,}
\mcm{\such}{0}{\:|\:} \mcm{\without}{0}{\setminus}

\mcm{\atsr}{0}{\Box} \mcm{\eqv}{0}{\,\simeq\,}
\mcm{\iso}{0}{\,\cong\,}
\mcm{\of}{0}{\raisebox{0.2mm}{\ensuremath{\scriptstyle\circ}}}

\mcm{\bdry}{0}{\partial}

\mcm{\Bee}{0}{\cat{B}} \mcm{\Beep}{0}{\cat{B'}}
\mcm{\Eee}{0}{\cat{E}} \mcm{\Eeep}{0}{\cat{E'}}

\mcm{\Ess}{0}{\cat{S}} \mcm{\Tee}{0}{\cat{T}}
\mcm{\Teep}{0}{\cat{T'}} \mcm{\Stee}{0}{\scat{T}}
\mcm{\Steep}{0}{\scat{T'}}

\mcm{\blbk}{0}{\blank^{\blob}}
\mcm{\blob}{0}{\scriptscriptstyle{\bullet}}
\mcm{\stbk}{0}{\blank^{*}} \mcm{\ubl}{0}{{}^{\blob}}
\mcm{\ust}{0}{{}^{*}}

\mcm{\Cartpr}{0}{\pr{\Eee}{T}} \mcm{\Cartprp}{0}{\pr{\Eeep}{T'}}
\mcm{\Mnd}{0}{\triple{T}{\eta}{\mu}}
\mcm{\Zeropr}{0}{\pr{\Set}{\id}}

\mcm{\dopset}{0}{\ftrcat{\Delta^{\op}}{\Set}}
\mcm{\tropset}{0}{\ftrcat{\fcat{TR}^{\op}}{\Set}}

\mcm{\cod}{0}{\mr{cod}} \mcm{\dom}{0}{\mr{dom}}
\mcm{\End}{0}{\mr{End}} \mcm{\Hom}{0}{\mr{Hom}}
\mcm{\ob}{0}{\mr{ob}\,} \mcm{\op}{0}{\mr{op}}

\mcm{\comp}{0}{\mi{comp}} \mcm{\id}{0}{\mi{id}}
\mcm{\ids}{0}{\mi{ids}} \mcm{\mult}{0}{\mi{mult}}
\mcm{\unit}{0}{\mi{unit}}

\mcm{\Ab}{0}{\fcat{Ab}} \mcm{\Alg}{0}{\fcat{Alg}}
\mcm{\Bim}{1}{\fcat{Bim}(#1)} \mcm{\Cat}{0}{\fcat{Cat}}
\mcm{\Cay}{0}{\fcat{Cay}} \mcm{\Cpn}{1}{\pr{\Set/S_{#1}}{T_{#1}}}
\mcm{\fc}{0}{\fcat{fc}} \mcm{\fm}{0}{\fcat{fm}}
\mcm{\Graph}{0}{\fcat{Graph}} \mcm{\Gy}{0}{\fcat{Gy}}
\mcm{\Hpn}{1}{\pr{\Eee_{#1}}{P_{#1}}} \mcm{\Mon}{0}{\mb{Mon}}
\mcm{\Multicat}{0}{\fcat{Multicat}} \mcm{\One}{0}{\fcat{1}}
\mcm{\PD}{1}{\fcat{PD}_{#1}} \mcm{\Prof}{0}{\fcat{Prof}}
\mcm{\Set}{0}{\fcat{Set}} \mcm{\Span}{0}{\fcat{Span}}
\mcm{\Ssq}{0}{\fcat{Ssq}} \mcm{\Struc}{0}{\fcat{Struc}}
\mcm{\Sym}{0}{\fcat{Sym}} \mcm{\TR}{1}{\fcat{TR}(#1)}
\mcm{\Tr}{0}{\fcat{Tr}} \mcm{\Twocat}{0}{\fcat{2\hyph\Cat}}

\mcm{\integers}{0}{\mathbb{Z}}

\mcm{\range}{2}{#1,\,\ldots\,,#2}
\mcm{\bftuple}{2}{\tuplebts{\range{#1}{#2}}}
\mcm{\tuple}{3}{\tuplebts{\range{#1,#2}{#3}}}
\mcm{\rttuple}{1}{\tuplebts{\,\ldots\,,#1}}
\mcm{\abftuple}{2}{\atuplebts{\range{#1}{#2}}}
\mcm{\atuple}{3}{\atuplebts{\range{#1,#2}{#3}}}
\mcm{\arttuple}{1}{\atuplebts{\,\ldots\,,#1}}
\mcm{\sqbftuple}{2}{\obt\range{#1}{#2}\cbt}
\mcm{\pr}{2}{\tuplebts{#1,#2}}
\mcm{\triple}{3}{\tuplebts{#1,#2,#3}}

\mcm{\eend}{2}{#1[#2]} \mcm{\ehom}{3}{#1[#2,#3]}
\mcm{\ftrcat}{2}{[#1,#2]} \mcm{\homset}{3}{#1(#2,#3)}
\mcm{\multihom}{3}{#1(#2;#3)}
\mcm{\relhom}{5}{#1_{#2}(\range{#3}{#4};#5)}

\mcm{\go}{0}{\rTo} \mcm{\goby}{1}{\rTo^{#1}}
\mcm{\goesto}{0}{\,\longmapsto\,} \mcm{\goiso}{0}{\goby{\diso}}
\mcm{\monic}{0}{\rMonic} \mcm{\og}{0}{\lTo}
\mcm{\ogby}{1}{\lTo^{#1}}

\mcm{\gph}{2}{\spn{#1}{T #2}{#2}} \mcm{\graph}{4}{\spaan{#1}{T
#2}{#2}{#3}{#4}} \mcm{\oppair}{2}{\stackrel{\rTo^{#1}}{\lTo_{#2}}}
\mcm{\parpair}{2}{\stackrel{\rTo^{#1}}{\rTo_{#2}}}
\mcm{\spn}{3}{#2 \og #1 \go #3} \mcm{\spaan}{5}{#2 \ogby{#4} #1
\goby{#5} #3}

\mcm{\bktdvslob}{3}
    {\left(
    \begin{diagram}[height=1.5em]
    #1      \\
    \dTo>{\,#2} \\
    #3      \\
    \end{diagram}
    \right)}
\mcm{\slob}{3}{(#1 \goby{#2} #3)} \mcm{\vslob}{3}
    {\left.
    \begin{diagram}[height=1.5em]
    #1      \\
    \dTo>{\,#2} \\
    #3      \\
    \end{diagram}
    \right.}

\newenvironment{tree}
    {\begin{diagram}[height=1em,width=.75em,abut,noPS,tight]}
    {\end{diagram}}

\mcm{\enode}{0}{\circ}

\mcm{\nl}{1}{\stackrel{\textstyle #1}{\node}}
\mcm{\node}{0}{\bullet}

\mcm{\utree}{0}{\node}

\mcm{\diso}{0}{\sim}

\mcm{\vdiso}{0}{\wr}

\mcm{\nat}{0}{\mathbb{N}}

\mcm{\Onepr}{0}{\pr{\Graph}{\fc}}
\newlength{\nllwidth}
\newlength{\nllheight}
\newcommand{\stackbelow}[2]{%
\settowidth{\nllwidth}{\ensuremath{#1}\ensuremath{#2}}%
\settoheight{\nllheight}{\ensuremath{#2}}%
\addtolength{\nllheight}{.3ex}%
\mbox{%
\ensuremath{#1}%
\hspace{-.5\nllwidth}%
\raisebox{-1\nllheight}{\ensuremath{#2}}}}
\mcm{\nlal}{2}{\stackbelow{\nl{#1}}{#2}}
\mcm{\nll}{1}{\stackbelow{\node}{#1}} \mcm{\wun}{0}{\fcat{1}}
\mcm{\atuplebts}{1}{\langle #1 \rangle} \mcm{\tuplebts}{1}{(#1)}
\mcm{\bo}{0}{(} \mcm{\bc}{0}{)}
\mcm{\UBilax}{0}{\fcat{UBicat}_\mr{lax}}
\mcm{\UBiwk}{0}{\fcat{UBicat}_\mr{wk}}
\mcm{\UBistr}{0}{\fcat{UBicat}_\mr{str}}
\mcm{\Bilax}{0}{\fcat{Bicat}_\mr{lax}}
\mcm{\Biwk}{0}{\fcat{Bicat}_\mr{wk}}
\mcm{\Bistr}{0}{\fcat{Bicat}_\mr{str}} \mcm{\rotsub}{0}{\cup
\raisebox{0.1em}{$\scriptstyle{|}$}} \mcm{\pd}{0}{\fcat{pd}}
\mcm{\rep}{1}{\widehat{#1}} \mcm{\ovln}{1}{\overline{#1}}
\mcm{\Gph}{0}{\fcat{Gph}} \mcm{\tr}{0}{\fcat{tr}}

\mcm{\ladj}{0}{\,\dashv\,} \mcm{\zeropd}{0}{\node}
    {\end{diagram}}
\mcm{\END}{0}{\fcat{End}} \mcm{\HOM}{0}{\fcat{Hom}}

\newlength{\gwidth} 
\newlength{\gvert}  
\newlength{\gdrop}  
\newlength{\gbaredrop}  
\newlength{\goffset}    
\newlength{\gtemp}  

\newcommand{\present}[1]{%
\makebox[1\gwidth]{%
\rule[-1\gdrop]{0ex}{1\gvert}%
\raisebox{-1\gbaredrop}{#1}}}

\newcommand{\presentl}[1]{%
\makebox[1\gwidth][l]{%
\rule[-1\gdrop]{0ex}{1\gvert}%
\raisebox{-1\gbaredrop}{#1}}}

\newcommand{\presentr}[1]{%
\makebox[1\gwidth][r]{%
\rule[-1\gdrop]{0ex}{1\gvert}%
\raisebox{-1\gbaredrop}{#1}}}

\newcommand{\ginitdims}[2]{
\setlength{\unitlength}{1em}
\setlength{\goffset}{.25\unitlength}
\setlength{\gwidth}{#1\unitlength}
\setlength{\gvert}{#2\unitlength}
\setlength{\gdrop}{.5\gvert}
\addtolength{\gdrop}{-1\goffset}
\setlength{\gbaredrop}{1\gdrop}
\addtolength{\gvert}{.6\unitlength}
\addtolength{\gdrop}{.3\unitlength}}    

\newcommand{\cinitdims}[2]{
\setlength{\unitlength}{1em}
\setlength{\goffset}{.35\unitlength}
\setlength{\gwidth}{#1\unitlength}
\setlength{\gvert}{#2\unitlength}
\setlength{\gdrop}{.5\gvert}
\addtolength{\gdrop}{-1\goffset}
\setlength{\gbaredrop}{1\gdrop}
\addtolength{\gvert}{.6\unitlength}
\addtolength{\gdrop}{.3\unitlength}}    

\newcommand{\gsinitdims}[2]{
\setlength{\unitlength}{0.5em}
\setlength{\goffset}{.25\unitlength}
\setlength{\gwidth}{#1\unitlength}
\setlength{\gvert}{#2\unitlength}
\setlength{\gdrop}{.5\gvert}
\addtolength{\gdrop}{-1\goffset}
\setlength{\gbaredrop}{1\gdrop}
\addtolength{\gvert}{.6\unitlength}
\addtolength{\gdrop}{.3\unitlength}}    

\newcommand{\sidespic}[1]{%
\settowidth{\gtemp}{\ensuremath{#1}}%
\addtolength{\gwidth}{1\gtemp}}

\newcommand{\abovepic}[1]{%
\settoheight{\gtemp}{\ensuremath{#1}}%
\addtolength{\gvert}{1\gtemp}%
\settodepth{\gtemp}{\ensuremath{#1}}%
\addtolength{\gvert}{1\gtemp}}

\newcommand{\belowpic}[1]{%
\settoheight{\gtemp}{\ensuremath{#1}}%
\addtolength{\gvert}{1\gtemp}%
\addtolength{\gdrop}{1\gtemp}%
\settodepth{\gtemp}{\ensuremath{#1}}%
\addtolength{\gvert}{1\gtemp}%
\addtolength{\gdrop}{1\gtemp}}

\newcommand{\cell}[4]{\put(#1,#2){\makebox(0,0)[#3]{\ensuremath{#4}}}}
\mcm{\zmark}{0}{\scriptstyle{\bullet}}

\newcommand{\pregfst}[1]{%
\begin{picture}(0.5,0.2)(-0.5,-0.2)%
\cell{-0.1}{-0.2}{tr}{#1}%
\cell{0}{0}{c}{\zmark}%
\end{picture}}

\mcm{\gfst}{1}{%
\ginitdims{0.5}{0.4}%
\sidespic{#1}%
\belowpic{#1}%
\presentr{\pregfst{#1}}}

\newcommand{\preglst}[1]{%
\begin{picture}(0.5,0.2)(0,-0.2)%
\cell{0.1}{-0.2}{tl}{#1}%
\cell{0.05}{0}{c}{\zmark}%
\end{picture}}

\mcm{\glst}{1}{%
\ginitdims{.5}{.4}%
\sidespic{#1}%
\belowpic{#1}%
\presentl{\preglst{#1}}}

\newcommand{\preglft}[1]{%
\begin{picture}(0,0.2)(0,-0.2)%
\cell{-0.1}{-0.2}{tr}{#1}%
\cell{0.05}{0}{c}{\zmark}%
\end{picture}}

\mcm{\glft}{1}{%
\ginitdims{0}{.4}%
\belowpic{#1}%
\present{\preglft{#1}}}

\newcommand{\pregrgt}[1]{%
\begin{picture}(0,0.2)(0,-0.2)%
\cell{0.1}{-0.2}{tl}{#1}%
\cell{0.05}{0}{c}{\zmark}%
\end{picture}}

\mcm{\grgt}{1}{%
\ginitdims{0}{.4}%
\belowpic{#1}%
\present{\pregrgt{#1}}}

\newcommand{\pregblw}[1]{%
\begin{picture}(0,0.3)(0,-0.3)
\cell{0}{-0.3}{t}{#1}%
\cell{0.05}{0}{c}{\zmark}%
\end{picture}}

\mcm{\gblw}{1}{%
\ginitdims{0}{.6}%
\belowpic{#1}%
\present{\pregblw{#1}}}

\newcommand{\pregfbw}[1]{%
\begin{picture}(0,0.65)(0,-0.65)
\cell{0}{-0.65}{t}{#1}%
\cell{0.05}{0}{c}{\zmark}%
\end{picture}}

\mcm{\gfbw}{1}{%
\ginitdims{0}{1.3}%
\belowpic{#1}%
\present{\pregfbw{#1}}}

\newcommand{\pregzero}[1]{%
\begin{picture}(0.8,0.4)(-0.4,-0.4)
\cell{0}{-0.4}{t}{#1}%
\cell{0}{0}{c}{\zmark}%
\end{picture}}

\mcm{\gzero}{1}{%
\ginitdims{0.8}{.6}%
\belowpic{#1}%
\sidespic{#1}%
\present{\pregzero{#1}}}

\newcommand{\pregone}[1]{%
\begin{picture}(5,0.4)(0,-0.2)%
\cell{2.5}{0.2}{b}{#1}%
\put(0,0){\vector(1,0){5}}%
\end{picture}}

\mcm{\gone}{1}{%
\ginitdims{5}{0.4}%
\abovepic{#1}%
\present{\pregone{#1}}}

\newcommand{\pregtwo}[3]{%
\begin{picture}(5,3.4)(0,-0.2)%
\cell{2.5}{3.2}{b}{#1}%
\cell{2.5}{-.2}{t}{#2}%
\cell{2.7}{1.5}{l}{#3}%
\qbezier(0,1.5)(2.5,4.5)(5,1.5)%
\qbezier(0,1.5)(2.5,-1.5)(5,1.5)%
\put(5,1.5){\vector(1,-1){0}}%
\put(5,1.5){\vector(1,1){0}}%
\put(2.5,2.5){\vector(0,-1){2}}%
\end{picture}}

\mcm{\gtwo}{3}{%
\ginitdims{5}{3.4}%
\abovepic{#1}%
\belowpic{#2}%
\present{\pregtwo{#1}{#2}{#3}}}

\newcommand{\pregthree}[5]{%
\begin{picture}(5,5.4)(0,-1.2)%
\cell{2.5}{4.2}{b}{#1}%
\cell{1.5}{1.7}{b}{#2}%
\cell{2.5}{-1.2}{t}{#3}%
\cell{2.7}{2.75}{l}{#4}%
\cell{2.7}{0.25}{l}{#5}%
\qbezier(0,1.5)(2.5,6.5)(5,1.5)%
\qbezier(0,1.5)(2.5,-3.5)(5,1.5)%
\put(0,1.5){\vector(1,0){5}}%
\put(2.5,3.5){\vector(0,-1){1.5}}%
\put(2.5,1){\vector(0,-1){1.5}}%
\put(5,1.5){\vector(1,-3){0}}%
\put(5,1.5){\vector(1,3){0}}%
\end{picture}}

\mcm{\gthree}{5}{%
\ginitdims{5}{5.4}%
\abovepic{#1}%
\belowpic{#3}%
\present{\pregthree{#1}{#2}{#3}{#4}{#5}}}

\newcommand{\pregfour}[7]{%
\begin{picture}(5,8.4)(0,-2.7)%
\cell{2.5}{5.7}{b}{#1}%
\cell{1.5}{2.8}{b}{#2}%
\cell{1.5}{0.2}{t}{#3}%
\cell{2.5}{-2.7}{t}{#4}%
\cell{2.7}{4.25}{l}{#5}%
\cell{2.7}{1.5}{l}{#6}%
\cell{2.7}{-1.25}{l}{#7}%
\qbezier(0,1.5)(2.5,9.5)(5,1.5)%
\qbezier(0,1.5)(2.5,4)(5,1.5)%
\qbezier(0,1.5)(2.5,-1)(5,1.5)%
\qbezier(0,1.5)(2.5,-6.5)(5,1.5)%
\put(2.5,5.25){\vector(0,-1){2}}%
\put(2.5,2.5){\vector(0,-1){2}}%
\put(2.5,-0.25){\vector(0,-1){2}}%
\put(5,1.5){\vector(1,-4){0}}%
\put(5,1.5){\vector(4,-3){0}}%
\put(5,1.5){\vector(4,3){0}}%
\put(5,1.5){\vector(1,4){0}}%
\end{picture}}

\mcm{\gfour}{7}{%
\ginitdims{5}{8.4}%
\abovepic{#1}%
\belowpic{#4}%
\present{\pregfour{#1}{#2}{#3}{#4}{#5}{#6}{#7}}}

\newcommand{\pregthreecell}[5]{%
\begin{picture}(8,5)(-4,-2.5)%
\cell{0}{2.5}{b}{#1}%
\cell{0}{-2.5}{t}{#2}%
\cell{-1.7}{0}{r}{#3}%
\cell{1.7}{0}{l}{#4}%
\cell{0}{0.2}{b}{#5}%
\qbezier(-4,0)(0,4.2)(4,0)%
\qbezier(-4,0)(0,-4.2)(4,0)%
\qbezier(-0.5,1.8)(-2.5,0)(-0.5,-1.8)%
\qbezier(0.5,1.8)(2.5,0)(0.5,-1.8)%
\put(-1,0){\vector(1,0){2}}%
\put(4,0){\vector(1,-1){0}}%
\put(4,0){\vector(1,1){0}}%
\put(-0.5,-1.8){\vector(1,-1){0}}%
\put(0.5,-1.8){\vector(-1,-1){0}}%
\end{picture}}

\mcm{\gthreecell}{5}{%
\ginitdims{8}{5}%
\abovepic{#1}%
\belowpic{#2}%
\present{\pregthreecell{#1}{#2}{#3}{#4}{#5}}}

\newcommand{\pregthreecellu}{%
\begin{picture}(5,3.4)(-0.5,-0.2)%
\qbezier(-.5,1.5)(2,4.5)(4.5,1.5)%
\qbezier(-.5,1.5)(2,-1.5)(4.5,1.5)%
\qbezier(1.5,2.7)(0.5,1.5)(1.5,0.3)%
\qbezier(2.5,2.7)(3.5,1.5)(2.5,0.3)%
\put(1.3,1.5){\vector(1,0){1.4}}%
\put(4.5,1.5){\vector(1,-1){0}}%
\put(4.5,1.5){\vector(1,1){0}}%
\put(1.5,0.3){\vector(2,-3){0}}%
\put(2.5,0.3){\vector(-2,-3){0}}%
\end{picture}}

\mcm{\gthreecellu}{0}{%
\ginitdims{5}{3.4}%
\present{\pregthreecellu}}

\newcommand{\pregtwocentre}[3]{%
\begin{picture}(5,3.4)(0,-0.2)%
\cell{2.5}{3.2}{b}{#1}%
\cell{2.5}{-.2}{t}{#2}%
\cell{2.5}{1.5}{c}{#3}%
\qbezier(0,1.5)(2.5,4.5)(5,1.5)%
\qbezier(0,1.5)(2.5,-1.5)(5,1.5)%
\put(5,1.5){\vector(1,-1){0}}%
\put(5,1.5){\vector(1,1){0}}%
\put(2.5,2.5){\vector(0,-1){2}}%
\end{picture}}

\mcm{\gtwocentre}{3}{%
\ginitdims{5}{3.4}%
\abovepic{#1}%
\belowpic{#2}%
\present{\pregtwocentre{#1}{#2}{#3}}}

\newcommand{\pregspecialone}[9]{%
\begin{picture}(8,8)(-4,-4)%
\cell{0}{3.9}{b}{#1}%
\cell{-2}{-0.2}{t}{#2}%
\cell{0}{-3.9}{t}{#3}%
\cell{-1.5}{1.1}{r}{#4}%
\cell{0.2}{1.5}{l}{#5}%
\cell{1.5}{1.1}{l}{#6}%
\cell{0.2}{-2}{l}{#7}%
\cell{-0.9}{2.3}{b}{#8}%
\cell{0.9}{2.3}{b}{#9}%
\qbezier(-4,0)(0,8)(4,0)%
\qbezier(-4,0)(0,-8)(4,0)%
\qbezier(-0.5,3.4)(-3.5,2)(-0.5,0.6)%
\qbezier(0.5,3.4)(3.5,2)(0.5,0.6)%
\put(-4,0){\vector(1,0){8}}%
\put(0,3.4){\vector(0,-1){2.8}}%
\put(0,-0.8){\vector(0,-1){2.4}}%
\put(-1.5,2.2){\vector(1,0){1.2}}%
\put(0.3,2.2){\vector(1,0){1.2}}%
\put(4,0){\vector(1,-2){0}}%
\put(4,0){\vector(1,2){0}}%
\put(-0.5,0.6){\vector(2,-1){0}}%
\put(0.5,0.6){\vector(-2,-1){0}}%
\end{picture}}

\mcm{\gspecialone}{9}{%
\ginitdims{8}{8}%
\abovepic{#1}%
\belowpic{#3}%
\present{\pregspecialone{#1}{#2}{#3}{#4}{#5}{#6}{#7}{#8}{#9}}}

\newcommand{\pregspecialtwo}{%
\begin{picture}(5,3.4)(0,-0.2)%
\qbezier(0,1.5)(2.5,4.5)(5,1.5)%
\qbezier(0,1.5)(2.5,-1.5)(5,1.5)%
\qbezier(1.7,2.5)(0,1.5)(1.7,0.5)%
\qbezier(3.3,2.5)(5,1.5)(3.3,0.5)%
\put(5,1.5){\vector(1,-1){0}}%
\put(5,1.5){\vector(1,1){0}}%
\put(1.7,0.5){\vector(3,-2){0}}%
\put(3.3,0.5){\vector(-3,-2){0}}%
\put(2.5,2.5){\vector(0,-1){2}}%
\put(1.2,1.5){\vector(1,0){1}}%
\put(2.8,1.5){\vector(1,0){1}}%
\end{picture}}

\mcm{\gspecialtwo}{0}{%
\ginitdims{5}{3.4}%
\present{\pregspecialtwo}}

\newcommand{\pregspecialthree}{%
\begin{picture}(5,5.4)(0,-1.2)%
\qbezier(0,1.5)(2.5,6.5)(5,1.5)%
\qbezier(0,1.5)(2.5,-3.5)(5,1.5)%
\qbezier(2,3.5)(1,2.75)(2,2)%
\qbezier(3,3.5)(4,2.75)(3,2)%
\qbezier(2,1)(1,0.25)(2,-0.5)%
\qbezier(3,1)(4,0.25)(3,-0.5)%
\put(0,1.5){\vector(1,0){5}}%
\put(1.5,2.75){\vector(1,0){2}}%
\put(1.5,0.25){\vector(1,0){2}}%
\put(5,1.5){\vector(1,-3){0}}%
\put(5,1.5){\vector(1,3){0}}%
\put(2,2){\vector(1,-1){0}}%
\put(3,2){\vector(-1,-1){0}}%
\put(2,-0.5){\vector(1,-1){0}}%
\put(3,-0.5){\vector(-1,-1){0}}%
\end{picture}}

\mcm{\gspecialthree}{0}{%
\ginitdims{5}{5.4}%
\present{\pregspecialthree}}

\newcommand{\pregonew}[1]{%
\begin{picture}(8,0.4)(0,-0.2)%
\cell{4}{0.2}{b}{#1}%
\put(0,0){\vector(1,0){8}}%
\end{picture}}

\mcm{\gonew}{1}{%
\ginitdims{8}{0.4}%
\abovepic{#1}%
\present{\pregonew{#1}}}

\mcm{\gzersu}{0}{%
\gsinitdims{0}{.6}%
\present{\pregblw{}}}

\mcm{\gonesu}{0}{%
\gsinitdims{5}{0.4}%
\present{\pregone{}}}

\mcm{\gtwosu}{0}{%
\gsinitdims{5}{3.4}%
\present{\pregtwo{}{}{}}}

\mcm{\gthreesu}{0}{%
\gsinitdims{5}{5.4}%
\present{\pregthree{}{}{}{}{}}}

\mcm{\gfoursu}{0}{%
\gsinitdims{5}{8.4}%
\present{\pregfour{}{}{}{}{}{}{}}}

\newcommand{\precone}[1]{%
\begin{picture}(4.2,0.4)(-0.3,-0.2)%
\cell{1.8}{0.2}{b}{#1}%
\put(0,0){\vector(1,0){3.6}}%
\end{picture}}

\mcm{\cone}{1}{%
\cinitdims{4.2}{0.4}%
\abovepic{#1}%
\present{\precone{#1}}}

\mcm{\gfstsu}{0}{%
\gsinitdims{0.5}{0.4}%
\presentr{\pregfst{}}}

\mcm{\glstsu}{0}{%
\gsinitdims{0.5}{0.4}%
\presentl{\preglst{}}}

\newcommand{\prectwodbl}[3]%
{\begin{picture}(4.2,3.4)(-0.1,-0.2)%
\cell{2}{3.2}{b}{#1}%
\cell{2}{-0.2}{t}{#2}%
\cell{2.3}{1.5}{l}{#3}%
\qbezier(0,2)(2,4)(4,2)%
\qbezier(0,1)(2,-1)(4,1)%
\put(4,2){\vector(1,-1){0}}%
\put(4,1){\vector(1,1){0}}%
\put(1.9,2.5){\line(0,-1){1.8}}%
\put(2.1,2.5){\line(0,-1){1.8}}%
\cell{2.01}{0.4}{b}{\vee}%
\end{picture}}

\mcm{\ctwodbl}{3}{%
\cinitdims{4.2}{3.4}%
\abovepic{#1}%
\belowpic{#2}%
\present{\prectwodbl{#1}{#2}{#3}}}

\newcommand{\precthreedbl}[5]{%
\begin{picture}(4.2,5.4)(-0.1,-0.2)%
\cell{2}{5.2}{b}{#1}%
\cell{1}{2.7}{b}{#2}%
\cell{2}{-.2}{t}{#3}%
\cell{2.3}{3.75}{l}{#4}%
\cell{2.3}{1.25}{l}{#5}%
\qbezier(0,3)(2,7)(4,3)%
\qbezier(0,2)(2,-2)(4,2)%
\put(0,2.5){\vector(1,0){4}}%
\put(1.9,4.5){\line(0,-1){1.3}}%
\put(2.1,4.5){\line(0,-1){1.3}}%
\cell{2.01}{2.9}{b}{\vee}%
\put(1.9,2){\line(0,-1){1.3}}%
\put(2.1,2){\line(0,-1){1.3}}%
\cell{2.01}{0.4}{b}{\vee}%
\put(4,3){\vector(1,-3){0}}%
\put(4,2){\vector(1,3){0}}%
\end{picture}}

\mcm{\cthreedbl}{5}{%
\cinitdims{4.2}{5.4}%
\abovepic{#1}%
\belowpic{#3}%
\present{\precthreedbl{#1}{#2}{#3}{#4}{#5}}}

\newcommand{\precthreecelltrp}[5]{%
\begin{picture}(8.2,5)(-4.1,-2.5)%
\cell{0}{2.5}{b}{#1}%
\cell{0}{-2.5}{t}{#2}%
\cell{-1.8}{0}{r}{#3}%
\cell{1.8}{0}{l}{#4}%
\cell{0}{0.3}{b}{#5}%
\qbezier(-4,0.5)(0,4)(4,0.5)%
\qbezier(-4,-0.5)(0,-4)(4,-0.5)%
\qbezier(-0.6,2)(-2.6,0)(-0.6,-2)%
\qbezier(-0.4,2)(-2.4,0)(-0.5,-1.9)%
\cell{-0.6}{-2}{b}{\lrcorner}%
\qbezier(0.4,2)(2.4,0)(0.5,-1.9)%
\qbezier(0.6,2)(2.6,0)(0.6,-2)%
\cell{0.65}{-2}{b}{\llcorner}%
\put(-1,0.15){\line(1,0){1.7}}%
\put(-1,0){\line(1,0){2}}%
\put(-1,-0.15){\line(1,0){1.7}}%
\cell{1.15}{0}{r}{>}%
\put(4,0.5){\vector(1,-1){0}}%
\put(4,-0.5){\vector(1,1){0}}%
\end{picture}}

\mcm{\cthreecelltrp}{5}{%
\cinitdims{8.2}{5}%
\abovepic{#1}%
\belowpic{#2}%
\present{\precthreecelltrp{#1}{#2}{#3}{#4}{#5}}}

\newcommand{\prectwo}[3]%
{\begin{picture}(4.2,3.4)(-0.1,-0.2)%
\cell{2}{3.2}{b}{#1}%
\cell{2}{-0.2}{t}{#2}%
\cell{2.2}{1.5}{l}{#3}%
\qbezier(0,2)(2,4)(4,2)%
\qbezier(0,1)(2,-1)(4,1)%
\put(4,2){\vector(1,-1){0}}%
\put(4,1){\vector(1,1){0}}%
\put(2,2.5){\vector(0,-1){2}}%
\end{picture}}

\mcm{\ctwo}{3}{%
\cinitdims{4.2}{3.4}%
\abovepic{#1}%
\belowpic{#2}%
\present{\prectwo{#1}{#2}{#3}}}

\newcommand{\precthree}[5]{%
\begin{picture}(4.2,5.4)(-0.1,-0.2)%
\cell{2}{5.2}{b}{#1}%
\cell{1}{2.7}{b}{#2}%
\cell{2}{-.2}{t}{#3}%
\cell{2.2}{3.75}{l}{#4}%
\cell{2.2}{1.25}{l}{#5}%
\qbezier(0,3)(2,7)(4,3)%
\qbezier(0,2)(2,-2)(4,2)%
\put(0,2.5){\vector(1,0){4}}%
\put(2,4.5){\vector(0,-1){1.5}}%
\put(2,2){\vector(0,-1){1.5}}%
\put(4,3){\vector(1,-3){0}}%
\put(4,2){\vector(1,3){0}}%
\end{picture}}

\mcm{\cthree}{5}{%
\cinitdims{4.2}{5.4}%
\abovepic{#1}%
\belowpic{#3}%
\present{\precthree{#1}{#2}{#3}{#4}{#5}}}

\newcommand{\prectwoop}[3]%
{\begin{picture}(4.2,3.4)(-0.1,-0.2)%
\cell{2}{3.2}{b}{#1}%
\cell{2}{-0.2}{t}{#2}%
\cell{2.2}{1.5}{l}{#3}%
\qbezier(0,2)(2,4)(4,2)%
\qbezier(0,1)(2,-1)(4,1)%
\put(0,2){\vector(-1,-1){0}}%
\put(0,1){\vector(-1,1){0}}%
\put(2,2.5){\vector(0,-1){2}}%
\end{picture}}

\mcm{\ctwoop}{3}{%
\cinitdims{4.2}{3.4}%
\abovepic{#1}%
\belowpic{#2}%
\present{\prectwoop{#1}{#2}{#3}}}

\newcommand{\prectwopar}[4]{%
\begin{picture}(4.2,3.4)(-0.1,-0.2)%
\cell{2}{3.2}{b}{#1}%
\cell{2}{-0.2}{t}{#2}%
\cell{1.6}{1.5}{r}{#3}%
\cell{2.4}{1.5}{l}{#4}%
\qbezier(0,2)(2,4)(4,2)%
\qbezier(0,1)(2,-1)(4,1)%
\put(4,2){\vector(1,-1){0}}%
\put(4,1){\vector(1,1){0}}%
\put(1.8,2.5){\vector(0,-1){2}}%
\put(2.2,2.5){\vector(0,-1){2}}%
\end{picture}}

\mcm{\ctwopar}{4}{%
\cinitdims{4.2}{3.4}%
\abovepic{#1}%
\belowpic{#2}%
\present{\prectwopar{#1}{#2}{#3}{#4}}}

\newcommand{\precthreein}[5]{%
\begin{picture}(4.2,5.4)(-0.1,-0.2)%
\cell{2}{5.2}{b}{#1}%
\cell{1}{2.7}{b}{#2}%
\cell{2}{-.2}{t}{#3}%
\cell{2.2}{3.75}{l}{#4}%
\cell{2.2}{1.25}{l}{#5}%
\qbezier(0,3)(2,7)(4,3)%
\qbezier(0,2)(2,-2)(4,2)%
\put(0,2.5){\vector(1,0){4}}%
\put(2,4.5){\vector(0,-1){1.5}}%
\put(2,0.5){\vector(0,1){1.5}}%
\put(4,3){\vector(1,-3){0}}%
\put(4,2){\vector(1,3){0}}%
\end{picture}}

\mcm{\cthreein}{5}{%
\cinitdims{4.2}{5.4}%
\abovepic{#1}%
\belowpic{#3}%
\present{\precthreein{#1}{#2}{#3}{#4}{#5}}}

\newcommand{\precthreecell}[5]{%
\begin{picture}(8.2,5)(-4.1,-2.5)%
\cell{0}{2.5}{b}{#1}%
\cell{0}{-2.5}{t}{#2}%
\cell{-1.7}{0}{r}{#3}%
\cell{1.7}{0}{l}{#4}%
\cell{0}{0.2}{b}{#5}%
\qbezier(-4,0.5)(0,4)(4,0.5)%
\qbezier(-4,-0.5)(0,-4)(4,-0.5)%
\qbezier(-0.5,2)(-2.5,0)(-0.5,-2)%
\qbezier(0.5,2)(2.5,0)(0.5,-2)%
\put(-1,0){\vector(1,0){2}}%
\put(4,0.5){\vector(1,-1){0}}%
\put(4,-0.5){\vector(1,1){0}}%
\put(-0.5,-2){\vector(1,-1){0}}%
\put(0.5,-2){\vector(-1,-1){0}}%
\end{picture}}

\mcm{\cthreecell}{5}{%
\cinitdims{8.2}{5}%
\abovepic{#1}%
\belowpic{#2}%
\present{\precthreecell{#1}{#2}{#3}{#4}{#5}}}

\newcommand{\precthreecellpar}[6]{%
\begin{picture}(8.2,5)(-4.1,-2.5)%
\cell{0}{2.5}{b}{#1}%
\cell{0}{-2.5}{t}{#2}%
\cell{-1.7}{0}{r}{#3}%
\cell{1.7}{0}{l}{#4}%
\cell{0}{0.4}{b}{#5}%
\cell{0}{-0.4}{t}{#6}%
\qbezier(-4,0.5)(0,4)(4,0.5)%
\qbezier(-4,-0.5)(0,-4)(4,-0.5)%
\qbezier(-0.5,2)(-2.5,0)(-0.5,-2)%
\qbezier(0.5,2)(2.5,0)(0.5,-2)%
\put(-1,0.2){\vector(1,0){2}}%
\put(-1,-0.2){\vector(1,0){2}}%
\put(4,0.5){\vector(1,-1){0}}%
\put(4,-0.5){\vector(1,1){0}}%
\put(-0.5,-2){\vector(1,-1){0}}%
\put(0.5,-2){\vector(-1,-1){0}}%
\end{picture}}

\mcm{\cthreecellpar}{6}{%
\cinitdims{8.2}{5}%
\abovepic{#1}%
\belowpic{#2}%
\present{\precthreecellpar{#1}{#2}{#3}{#4}{#5}{#6}}}

\newcommand{\prectwov}[5]{%
\begin{picture}(3.4,4.2)(0.8,0.9)%
\cell{2.5}{5.1}{b}{#1}%
\cell{2.5}{0.9}{t}{#2}%
\cell{0.8}{3}{r}{#3}%
\cell{4.2}{3}{l}{#4}%
\cell{2.5}{3.2}{b}{#5}%
\qbezier(2,5)(0,3)(2,1)%
\qbezier(3,5)(5,3)(3,1)%
\put(2,1){\vector(1,-1){0}}%
\put(3,1){\vector(-1,-1){0}}%
\put(1.5,3){\vector(1,0){2}}%
\end{picture}}

\mcm{\ctwov}{5}{%
\cinitdims{3.4}{4.2}%
\abovepic{#1}%
\belowpic{#2}%
\sidespic{#3}%
\sidespic{#4}%
\present{\prectwov{#1}{#2}{#3}{#4}{#5}}}

\newcommand{\precthreecellv}[7]{%
\begin{picture}(5,8.2)(0.5,-1.6)%
\cell{3}{6.6}{b}{#1}%
\cell{3}{-1.6}{t}{#2}%
\cell{0.5}{2.5}{r}{#3}%
\cell{5.5}{2.5}{l}{#4}%
\cell{3}{4.2}{b}{#5}%
\cell{3}{0.8}{t}{#6}%
\cell{3.2}{2.5}{l}{#7}%
\qbezier(3.5,6.5)(7,2.5)(3.5,-1.5)%
\qbezier(2.5,6.5)(-1,2.5)(2.5,-1.5)%
\put(2.5,-1.5){\vector(1,-1){0}}%
\put(3.5,-1.5){\vector(-1,-1){0}}%
\qbezier(1,3)(3,5)(5,3)%
\qbezier(1,2)(3,0)(5,2)%
\put(5,3){\vector(1,-1){0}}%
\put(5,2){\vector(1,1){0}}%
\put(3,3.5){\vector(0,-1){2}}%
\end{picture}}

\mcm{\cthreecellv}{7}{%
\cinitdims{5}{8.2}%
\abovepic{#1}%
\belowpic{#2}%
\sidespic{#3}%
\sidespic{#4}%
\present{\precthreecellv{#1}{#2}{#3}{#4}{#5}{#6}{#7}}}

\newcommand{\pretopez}[2]{%
\begin{picture}(2.6,2.3)(-1.3,-2.2)%
\cell{0}{-2.2}{t}{#1}%
\cell{0}{-1.2}{c}{#2}%
\qbezier(0,0)(-2,-2)(0,-2)%
\qbezier(0,0)(2,-2)(0,-2)%
\put(0,0){\vector(-1,1){0}}%
\end{picture}}

\mcm{\topez}{2}{%
\ginitdims{2.6}{2.3}%
\belowpic{#1}%
\present{\pretopez{#1}{#2}}}

\newcommand{\pretopea}[3]{%
\begin{picture}(4,1.9)(-2,-0,2)%
\cell{0}{1.7}{b}{#1}%
\cell{0}{-0.2}{t}{#2}%
\cell{0}{0.7}{c}{#3}%
\qbezier(-2,0)(0,3)(2,0)%
\put(-2,0){\vector(1,0){4}}%
\put(2,0){\vector(2,-3){0}}%
\end{picture}}

\mcm{\topea}{3}{%
\ginitdims{4}{1.9}%
\abovepic{#1}%
\belowpic{#2}%
\present{\pretopea{#1}{#2}{#3}}}

\newcommand{\pretopeb}[4]{%
\begin{picture}(4,2.2)(-2,-0.2)%
\cell{-1.1}{1}{br}{#1}%
\cell{1.1}{1}{bl}{#2}%
\cell{0}{-0.2}{t}{#3}%
\cell{0}{0.8}{c}{#4}%
\put(-2,0){\vector(1,1){2}}%
\put(0,2){\vector(1,-1){2}}%
\put(-2,0){\vector(1,0){4}}%
\end{picture}}

\mcm{\topeb}{4}{%
\ginitdims{4}{2.2}%
\belowpic{#3}%
\present{\pretopeb{#1}{#2}{#3}{#4}}}

\newcommand{\pretopec}[5]{%
\begin{picture}(4,2.2)(-2,-0.2)%
\cell{-1.8}{1}{br}{#1}%
\cell{0}{2.2}{b}{#2}%
\cell{1.8}{1}{bl}{#3}%
\cell{0}{-0.2}{t}{#4}%
\cell{0}{0.8}{c}{#5}%
\put(-2,0){\vector(1,2){1}}%
\put(-1,2){\vector(1,0){2}}%
\put(1,2){\vector(1,-2){1}}%
\put(-2,0){\vector(1,0){4}}%
\end{picture}}

\mcm{\topec}{5}{%
\ginitdims{4}{2.2}%
\sidespic{#1}%
\abovepic{#2}%
\sidespic{#3}%
\belowpic{#4}%
\present{\pretopec{#1}{#2}{#3}{#4}{#5}}}

\newcommand{\pretoped}[6]{%
\begin{picture}(4,2.5)(-2,-0.2)%
\cell{-2}{0.6}{br}{#1}%
\cell{-0.7}{2.2}{br}{#2}%
\cell{0.7}{2.2}{bl}{#3}%
\cell{2}{0.6}{bl}{#4}%
\cell{0}{-0.2}{t}{#5}%
\cell{0}{0.8}{c}{#6}%
\put(-2,0){\vector(1,3){0.5}}%
\put(-1.5,1.5){\vector(3,2){1.5}}%
\put(0,2.5){\vector(3,-2){1.5}}%
\put(1.5,1.5){\vector(1,-3){0.5}}%
\put(-2,0){\vector(1,0){4}}%
\end{picture}}

\mcm{\toped}{6}{%
\ginitdims{4}{2.5}%
\sidespic{#1}%
\abovepic{#2}%
\abovepic{#3}%
\sidespic{#4}%
\belowpic{#5}%
\present{\pretoped{#1}{#2}{#3}{#4}{#5}{#6}}}

\newcommand{\pretopeq}[5]{%
\begin{picture}(4,2.5)(-2,-0.2)%
\cell{-2}{0.6}{br}{#1}%
\cell{-1}{2.2}{br}{#2}%
\cell{2}{0.6}{bl}{#3}%
\cell{0}{-0.2}{t}{#4}%
\cell{0}{0.8}{c}{#5}%
\put(-2,0){\vector(1,3){0.5}}%
\put(-1.5,1.5){\vector(1,1){1}}%
\cell{0.9}{2.3}{c}{\ddots}
\put(1.5,1.5){\vector(1,-3){0.5}}%
\put(-2,0){\vector(1,0){4}}%
\end{picture}}

\mcm{\topeq}{5}{%
\ginitdims{4}{2.5}%
\sidespic{#1}%
\abovepic{#2}%
\sidespic{#3}%
\belowpic{#4}%
\present{\pretopeq{#1}{#2}{#3}{#4}{#5}}}

\newcommand{\pretopebase}[1]{%
\begin{picture}(4,0.4)(0,-0.2)%
\cell{2}{0.2}{b}{#1}%
\put(0,0){\vector(1,0){4}}%
\end{picture}}

\mcm{\topebase}{1}{%
\ginitdims{4}{0.4}%
\abovepic{#1}%
\present{\pretopebase{#1}}}

\newcommand{\pretopezs}[2]{%
\begin{picture}(2.6,2.3)(-1.3,-2.2)%
\cell{0}{-2.2}{t}{#1}%
\cell{0}{-1.2}{c}{#2}%
\qbezier(0,0)(-2,-2)(0,-2)%
\qbezier(0,0)(2,-2)(0,-2)%
\end{picture}}

\mcm{\topezs}{2}{%
\ginitdims{2.6}{2.3}%
\belowpic{#1}%
\present{\pretopezs{#1}{#2}}}

\newcommand{\pretopeas}[3]{%
\begin{picture}(4,1.9)(-2,-0,2)%
\cell{0}{1.7}{b}{#1}%
\cell{0}{-0.2}{t}{#2}%
\cell{0}{0.7}{c}{#3}%
\qbezier(-2,0)(0,3)(2,0)%
\put(-2,0){\line(1,0){4}}%
\end{picture}}

\mcm{\topeas}{3}{%
\ginitdims{4}{1.9}%
\abovepic{#1}%
\belowpic{#2}%
\present{\pretopeas{#1}{#2}{#3}}}

\newcommand{\pretopebs}[4]{%
\begin{picture}(4,2.2)(-2,-0.2)%
\cell{-1.1}{1}{br}{#1}%
\cell{1.1}{1}{bl}{#2}%
\cell{0}{-0.2}{t}{#3}%
\cell{0}{0.8}{c}{#4}%
\put(-2,0){\line(1,1){2}}%
\put(0,2){\line(1,-1){2}}%
\put(-2,0){\line(1,0){4}}%
\end{picture}}

\mcm{\topebs}{4}{%
\ginitdims{4}{2.2}%
\belowpic{#3}%
\present{\pretopebs{#1}{#2}{#3}{#4}}}

\newcommand{\pretopecs}[5]{%
\begin{picture}(4,2.2)(-2,-0.2)%
\cell{-1.8}{1}{br}{#1}%
\cell{0}{2.2}{b}{#2}%
\cell{1.8}{1}{bl}{#3}%
\cell{0}{-0.2}{t}{#4}%
\cell{0}{0.8}{c}{#5}%
\put(-2,0){\line(1,2){1}}%
\put(-1,2){\line(1,0){2}}%
\put(1,2){\line(1,-2){1}}%
\put(-2,0){\line(1,0){4}}%
\end{picture}}

\mcm{\topecs}{5}{%
\ginitdims{4}{2.2}%
\sidespic{#1}%
\abovepic{#2}%
\sidespic{#3}%
\belowpic{#4}%
\present{\pretopecs{#1}{#2}{#3}{#4}{#5}}}

\newcommand{\pretopeds}[6]{%
\begin{picture}(4,2.5)(-2,-0.2)%
\cell{-2}{0.6}{br}{#1}%
\cell{-0.7}{2.2}{br}{#2}%
\cell{0.7}{2.2}{bl}{#3}%
\cell{2}{0.6}{bl}{#4}%
\cell{0}{-0.2}{t}{#5}%
\cell{0}{0.8}{c}{#6}%
\put(-2,0){\line(1,3){0.5}}%
\put(-1.5,1.5){\line(3,2){1.5}}%
\put(0,2.5){\line(3,-2){1.5}}%
\put(1.5,1.5){\line(1,-3){0.5}}%
\put(-2,0){\line(1,0){4}}%
\end{picture}}

\mcm{\topeds}{6}{%
\ginitdims{4}{2.5}%
\sidespic{#1}%
\abovepic{#2}%
\abovepic{#3}%
\sidespic{#4}%
\belowpic{#5}%
\present{\pretopeds{#1}{#2}{#3}{#4}{#5}{#6}}}

\newcommand{\pretopeqs}[5]{%
\begin{picture}(4,2.5)(-2,-0.2)%
\cell{-2}{0.6}{br}{#1}%
\cell{-1}{2.2}{br}{#2}%
\cell{2}{0.6}{bl}{#3}%
\cell{0}{-0.2}{t}{#4}%
\cell{0}{0.8}{c}{#5}%
\put(-2,0){\line(1,3){0.5}}%
\put(-1.5,1.5){\line(1,1){1}}%
\cell{0.9}{2.3}{c}{\ddots}
\put(1.5,1.5){\line(1,-3){0.5}}%
\put(-2,0){\line(1,0){4}}%
\end{picture}}

\mcm{\topeqs}{5}{%
\ginitdims{4}{2.5}%
\sidespic{#1}%
\abovepic{#2}%
\sidespic{#3}%
\belowpic{#4}%
\present{\pretopeqs{#1}{#2}{#3}{#4}{#5}}}

\newcommand{\pretopebases}[1]{%
\begin{picture}(4,0.4)(0,-0.2)%
\cell{2}{0.2}{b}{#1}%
\put(0,0){\line(1,0){4}}%
\end{picture}}

\mcm{\topebases}{1}{%
\ginitdims{4}{0.4}%
\abovepic{#1}%
\present{\pretopebases{#1}}}

\newcommand{\pregdots}[6]{%
\begin{picture}(5,8.4)(0,-2.7)%
\cell{2.5}{5.7}{b}{#1}%
\cell{1.5}{2.8}{b}{#2}%
\cell{1.5}{0.2}{t}{#3}%
\cell{2.5}{-2.7}{t}{#4}%
\cell{2.7}{4.25}{l}{#5}%
\cell{2.7}{-1.25}{l}{#6}%
\qbezier(0,1.5)(2.5,9.5)(5,1.5)%
\qbezier(0,1.5)(2.5,4)(5,1.5)%
\qbezier(0,1.5)(2.5,-1)(5,1.5)%
\qbezier(0,1.5)(2.5,-6.5)(5,1.5)%
\put(2.5,5.25){\vector(0,-1){2}}%
\put(2.5,-0.25){\vector(0,-1){2}}%
\cell{2.5}{1.7}{c}{\vdots}%
\put(5,1.5){\vector(1,-4){0}}%
\put(5,1.5){\vector(4,-3){0}}%
\put(5,1.5){\vector(4,3){0}}%
\put(5,1.5){\vector(1,4){0}}%
\end{picture}}

\mcm{\gdots}{6}{%
\ginitdims{5}{8.4}%
\abovepic{#1}%
\belowpic{#4}%
\present{\pregdots{#1}{#2}{#3}{#4}{#5}{#6}}}

\newlength{\volt}
\makeatletter

\def\diagram{\m@th\leftwidth=\z@ \rightwidth=\z@ \topheight=\z@
\botheight=\z@ \setbox\@picbox\hbox\bgroup}

\def\enddiagram{\egroup\wd\@picbox\rightwidth\unitlength
\ht\@picbox\topheight\unitlength \dp\@picbox\botheight\unitlength
\hskip\leftwidth\unitlength\box\@picbox}

\def\bfig{\begin{diagram}}
\def\efig{\end{diagram}}
\newcount\wideness \newcount\leftwidth \newcount\rightwidth
\newcount\highness \newcount\topheight \newcount\botheight

\def\ratchet#1#2{\ifnum#1<#2 \global #1=#2 \fi}

\def\putbox(#1,#2)#3{%
\horsize{\wideness}{#3} \divide\wideness by 2 {\advance\wideness
by #1 \ratchet{\rightwidth}{\wideness}} {\advance\wideness by -#1
\ratchet{\leftwidth}{\wideness}} \vertsize{\highness}{#3}
\divide\highness by 2 {\advance\highness by #2
\ratchet{\topheight}{\highness}} {\advance\highness by -#2
\ratchet{\botheight}{\highness}} \put(#1,#2){\makebox(0,0){$#3$}}}

\def\putlbox(#1,#2)#3{%
\horsize{\wideness}{#3} {\advance\wideness by #1
\ratchet{\rightwidth}{\wideness}} {\ratchet{\leftwidth}{-#1}}
\vertsize{\highness}{#3} \divide\highness by 2 {\advance\highness
by #2 \ratchet{\topheight}{\highness}} {\advance\highness by -#2
\ratchet{\botheight}{\highness}}
\put(#1,#2){\makebox(0,0)[l]{$#3$}}}

\def\putrbox(#1,#2)#3{%
\horsize{\wideness}{#3} {\ratchet{\rightwidth}{#1}}
{\advance\wideness by -#1 \ratchet{\leftwidth}{\wideness}}
\vertsize{\highness}{#3} \divide\highness by 2 {\advance\highness
by #2 \ratchet{\topheight}{\highness}} {\advance\highness by -#2
\ratchet{\botheight}{\highness}}
\put(#1,#2){\makebox(0,0)[r]{$#3$}}}

\def\adjust[#1]{} 

\newcount \coefa
\newcount \coefb
\newcount \coefc
\newcount\tempcounta
\newcount\tempcountb
\newcount\tempcountc
\newcount\tempcountd
\newcount\xext
\newcount\yext
\newcount\xoff
\newcount\yoff
\newcount\gap%
\newcount\arrowtypea
\newcount\arrowtypeb
\newcount\arrowtypec
\newcount\arrowtyped
\newcount\arrowtypee
\newcount\height
\newcount\width
\newcount\xpos
\newcount\ypos
\newcount\run
\newcount\rise
\newcount\arrowlength
\newcount\halflength
\newcount\arrowtype
\newdimen\tempdimen
\newdimen\xlen
\newdimen\ylen
\newsavebox{\tempboxa}%
\newsavebox{\tempboxb}%
\newsavebox{\tempboxc}%

\newdimen\w@dth

\def\setw@dth#1#2{\setbox\z@\hbox{\m@th$#1$}\w@dth=\wd\z@
\setbox\@ne\hbox{\m@th$#2$}\ifnum\w@dth<\wd\@ne \w@dth=\wd\@ne \fi
\advance\w@dth by 1.2em}

\def\t@^#1_#2{\allowbreak\def\n@one{#1}\def\n@two{#2}\mathrel
{\setw@dth{#1}{#2} \mathop{\hbox to
\w@dth{\rightarrowfill}}\limits \ifx\n@one\empty\else
^{\box\z@}\fi \ifx\n@two\empty\else _{\box\@ne}\fi}}
\def\t@@^#1{\@ifnextchar_{\t@^{#1}}{\t@^{#1}_{}}}
\def\to{\@ifnextchar^{\t@@}{\t@@^{}}}

\def\t@left^#1_#2{\def\n@one{#1}\def\n@two{#2}\mathrel{\setw@dth{#1}{#2}
\mathop{\hbox to \w@dth{\leftarrowfill}}\limits
\ifx\n@one\empty\else ^{\box\z@}\fi \ifx\n@two\empty\else
_{\box\@ne}\fi}}
\def\t@@left^#1{\@ifnextchar_{\t@left^{#1}}{\t@left^{#1}_{}}}
\def\toleft{\@ifnextchar^{\t@@left}{\t@@left^{}}}

\def\two@^#1_#2{\allowbreak
\def\n@one{#1}\def\n@two{#2}\mathrel{\setw@dth{#1}{#2}
\mathop{\vcenter{\lineskip\z@\baselineskip\z@
                 \hbox to \w@dth{\rightarrowfill}%
                 \hbox to \w@dth{\rightarrowfill}}%
       }\limits
\ifx\n@one\empty\else ^{\box\z@}\fi \ifx\n@two\empty\else
_{\box\@ne}\fi}}
\def\tw@@^#1{\@ifnextchar _{\two@^{#1}}{\two@^{#1}_{}}}
\def\two{\@ifnextchar ^{\tw@@}{\tw@@^{}}}

\def\tofr@^#1_#2{\def\n@one{#1}\def\n@two{#2}\mathrel{\setw@dth{#1}{#2}
\mathop{\vcenter{\hbox to \w@dth{\rightarrowfill}\kern-1.7ex
                 \hbox to \w@dth{\leftarrowfill}}%
       }\limits
\ifx\n@one\empty\else ^{\box\z@}\fi \ifx\n@two\empty\else
_{\box\@ne}\fi}}
\def\t@fr@^#1{\@ifnextchar_ {\tofr@^{#1}}{\tofr@^{#1}_{}}}
\def\tofro{\@ifnextchar^ {\t@fr@}{\t@fr@^{}}}

\def\mon{\mathop{\m@th\hbox to
      14.6\P@{\lasyb\char'51\hskip-2.1\P@$\arrext$\hss
$\mathord\rightarrow$}}\limits} 
\def\leftmono{\mathrel{\m@th\hbox to
14.6\P@{$\mathord\leftarrow$\hss$\arrext$\hskip-2.1\P@\lasyb\char'50%
}}\limits} 
\mathchardef\arrext="0200       

\def\settypes(#1,#2,#3){\arrowtypea#1 \arrowtypeb#2 \arrowtypec#3}
\def\settoheight#1#2{\setbox\@tempboxa\hbox{#2}#1\ht\@tempboxa\relax}%
\def\settodepth#1#2{\setbox\@tempboxa\hbox{#2}#1\dp\@tempboxa\relax}%
\def\settokens`#1`#2`#3`#4`{%
     \def\tokena{#1}\def\tokenb{#2}\def\tokenc{#3}\def\tokend{#4}}
\def\setsqparms[#1`#2`#3`#4;#5`#6]{%
\arrowtypea #1 \arrowtypeb #2 \arrowtypec #3 \arrowtyped #4
\width #5 \height #6 }
\def\setpos(#1,#2){\xpos=#1 \ypos#2}

\def\settriparms[#1`#2`#3;#4]{\settripairparms[#1`#2`#3`1`1;#4]}%

\def\settripairparms[#1`#2`#3`#4`#5;#6]{%
\arrowtypea #1 \arrowtypeb #2 \arrowtypec #3 \arrowtyped #4
\arrowtypee #5 \width #6 \height #6 }

\def\resetparms{\settripairparms[1`1`1`1`1;500]\width 500}

\resetparms

\def\mvector(#1,#2)#3{
\put(0,0){\vector(#1,#2){#3}}%
\put(0,0){\vector(#1,#2){26}}%
}
\def\evector(#1,#2)#3{{
\arrowlength #3
\put(0,0){\vector(#1,#2){\arrowlength}}%
\advance \arrowlength by-30
\put(0,0){\vector(#1,#2){\arrowlength}}%
}}

\def\horsize#1#2{%
\settowidth{\tempdimen}{$#2$}%
#1=\tempdimen \divide #1 by\unitlength }

\def\vertsize#1#2{%
\settoheight{\tempdimen}{$#2$}%
#1=\tempdimen
\settodepth{\tempdimen}{$#2$}%
\advance #1 by\tempdimen \divide #1 by\unitlength }

\def\putvector(#1,#2)(#3,#4)#5#6{{%
\ifnum3<\arrowtype \putdashvector(#1,#2)(#3,#4)#5\arrowtype \else
\ifnum\arrowtype<-3 \putdashvector(#1,#2)(#3,#4)#5\arrowtype \else
\xpos=#1 \ypos=#2 \run=#3 \rise=#4 \arrowlength=#5 \ifnum
\arrowtype<0
    \ifnum \run=0
        \advance \ypos by-\arrowlength
    \else
        \tempcounta \arrowlength
        \multiply \tempcounta by\rise
        \divide \tempcounta by\run
        \ifnum\run>0
            \advance \xpos by\arrowlength
            \advance \ypos by\tempcounta
        \else
            \advance \xpos by-\arrowlength
            \advance \ypos by-\tempcounta
        \fi
    \fi
    \multiply \arrowtype by-1
    \multiply \rise by-1
    \multiply \run by-1
\fi \ifcase \arrowtype
\or \put(\xpos,\ypos){\vector(\run,\rise){\arrowlength}}%
\or \put(\xpos,\ypos){\mvector(\run,\rise)\arrowlength}%
\or \put(\xpos,\ypos){\evector(\run,\rise){\arrowlength}}%
\fi\fi\fi }}

\def\putsplitvector(#1,#2)#3#4{
\xpos #1 \ypos #2 \arrowtype #4 \halflength #3 \arrowlength #3
\gap 140 \advance \halflength by-\gap \divide \halflength by2
\ifnum\arrowtype>0
   \ifcase \arrowtype
   \or \put(\xpos,\ypos){\line(0,-1){\halflength}}%
       \advance\ypos by-\halflength
       \advance\ypos by-\gap
       \put(\xpos,\ypos){\vector(0,-1){\halflength}}%
   \or \put(\xpos,\ypos){\line(0,-1)\halflength}%
       \put(\xpos,\ypos){\vector(0,-1)3}%
       \advance\ypos by-\halflength
       \advance\ypos by-\gap
       \put(\xpos,\ypos){\vector(0,-1){\halflength}}%
   \or \put(\xpos,\ypos){\line(0,-1)\halflength}%
       \advance\ypos by-\halflength
       \advance\ypos by-\gap
       \put(\xpos,\ypos){\evector(0,-1){\halflength}}%
   \fi
\else \arrowtype=-\arrowtype
   \ifcase\arrowtype
   \or \advance \ypos by-\arrowlength
       \put(\xpos,\ypos){\line(0,1){\halflength}}%
       \advance\ypos by\halflength
       \advance\ypos by\gap
       \put(\xpos,\ypos){\vector(0,1){\halflength}}%
   \or \advance \ypos by-\arrowlength
       \put(\xpos,\ypos){\line(0,1)\halflength}%
       \put(\xpos,\ypos){\vector(0,1)3}%
       \advance\ypos by\halflength
       \advance\ypos by\gap
       \put(\xpos,\ypos){\vector(0,1){\halflength}}%
   \or \advance \ypos by-\arrowlength
       \put(\xpos,\ypos){\line(0,1)\halflength}%
       \advance\ypos by\halflength
       \advance\ypos by\gap
       \put(\xpos,\ypos){\evector(0,1){\halflength}}%
   \fi
\fi }

\def\putmorphism(#1)(#2,#3)[#4`#5`#6]#7#8#9{{%
\run #2 \rise #3 \ifnum\rise=0
  \puthmorphism(#1)[#4`#5`#6]{#7}{#8}#9%
\else\ifnum\run=0
  \putvmorphism(#1)[#4`#5`#6]{#7}{#8}#9%
\else
\setpos(#1)%
\arrowlength #7 \arrowtype #8 \ifnum\run=0 \else\ifnum\rise=0
\else \ifnum\run>0
    \coefa=1
\else
   \coefa=-1
\fi \ifnum\arrowtype>0
   \coefb=0
   \coefc=-1
\else
   \coefb=\coefa
   \coefc=1
   \arrowtype=-\arrowtype
\fi \width=2 \multiply \width by\run \divide \width by\rise
\ifnum \width<0  \width=-\width\fi \advance\width by60 \if l#9
\width=-\width\fi
\putbox(\xpos,\ypos){#4}
{\multiply \coefa by\arrowlength
\advance\xpos by\coefa \multiply \coefa by\rise \divide \coefa
by\run \advance \ypos by\coefa
\putbox(\xpos,\ypos){#5} }%
{\multiply \coefa by\arrowlength
\divide \coefa by2 \advance \xpos by\coefa \advance \xpos by\width
\multiply \coefa by\rise \divide \coefa by\run \advance \ypos
by\coefa
\if l#9%
   \putrbox(\xpos,\ypos){#6}%
\else\if r#9%
   \putlbox(\xpos,\ypos){#6}%
\fi\fi }%
{\multiply \rise by-\coefc
\multiply \run by-\coefc \multiply \coefb by\arrowlength \advance
\xpos by\coefb \multiply \coefb by\rise \divide \coefb by\run
\advance \ypos by\coefb \multiply \coefc by70 \advance \ypos
by\coefc \multiply \coefc by\run \divide \coefc by\rise \advance
\xpos by\coefc \multiply \coefa by140 \multiply \coefa by\run
\divide \coefa by\rise \advance \arrowlength by\coefa
\ifcase\arrowtype
\or \put(\xpos,\ypos){\vector(\run,\rise){\arrowlength}}%
\or \put(\xpos,\ypos){\mvector(\run,\rise){\arrowlength}}%
\or \put(\xpos,\ypos){\evector(\run,\rise){\arrowlength}}%
\fi}\fi\fi\fi\fi}}

\newcount\numbdashes \newcount\lengthdash \newcount\increment

\def\howmanydashes{
\numbdashes=\arrowlength \lengthdash=40 \divide\numbdashes by
\lengthdash \lengthdash=\arrowlength \divide\lengthdash by
\numbdashes
\increment=\lengthdash \multiply\lengthdash by 3
\divide\lengthdash by 5 }

\def\putdashvector(#1)(#2,#3)#4#5{%
\ifnum#3=0 \putdashhvector(#1){#4}#5 \else \ifnum#2=0
\putdashvvector(#1){#4}#5\fi\fi}

\def\putdashhvector(#1,#2)#3#4{{%
\arrowlength=#3 \howmanydashes
\multiput(#1,#2)(\increment,0){\numbdashes}%
{\vrule height .4pt width \lengthdash\unitlength} \arrowtype=#4
\xpos=#1 \ifnum\arrowtype<0 \advance\arrowtype by 7 \fi
\ifcase\arrowtype \or \advance\xpos by 10
    \put(\xpos,#2){\vector(-1,0){\lengthdash}}
    \advance\xpos by 40
    \put(\xpos,#2){\vector(-1,0){\lengthdash}}
\or \advance \xpos by 10
    \put(\xpos,#2){\vector(-1,0){\lengthdash}}
    \advance\xpos by  \arrowlength
    \advance\xpos by  -50
    \put(\xpos,#2){\vector(-1,0){\lengthdash}}
\or \advance\xpos by 10
    \put(\xpos,#2){\vector(-1,0){\lengthdash}}
\or \advance\xpos by \arrowlength
    \advance\xpos by -\lengthdash
    \put(\xpos,#2){\vector(1,0){\lengthdash}}
\or {\advance\xpos by 10
    \put(\xpos,#2){\vector(1,0){\lengthdash}}}
    \advance\xpos by \arrowlength
    \advance\xpos by -\lengthdash
    \put(\xpos,#2){\vector(1,0){\lengthdash}}
\or \advance\xpos by \arrowlength
    \advance\xpos by -\lengthdash
    \put(\xpos,#2){\vector(1,0){\lengthdash}}
    \advance\xpos by -40
    \put(\xpos,#2){\vector(1,0){\lengthdash}}
   \fi
}}

\def\putdashvvector(#1,#2)#3#4{{%
\arrowlength=#3 \howmanydashes \ypos=#2 \advance\ypos by
-\arrowlength
\multiput(#1,#2)(0,\increment){\numbdashes}%
    {\vrule width .4pt height \lengthdash\unitlength}
\arrowtype=#4 \ypos=#2 \ifnum\arrowtype<0 \advance\arrowtype by 7
\fi \ifcase\arrowtype \or \advance\ypos by \arrowlength
\advance\ypos by -40
    \put(#1,\ypos){\vector(0,1){\lengthdash}}
    \advance\ypos by -40
    \put(#1,\ypos){\vector(0,1){\lengthdash}}
\or \advance\ypos by 10
    \put(#1,\ypos){\vector(0,1){\lengthdash}}
    \advance\ypos by \arrowlength \advance\ypos by -40
    \put(#1,\ypos){\vector(0,1){\lengthdash}}
\or \advance\ypos by \arrowlength \advance\ypos by -40
    \put(#1,\ypos){\vector(0,1){\lengthdash}}
\or \advance\ypos by 10
    \put(#1,\ypos){\vector(0,-1){\lengthdash}}
\or \advance\ypos by 10
    \put(#1,\ypos){\vector(0,-1){\lengthdash}}
    \advance\ypos by \arrowlength \advance\ypos by -40
    \put(#1,\ypos){\vector(0,-1){\lengthdash}}
\or \advance\ypos by 10
    \put(#1,\ypos){\vector(0,-1){\lengthdash}}
    \advance\ypos by 40
    \put(#1,\ypos){\vector(0,-1){\lengthdash}}
\fi }}

\def\puthmorphism(#1,#2)[#3`#4`#5]#6#7#8{{%
\xpos #1 \ypos #2 \width #6 \arrowlength #6 \arrowtype=#7
\putbox(\xpos,\ypos){#3\vphantom{#4}}%
{\advance \xpos by\arrowlength
\putbox(\xpos,\ypos){\vphantom{#3}#4}}%
\horsize{\tempcounta}{#3}%
\horsize{\tempcountb}{#4}%
\divide \tempcounta by2 \divide \tempcountb by2 \advance
\tempcounta by30 \advance \tempcountb by30 \advance \xpos
by\tempcounta \advance \arrowlength by-\tempcounta \advance
\arrowlength by-\tempcountb
\putvector(\xpos,\ypos)(1,0)\arrowlength\arrowtype \divide
\arrowlength by2 \advance \xpos by\arrowlength
\vertsize{\tempcounta}{#5}%
\divide\tempcounta by2 \advance \tempcounta by20
\if a#8 %
   \advance \ypos by\tempcounta
   \putbox(\xpos,\ypos){#5}%
\else
   \advance \ypos by-\tempcounta
   \putbox(\xpos,\ypos){#5}%
\fi}}

\def\putvmorphism(#1,#2)[#3`#4`#5]#6#7#8{{%
\xpos #1 \ypos #2 \arrowlength #6 \arrowtype #7
\settowidth{\xlen}{$#5$}%
\putbox(\xpos,\ypos){#3}%
{\advance \ypos by-\arrowlength
\putbox(\xpos,\ypos){#4}}%
{\advance\arrowlength by-140 \advance \ypos by-70 \ifdim\xlen>0pt
   \if m#8%
      \putsplitvector(\xpos,\ypos)\arrowlength\arrowtype
   \else
   \putvector(\xpos,\ypos)(0,-1)\arrowlength\arrowtype
   \fi
\else
   \putvector(\xpos,\ypos)(0,-1)\arrowlength\arrowtype
\fi}%
\ifdim\xlen>0pt
   \divide \arrowlength by2
   \advance\ypos by-\arrowlength
   \if l#8%
      \advance \xpos by-40
      \putrbox(\xpos,\ypos){#5}%
   \else\if r#8%
      \advance \xpos by40
      \putlbox(\xpos,\ypos){#5}%
   \else
      \putbox(\xpos,\ypos){#5}%
   \fi\fi
\fi }}

\def\putsquarep<#1>(#2)[#3;#4`#5`#6`#7]{{%
\setsqparms[#1]%
\setpos(#2)%
\settokens`#3`%
\puthmorphism(\xpos,\ypos)[\tokenc`\tokend`{#7}]{\width}{\arrowtyped}b%
\advance\ypos by \height
\puthmorphism(\xpos,\ypos)[\tokena`\tokenb`{#4}]{\width}{\arrowtypea}a%
\putvmorphism(\xpos,\ypos)[``{#5}]{\height}{\arrowtypeb}l%
\advance\xpos by \width
\putvmorphism(\xpos,\ypos)[``{#6}]{\height}{\arrowtypec}r%
}}

\def\putsquare{\@ifnextchar <{\putsquarep}{\putsquarep%
   <\arrowtypea`\arrowtypeb`\arrowtypec`\arrowtyped;\width`\height>}}
\def\square{\@ifnextchar< {\squarep}{\squarep
   <\arrowtypea`\arrowtypeb`\arrowtypec`\arrowtyped;\width`\height>}}
\def\squarep<#1>[#2`#3`#4`#5;#6`#7`#8`#9]{{
\setsqparms[#1]
\diagram
\putsquarep<\arrowtypea`\arrowtypeb`\arrowtypec`
\arrowtyped;\width`\height>
(0,0)[#2`#3`#4`{#5};#6`#7`#8`{#9}]
\enddiagram
}}                                                 
\def\putptrianglep<#1>(#2,#3)[#4`#5`#6;#7`#8`#9]{{%
\settriparms[#1]%
\xpos=#2 \ypos=#3 \advance\ypos by \height
\puthmorphism(\xpos,\ypos)[#4`#5`{#7}]{\height}{\arrowtypea}a%
\putvmorphism(\xpos,\ypos)[`#6`{#8}]{\height}{\arrowtypeb}l%
\advance\xpos by\height
\putmorphism(\xpos,\ypos)(-1,-1)[``{#9}]{\height}{\arrowtypec}r%
}}

\def\putptriangle{\@ifnextchar <{\putptrianglep}{\putptrianglep
   <\arrowtypea`\arrowtypeb`\arrowtypec;\height>}}
\def\ptriangle{\@ifnextchar <{\ptrianglep}{\ptrianglep
   <\arrowtypea`\arrowtypeb`\arrowtypec;\height>}}
\def\ptrianglep<#1>[#2`#3`#4;#5`#6`#7]{{
\settriparms[#1]
\diagram
\putptrianglep<\arrowtypea`\arrowtypeb`
\arrowtypec;\height>
(0,0)[#2`#3`#4;#5`#6`{#7}]
\enddiagram
}}                                            

\def\putqtrianglep<#1>(#2,#3)[#4`#5`#6;#7`#8`#9]{{%
\settriparms[#1]%
\xpos=#2 \ypos=#3 \advance\ypos by\height
\puthmorphism(\xpos,\ypos)[#4`#5`{#7}]{\height}{\arrowtypea}a%
\putmorphism(\xpos,\ypos)(1,-1)[``{#8}]{\height}{\arrowtypeb}l%
\advance\xpos by\height
\putvmorphism(\xpos,\ypos)[`#6`{#9}]{\height}{\arrowtypec}r%
}}

\def\putqtriangle{\@ifnextchar <{\putqtrianglep}{\putqtrianglep
   <\arrowtypea`\arrowtypeb`\arrowtypec;\height>}}
\def\qtriangle{\@ifnextchar <{\qtrianglep}{\qtrianglep
   <\arrowtypea`\arrowtypeb`\arrowtypec;\height>}}
\def\qtrianglep<#1>[#2`#3`#4;#5`#6`#7]{{
\settriparms[#1]
\width=\height                                
\diagram
\putqtrianglep<\arrowtypea`\arrowtypeb`
\arrowtypec;\height>
(0,0)[#2`#3`#4;#5`#6`{#7}]
\enddiagram
}}

\def\putdtrianglep<#1>(#2,#3)[#4`#5`#6;#7`#8`#9]{{%
\settriparms[#1]%
\xpos=#2 \ypos=#3
\puthmorphism(\xpos,\ypos)[#5`#6`{#9}]{\height}{\arrowtypec}b%
\advance\xpos by \height \advance\ypos by\height
\putmorphism(\xpos,\ypos)(-1,-1)[``{#7}]{\height}{\arrowtypea}l%
\putvmorphism(\xpos,\ypos)[#4``{#8}]{\height}{\arrowtypeb}r%
}}

\def\putdtriangle{\@ifnextchar <{\putdtrianglep}{\putdtrianglep
   <\arrowtypea`\arrowtypeb`\arrowtypec;\height>}}
\def\dtriangle{\@ifnextchar <{\dtrianglep}{\dtrianglep
   <\arrowtypea`\arrowtypeb`\arrowtypec;\height>}}
\def\dtrianglep<#1>[#2`#3`#4;#5`#6`#7]{{
\settriparms[#1]
\width=\height                                
\diagram
\putdtrianglep<\arrowtypea`\arrowtypeb`
\arrowtypec;\height>
(0,0)[#2`#3`#4;#5`#6`{#7}]
\enddiagram
}}

\def\putbtrianglep<#1>(#2,#3)[#4`#5`#6;#7`#8`#9]{{%
\settriparms[#1]%
\xpos=#2 \ypos=#3
\puthmorphism(\xpos,\ypos)[#5`#6`{#9}]{\height}{\arrowtypec}b%
\advance\ypos by\height
\putmorphism(\xpos,\ypos)(1,-1)[``{#8}]{\height}{\arrowtypeb}r%
\putvmorphism(\xpos,\ypos)[#4``{#7}]{\height}{\arrowtypea}l%
}}

\def\putbtriangle{\@ifnextchar <{\putbtrianglep}{\putbtrianglep
   <\arrowtypea`\arrowtypeb`\arrowtypec;\height>}}
\def\btriangle{\@ifnextchar <{\btrianglep}{\btrianglep
   <\arrowtypea`\arrowtypeb`\arrowtypec;\height>}}
\def\btrianglep<#1>[#2`#3`#4;#5`#6`#7]{{
\settriparms[#1]
\width=\height                               
\diagram
\putbtrianglep<\arrowtypea`\arrowtypeb`
\arrowtypec;\height>
(0,0)[#2`#3`#4;#5`#6`{#7}]
\enddiagram
}}

\def\putAtrianglep<#1>(#2,#3)[#4`#5`#6;#7`#8`#9]{{%
\settriparms[#1]%
\xpos=#2 \ypos=#3 {\multiply \height by2
\puthmorphism(\xpos,\ypos)[#5`#6`{#9}]{\height}{\arrowtypec}b}%
\advance\xpos by\height \advance\ypos by\height
\putmorphism(\xpos,\ypos)(-1,-1)[#4``{#7}]{\height}{\arrowtypea}l%
\putmorphism(\xpos,\ypos)(1,-1)[``{#8}]{\height}{\arrowtypeb}r%
}}

\def\putAtriangle{\@ifnextchar <{\putAtrianglep}{\putAtrianglep
   <\arrowtypea`\arrowtypeb`\arrowtypec;\height>}}
\def\Atriangle{\@ifnextchar <{\Atrianglep}{\Atrianglep
   <\arrowtypea`\arrowtypeb`\arrowtypec;\height>}}
\def\Atrianglep<#1>[#2`#3`#4;#5`#6`#7]{{
\settriparms[#1]
\width=\height                                     
\diagram
\putAtrianglep<\arrowtypea`\arrowtypeb`
\arrowtypec;\height>
(0,0)[#2`#3`#4;#5`#6`{#7}]
\enddiagram
}}

\def\putAtrianglepairp<#1>(#2)[#3;#4`#5`#6`#7`#8]{{%
\settripairparms[#1]%
\setpos(#2)%
\settokens`#3`%
\puthmorphism(\xpos,\ypos)[\tokenb`\tokenc`{#7}]{\height}{\arrowtyped}b%
\advance\xpos by\height
\puthmorphism(\xpos,\ypos)[\phantom{\tokenc}`\tokend`{#8}]%
{\height}{\arrowtypee}b%
\advance\ypos by\height
\putmorphism(\xpos,\ypos)(-1,-1)[\tokena``{#4}]{\height}{\arrowtypea}l%
\putvmorphism(\xpos,\ypos)[``{#5}]{\height}{\arrowtypeb}m%
\putmorphism(\xpos,\ypos)(1,-1)[``{#6}]{\height}{\arrowtypec}r%
}}

\def\putAtrianglepair{\@ifnextchar <{\putAtrianglepairp}{\putAtrianglepairp%
   <\arrowtypea`\arrowtypeb`\arrowtypec`\arrowtyped`\arrowtypee;\height>}}
\def\Atrianglepair{\@ifnextchar <{\Atrianglepairp}{\Atrianglepairp%
   <\arrowtypea`\arrowtypeb`\arrowtypec`\arrowtyped`\arrowtypee;\height>}}

\def\Atrianglepairp<#1>[#2;#3`#4`#5`#6`#7]{{
\settripairparms[#1]
\settokens`#2`
\width=\height                                
\diagram
\putAtrianglepairp                            
<\arrowtypea`\arrowtypeb`\arrowtypec`
\arrowtyped`\arrowtypee;\height>
(0,0)[{#2};#3`#4`#5`#6`{#7}]
\enddiagram
}}

\def\putVtrianglep<#1>(#2,#3)[#4`#5`#6;#7`#8`#9]{{%
\settriparms[#1]%
\xpos=#2 \ypos=#3 \advance\ypos by\height {\multiply\height by2
\puthmorphism(\xpos,\ypos)[#4`#5`{#7}]{\height}{\arrowtypea}a}%
\putmorphism(\xpos,\ypos)(1,-1)[`#6`{#8}]{\height}{\arrowtypeb}l%
\advance\xpos by\height \advance\xpos by\height
\putmorphism(\xpos,\ypos)(-1,-1)[``{#9}]{\height}{\arrowtypec}r%
}}

\def\putVtriangle{\@ifnextchar <{\putVtrianglep}{\putVtrianglep
   <\arrowtypea`\arrowtypeb`\arrowtypec;\height>}}
\def\Vtriangle{\@ifnextchar <{\Vtrianglep}{\Vtrianglep
   <\arrowtypea`\arrowtypeb`\arrowtypec;\height>}}
\def\Vtrianglep<#1>[#2`#3`#4;#5`#6`#7]{{
\settriparms[#1]
\width=\height                                 
\diagram
\putVtrianglep<\arrowtypea`\arrowtypeb`
\arrowtypec;\height>
(0,0)[#2`#3`#4;#5`#6`{#7}]
\enddiagram
}}

\def\putVtrianglepairp<#1>(#2)[#3;#4`#5`#6`#7`#8]{{
\settripairparms[#1]%
\setpos(#2)%
\settokens`#3`%
\advance\ypos by\height
\putmorphism(\xpos,\ypos)(1,-1)[`\tokend`{#6}]{\height}{\arrowtypec}l%
\puthmorphism(\xpos,\ypos)[\tokena`\tokenb`{#4}]{\height}{\arrowtypea}a%
\advance\xpos by\height
\puthmorphism(\xpos,\ypos)[\phantom{\tokenb}`\tokenc`{#5}]%
{\height}{\arrowtypeb}a%
\putvmorphism(\xpos,\ypos)[``{#7}]{\height}{\arrowtyped}m%
\advance\xpos by\height
\putmorphism(\xpos,\ypos)(-1,-1)[``{#8}]{\height}{\arrowtypee}r%
}}

\def\putVtrianglepair{\@ifnextchar <{\putVtrianglepairp}{\putVtrianglepairp%
    <\arrowtypea`\arrowtypeb`\arrowtypec`\arrowtyped`\arrowtypee;\height>}}
\def\Vtrianglepair{\@ifnextchar <{\Vtrianglepairp}{\Vtrianglepairp%
    <\arrowtypea`\arrowtypeb`\arrowtypec`\arrowtyped`\arrowtypee;\height>}}
\def\Vtrianglepairp<#1>[#2;#3`#4`#5`#6`#7]{{
\settripairparms[#1]
\settokens`#2`
\diagram
\putVtrianglepairp                             
<\arrowtypea`\arrowtypeb`\arrowtypec`
\arrowtyped`\arrowtypee;\height>
(0,0)[{#2};#3`#4`#5`#6`{#7}]
\enddiagram
}}

\def\putCtrianglep<#1>(#2,#3)[#4`#5`#6;#7`#8`#9]{{%
\settriparms[#1]%
\xpos=#2 \ypos=#3 \advance\ypos by\height
\putmorphism(\xpos,\ypos)(1,-1)[``{#9}]{\height}{\arrowtypec}l%
\advance\xpos by\height \advance\ypos by\height
\putmorphism(\xpos,\ypos)(-1,-1)[#4`#5`{#7}]{\height}{\arrowtypea}l%
{\multiply\height by 2
\putvmorphism(\xpos,\ypos)[`#6`{#8}]{\height}{\arrowtypeb}r}%
}}

\def\putCtriangle{\@ifnextchar <{\putCtrianglep}{\putCtrianglep
    <\arrowtypea`\arrowtypeb`\arrowtypec;\height>}}
\def\Ctriangle{\@ifnextchar <{\Ctrianglep}{\Ctrianglep
    <\arrowtypea`\arrowtypeb`\arrowtypec;\height>}}
\def\Ctrianglep<#1>[#2`#3`#4;#5`#6`#7]{{
\settriparms[#1]
\width=\height                               
\diagram
\putCtrianglep<\arrowtypea`\arrowtypeb`
\arrowtypec;\height>
(0,0)[#2`#3`#4;#5`#6`{#7}]
\enddiagram
}}                                           
\def\putDtrianglep<#1>(#2,#3)[#4`#5`#6;#7`#8`#9]{{%
\settriparms[#1]%
\xpos=#2 \ypos=#3 \advance\xpos by\height \advance\ypos by\height
\putmorphism(\xpos,\ypos)(-1,-1)[``{#9}]{\height}{\arrowtypec}r%
\advance\xpos by-\height \advance\ypos by\height
\putmorphism(\xpos,\ypos)(1,-1)[`#5`{#8}]{\height}{\arrowtypeb}r%
{\multiply\height by 2
\putvmorphism(\xpos,\ypos)[#4`#6`{#7}]{\height}{\arrowtypea}l}%
}}

\def\putDtriangle{\@ifnextchar <{\putDtrianglep}{\putDtrianglep
    <\arrowtypea`\arrowtypeb`\arrowtypec;\height>}}
\def\Dtriangle{\@ifnextchar <{\Dtrianglep}{\Dtrianglep
   <\arrowtypea`\arrowtypeb`\arrowtypec;\height>}}
\def\Dtrianglep<#1>[#2`#3`#4;#5`#6`#7]{{
\settriparms[#1]
\width=\height                              
\diagram
\putDtrianglep<\arrowtypea`\arrowtypeb`
\arrowtypec;\height>
(0,0)[#2`#3`#4;#5`#6`{#7}]
\enddiagram
}}                                          
\def\setrecparms[#1`#2]{\width=#1 \height=#2}%

\def\recursep<#1`#2>[#3;#4`#5`#6`#7`#8]{{\m@th
\width=#1 \height=#2 \settokens`#3`
\settowidth{\tempdimen}{$\tokena$} \ifdim\tempdimen=0pt
  \savebox{\tempboxa}{\hbox{$\tokenb$}}%
  \savebox{\tempboxb}{\hbox{$\tokend$}}%
  \savebox{\tempboxc}{\hbox{$#6$}}%
\else
  \savebox{\tempboxa}{\hbox{$\hbox{$\tokena$}\times\hbox{$\tokenb$}$}}%
  \savebox{\tempboxb}{\hbox{$\hbox{$\tokena$}\times\hbox{$\tokend$}$}}%
  \savebox{\tempboxc}{\hbox{$\hbox{$\tokena$}\times\hbox{$#6$}$}}%
\fi \ypos=\height \divide\ypos by 2 \xpos=\ypos \advance\xpos by
\width \bfig
\putCtrianglep<-1`1`1;\ypos>(0,0)[`\tokenc`;#5`#6`{#7}]%
\puthmorphism(\ypos,0)[\tokend`\usebox{\tempboxb}`{#8}]{\width}{-1}b%
\puthmorphism(\ypos,\height)[\tokenb`\usebox{\tempboxa}`{#4}]{\width}{-1}a%
\advance\ypos by \width
\putvmorphism(\ypos,\height)[``\usebox{\tempboxc}]{\height}1r%
\efig }}

\def\recurse{\@ifnextchar <{\recursep}{\recursep<\width`\height>}}

\def\puttwohmorphisms(#1,#2)[#3`#4;#5`#6]#7#8#9{{%
\puthmorphism(#1,#2)[#3`#4`]{#7}0a \ypos=#2 \advance\ypos by 20
\puthmorphism(#1,\ypos)[\phantom{#3}`\phantom{#4}`#5]{#7}{#8}a
\advance\ypos by -40
\puthmorphism(#1,\ypos)[\phantom{#3}`\phantom{#4}`#6]{#7}{#9}b }}

\def\puttwovmorphisms(#1,#2)[#3`#4;#5`#6]#7#8#9{{%
\putvmorphism(#1,#2)[#3`#4`]{#7}0a \xpos=#1 \advance\xpos by -20
\putvmorphism(\xpos,#2)[\phantom{#3}`\phantom{#4}`#5]{#7}{#8}l
\advance\xpos by 40
\putvmorphism(\xpos,#2)[\phantom{#3}`\phantom{#4}`#6]{#7}{#9}r }}

\def\puthcoequalizer(#1)[#2`#3`#4;#5`#6`#7]#8#9{{%
\setpos(#1)%
\puttwohmorphisms(\xpos,\ypos)[#2`#3;#5`#6]{#8}11%
\advance\xpos by #8
\puthmorphism(\xpos,\ypos)[\phantom{#3}`#4`#7]{#8}1{#9} }}

\def\putvcoequalizer(#1)[#2`#3`#4;#5`#6`#7]#8#9{{%
\setpos(#1)%
\puttwovmorphisms(\xpos,\ypos)[#2`#3;#5`#6]{#8}11%
\advance\ypos by -#8
\putvmorphism(\xpos,\ypos)[\phantom{#3}`#4`#7]{#8}1{#9} }}

\def\putthreehmorphisms(#1)[#2`#3;#4`#5`#6]#7(#8)#9{{%
\setpos(#1) \settypes(#8)
\if a#9 %
     \vertsize{\tempcounta}{#5}%
     \vertsize{\tempcountb}{#6}%
     \ifnum \tempcounta<\tempcountb \tempcounta=\tempcountb \fi
\else
     \vertsize{\tempcounta}{#4}%
     \vertsize{\tempcountb}{#5}%
     \ifnum \tempcounta<\tempcountb \tempcounta=\tempcountb \fi
\fi \advance \tempcounta by 60
\puthmorphism(\xpos,\ypos)[#2`#3`#5]{#7}{\arrowtypeb}{#9}
\advance\ypos by \tempcounta
\puthmorphism(\xpos,\ypos)[\phantom{#2}`\phantom{#3}`#4]{#7}{\arrowtypea}{#9}
\advance\ypos by -\tempcounta \advance\ypos by -\tempcounta
\puthmorphism(\xpos,\ypos)[\phantom{#2}`\phantom{#3}`#6]{#7}{\arrowtypec}{#9}
}}

\def\setarrowtoks[#1`#2`#3`#4`#5`#6]{%
\def\toka{#1}
\def\tokb{#2}
\def\tokc{#3}
\def\tokd{#4}
\def\toke{#5}
\def\tokf{#6}
}
\def\hex{\@ifnextchar <{\hexp}{\hexp<1000`400>}}
\def\hexp<#1`#2>[#3`#4`#5`#6`#7`#8;#9]{%
\setarrowtoks[#9] \yext=#2 \advance \yext by #2 \xext=#1
\advance\xext by \yext \bfig
\putCtriangle<-1`0`1;#2>(0,0)[`#5`;\tokb``\tokd] \xext=#1
\yext=#2 \advance \yext by #2
\putsquare<1`0`0`1;\xext`\yext>(#2,0)[#3`#4`#7`#8;\toka```\tokf]
\advance \xext by #2
\putDtriangle<0`1`-1;#2>(\xext,0)[`#6`;`\tokc`\toke] \efig }

\makeatother

\input{tcilatex}
\begin{document}

\title{Undergraduate Lecture Notes in\\ Topological Quantum Field Theory}
\author{Vladimir G. Ivancevic\thanks{%
Land Operations Division, Defence Science \& Technology Organisation, P.O.
Box 1500, Edinburgh SA 5111, Australia
(Vladimir.Ivancevic@dsto.defence.gov.au)} \and Tijana T. Ivancevic\thanks{%
School of Electrical and Information Engineering, University of South
Australia, Mawson Lakes, S.A. 5095, Australia (Tijana.Ivancevic@unisa.edu.au)%
}}
\date{}
\maketitle

\begin{abstract}
These third--year lecture notes are designed for a 1--semester
course in topological quantum field theory (TQFT). Assumed background in
mathematics and physics are only standard second--year subjects: multivariable
calculus, introduction to quantum mechanics and basic electromagnetism.\bigskip

\noindent{\bf Keywords:} quantum mechanics/field theory, path integral, Hodge decomposition,\\ Chern--Simons and Yang--Mills gauge theories, Conformal field theory
\end{abstract}

\tableofcontents

\bigbreak\bigbreak

\section{Introduction}

There is a number of good textbooks in quantum field theory (QFT, see \cite%
{bjorken,itzykson,ramond,peskin,weinberg,deligne,zee,dewitt,nair,srednicki}.
However, they are all designed for the graduate-level study and we can only
hope that undergraduate students can read some easy parts of them. Moreover,
there are certainly no undergraduate-level textbooks for TQFT, so pure
students are forced to \emph{try to read} the original papers from its
inventors, Ed Witten \cite{WittenTQFT,Witten89} and Michael Atiyah \cite{Atiyah}. The
goal of the present lecture notes is to try to fill in this gap, to give the
talented undergraduates the very first glimpse of the mathematical physics
of the XXI Century.

Throughout these lecture notes we will use the following \emph{conventions}:\\ (i) natural units, in which (some or all of) the following definitions are used: $c=\hbar =m=1$;\\ (ii) $\mathrm{i}=\sqrt{-1}$, ~~$\dot{z}=dz/dt$, ~~$\partial
_{z}=\partial /\partial z$;\\ (iii) Einstein's summation convention
over repeated indices, ~while $n$D means $n-$dimensional.

\subsection{Basics of non-relativistic quantum mechanics}

Recall that quantum theory was born with Max Planck's 1900 paper, in which he derived the correct shape of the
black-body spectrum which now bears his name, eliminating the ultraviolet
catastrophe -- with the price of introducing a `bizarre assumption' (today called \textit{Planck's quantum hypothesis}) that energy was only emitted in certain finite chunks, or `quanta'. In 1905, Albert Einstein took this bold idea one step further. Assuming that radiation could only transport energy in such chunks, `photons', he was able to explain the so-called \textit{photoelectric effect}. In 1913, Niels Bohr made a new breakthrough by postulating that the amount of angular momentum in
an atom was quantized, so that the electrons were confined to a discrete set of
orbits, each with a definite energy. If the electron jumped from one orbit
to a lower one, the energy difference was sent off in the form of a photon.
If the electron was in the innermost allowed orbit, there were no orbits
with less energy to jump to, so the atom was stable. In addition, Bohr's
theory successfully explained a slew of spectral lines that had been
measured for Hydrogen. The famous \textit{wave--particle duality of matter} was proposed by French prince Louis de Broglie in 1923 in his Ph.D. thesis: that
electrons and other particles acted like standing waves. Such waves, like
vibrations of a guitar string, can only occur with certain discrete
(quantized) frequencies.\footnote{The idea was so new that the examining committee
went outside its circle for advice on the acceptability of the thesis.
Einstein gave a favorable opinion and the thesis was accepted.} In November 1925, Erwin Schr\"{o}dinger gave a seminar on de Broglie's work in Zurich.
When he was finished, P. Debye said in effect, \textquotedblleft You speak
about waves. But where is the wave equation?\textquotedblright\ Schr\"{o}%
dinger went on to produce and publish his famous wave equation, the master
key for so much of modern physics. An equivalent formulation involving infinite
matrices was provided by Werner Heisenberg, Max Born and Pasquale Jordan
around the same time. With this new powerful mathematical underpinning,
quantum theory made explosive progress. Within a few years, a host of
hitherto unexplained measurements had been successfully explained, including
spectra of more complicated atoms and various numbers describing properties
of chemical reactions. For more details, see, e.g.
\cite{TegmarkWheeler}.

\subsubsection{Quantum states and operators}

Non-relativistic quantum-mechanical systems have two modes of evolution in
time \cite{DiracBook,QuLeap}. The first, governed by standard, \textit{%
time--dependent Schr\"{o}dinger equation}:
\begin{equation}
\mathrm{i}\hbar \,\partial _{t}\left\vert \psi \rangle \right. =\hat{H}%
\left\vert \psi \rangle \right.,  \label{Sch}
\end{equation}%
describes the time evolution of quantum systems when they are undisturbed by
measurements. `Measurements' are defined as \emph{interactions} of the
quantum system with its classical environment. As long as the system is
sufficiently isolated from the environment, it follows Schr\"{o}dinger
equation. If an interaction with the environment takes place, i.e., a
measurement is performed, the system abruptly \emph{decoheres} i.e.,
collapses or reduces to one of its classically allowed states.

A \emph{time--dependent state of a quantum system} is determined by a
normalized, complex, \textit{wave psi--function} $\psi =\psi (t)$. In
Dirac's words \cite{DiracBook}, this is a unit \emph{ket} vector $|\psi
\rangle $, which is an element of the \textit{Hilbert space} $L^{2}(\psi
)\equiv \mathcal{H}$, with a coordinate basis $(q^{i})$.\footnote{%
The family of all possible states ($|\psi \rangle ,|\phi \rangle $, etc.) of
a quantum system confiture what is known as a \textit{Hilbert space}. It is
a \textit{complex vector space}, which means that can perform the
complex--number--weighted combinations that we considered before for quantum
states. If $|\psi \rangle $ and $|\phi \rangle $ are both elements of the
Hilbert space, then so also is $w|\psi \rangle +~z|\phi \rangle ,$ for any
pair of complex numbers $w$ and $\ z.$ Here, we even alow $w=z=0$, to give
the element $\mathbf{0}$ of the Hilbert space, which does not represent a
possible physical state. We have the normal algebraic rules for a vector
space:
\begin{eqnarray*}
&&|\psi \rangle +|\phi \rangle =|\phi \rangle +|\psi \rangle , \\
&&|\psi \rangle +(|\phi \rangle +|\chi \rangle )=(|\psi \rangle +|\phi
\rangle )+|\chi \rangle , \\
&&w(z|\psi \rangle )=(wz)|\psi \rangle , \\
&&(w+z)|\psi \rangle =w|\psi \rangle +z|\psi \rangle , \\
&&z(|\psi \rangle +|\phi \rangle )=z|\psi \rangle +z|\phi \rangle \\
&&0|\psi \rangle =\mathbf{0},\qquad z\mathbf{0}=\mathbf{0}.
\end{eqnarray*}%
\par
A Hilbert space can sometimes have a finite number of dimensions, as in the
case of the spin states of a particle. For spin $\frac{1}{2},$ the Hilbert
space is just 2D, its elements being the complex linear combinations of the
two states $|\uparrow \rangle $ and $|\downarrow \rangle .$ For spin $\frac{1%
}{2}n,$ the Hilbert space is $(n+1)$D. However, sometimes the Hilbert space
can have an infinite number of dimensions, as e.g., the states of position
or momentum of a particle. Here, each alternative position (or momentum)
that the particle might have counts as providing a separate dimension for
the Hilbert space. The general state describing the quantum location (or
momentum) of the particle is a complex--number superposition of all these
different individual positions (or momenta), which is the wave $\psi -$%
function for the particle.
\par
Another property of the Hilbert space, crucial for quantum mechanics, is the
\textit{Hermitian inner (scalar) product}, which can be applied to any pair
of Hilbert--space vectors to produce a single complex number. To understand
how important the Hermitian inner product is for quantum mechanics, recall
that the Dirac's `bra--ket' notation is formulated on the its basis. If we
have the two quantum states (i.e., Hilbert--space vectors) are $|\psi
\rangle $ and $|\phi \rangle ,$ then their Hermitian scalar product is
denoted $\langle \psi |\phi \rangle ,$ and it satisfies a number of simple
algebraic properties:
\begin{eqnarray*}
&&\overline{\langle \psi |\phi \rangle }=\langle \phi |\psi \rangle ,\qquad
\text{(bar denotes complex--conjugate)} \\
&&\langle \psi |(|\phi \rangle +|\chi \rangle )=\langle \psi |\phi \rangle
+\langle \psi |\chi \rangle , \\
&&(z\langle \psi |)|\phi \rangle =z\langle \psi |\phi \rangle , \\
&&\langle \psi |\phi \rangle \,\geq \,0,\qquad \langle \psi |\phi \rangle
=0\quad \text{if\quad }|\psi \rangle =\mathbf{0}.
\end{eqnarray*}%
For example, probability of finding a quantum particle at a given location
is a \emph{squared length} $|\psi |^{2}$ of a Hilbert--space position vector
$|\psi \rangle $, which is the scalar product $\langle \psi |\psi \rangle $ $%
\ $of the vector $|\psi \rangle $ with itself. A \textit{normalized state}
is given by a Hilbert--space vector whose squared length is \emph{unity}.
\par
The second important thing that the Hermitian scalar product gives us is the
notion of \textit{orthogonality} between Hilbert--space vectors, which
occurs when the scalar product of the two vectors is \emph{zero}. In
ordinary terms, orthogonal states are things that are independent of one
another. The importance of this concept for quantum physics is that the
different alternative outcomes of any measurement are always orthogonal to
each other. For example, states $|\uparrow \rangle $ and $|\downarrow
\rangle $ are mutually orthogonal. Also, orthogonal are \emph{all} different
possible \emph{positions} that a quantum particle might be located in.} The
state ket--vector $|\psi (t)\rangle $ is subject to action of the Hermitian
operators, obtained by the procedure of \emph{quantization} of classical
biodynamic quantities, and whose real eigenvalues are being measured.

\emph{Quantum superposition} is a generalization of the algebraic principle
of linear combination of vectors. The Hilbert space has a set of states $%
\left\vert \varphi _{i}\rangle \right. $ (where the index i runs over the
degrees--of--freedom of the system) that form a basis and the most general
state of such a system can be written as $\left\vert \psi \rangle \right. $$%
=\sum_{i}c_{i}$ $\left\vert {\varphi _{i}}\rangle \right. $. The system is
said to be in a state $|\psi (t)\rangle ,$ describing the motion of the
\emph{de Broglie waves}, which is a linear superposition of the basis states
$\left\vert \varphi _{i}\rangle \right. $ with weighting coefficients $c_{i}$
that can in general be complex. At the microscopic or quantum level, the
state of the system is described by the wave function $\left\vert \psi
\rangle \right. $, which in general appears as a linear superposition of all
basis states. This can be interpreted as the system being in all these
states at once. The coefficients $c_{i}$ are called the \textit{probability
amplitudes} and $\left\vert c_{i}\right\vert ^{2}$ gives the probability
that $\left\vert \psi \rangle \right. $ will collapse into state $\left\vert
\varphi \rangle \right. $ when it decoheres (interacts with the
environment). By simple normalization we have the constraint that $%
\sum_{i}\left\vert c_{i}\right\vert ^{2}=1$. This emphasizes the fact that
the wavefunction describes a \emph{real, physical system}, which must be in
one of its allowable classical states and therefore by summing over all the
possibilities, weighted by their corresponding probabilities, one must get
unity. In other words, we have the \textit{normalization condition} for the
psi--function, determining the unit length of the state ket--vector
\begin{equation*}
\langle \psi (t)|\psi (t)\rangle \,=\int \psi ^{\ast }\psi \,\,dV=\int |\psi
|^{2}\,dV=1,
\end{equation*}%
where $\psi ^{\ast }=\,\langle \psi (t)|$ denotes the \emph{bra} vector, the
complex--conjugate to the ket $\psi =|\psi (t)\rangle ,$ and $\langle \psi
(t)|\psi (t)\rangle $ is their scalar product, i.e., Dirac \emph{bracket}.
For this reason the scene of quantum mechanics is the functional space of
square--integrable complex psi--functions, i.e., the Hilbert space $%
L^{2}(\psi )$.

When the system is in the state $|\psi (t)\rangle $, the average value $%
\langle f\rangle $ of any physical observable $f$ is equal to
\begin{equation*}
\langle f\rangle \,=\,\langle \psi (t)|\,\hat{f}\,|\psi (t)\rangle ,
\end{equation*}
where $\hat{f}$ is the Hermitian operator corresponding to $f$.

A quantum system is \emph{coherent} if it is in a linear superposition of
its basis states. If a measurement is performed on the system and this means
that the system must somehow interact with its environment, the
superposition is destroyed and the system is observed to be in only one
basis state, as required classically. This process is called \emph{reduction}
or \emph{collapse} of the wavefunction or simply \textit{decoherence} and is
governed by the form of the wavefunction $\left| \psi \rangle \right. $.

\emph{Entanglement}, on the other hand, is a purely quantum phenomenon and
has no classical analogue. It accounts for the ability of quantum systems to
exhibit correlations in counterintuitive `action--at--a--distance' ways.
Entanglement is what makes all the difference in the operation of quantum
computers versus classical ones. Entanglement gives `special powers' to
quantum computers because it gives quantum states the potential to exhibit
and maintain correlations that cannot be accounted for classically.
Correlations between bits are what make information encoding possible in
classical computers. For instance, we can require two bits to have the same
value thus encoding a relationship. If we are to subsequently change the
encoded information, we must change the correlated bits in tandem by
explicitly accessing each bit. Since quantum bits exist as superpositions,
\emph{correlations} between them also exist in superposition. When the
superposition is destroyed (e.g., one qubit is measured), the correct
correlations are \emph{instantaneously} `communicated' between the qubits
and this communication allows \emph{many qubits} to be accessed \emph{at once%
}, preserving their correlations, something that is absolutely impossible
classically.

More precisely, the \textit{first quantization} is a \emph{linear
representation} of all classical dynamical variables (like coordinate,
momentum, energy, or angular momentum) by linear \emph{Hermitian} operators
acting on the associated Hilbert state--space $\mathcal{H}$, which has the
following properties \cite{DiracBook}:

\begin{enumerate}
\item \emph{Linearity:}
~~~$
\alpha f+\beta g\rightarrow \alpha \,\hat{f}+\beta \,\hat{g},
$~~
(for all constants $\alpha ,\beta \in \mathbb{C}$);

\item A `dynamical' variable, equal to unity everywhere in the phase--space,
corresponds to unit operator: $1\rightarrow \hat{I}$; and

\item \emph{Classical Poisson brackets}
\begin{equation*}
\{f,g\}=\frac{{\partial }f}{{\partial }q^{i}}\frac{{\partial }g}{{\partial }%
p_{i}}-\frac{{\partial }f}{{\partial }p_{i}}\frac{{\partial }g}{{\partial }%
q^{i}}
\end{equation*}%
\emph{quantize} to the corresponding \emph{commutators}%
\begin{equation*}
\{f,g\}\rightarrow -\mathrm{i}\hbar \lbrack \hat{f},\hat{g}],\qquad \lbrack
\hat{f},\hat{g}]=\hat{f}\hat{g}-\hat{g}\hat{f}.
\end{equation*}
\end{enumerate}

Like Poisson bracket, commutator is bilinear and skew--symmetric operation,
satisfying Jacobi identity. For Hermitian operators $\hat{f},\hat{g}$ their
commutator $[\hat{f},\hat{g}]$ is anti--Hermitian; for this reason $i$ is
required in $\{f,g\}\rightarrow -\mathrm{i}\hbar \lbrack \hat{f},\hat{g}].$

Property (2) is introduced for the following reason. In Hamiltonian
mechanics each dynamical variable $f$ generates some transformations in the
phase--space via Poisson brackets. In quantum mechanics it generates
transformations in the state--space by direct application to a state, i.e.,
\begin{equation}
\dot{u}=\{u,f\},\qquad \partial _{t}|\psi \rangle =\frac{\mathrm{i}}{\hbar }%
\hat{f}|\psi \rangle .  \label{pois}
\end{equation}

Exponent of anti--Hermitian operator is unitary. Due to this fact,
transformations, generated by Hermitian operators
\begin{equation*}
\hat{U}=\exp \frac{\mathrm{i}\hat{f}t}{\hbar },
\end{equation*}%
are unitary. They are \emph{motions} -- scalar product preserving
transformations in the Hilbert state--space $\mathcal{H}$. For this property
$i$ is needed in (\ref{pois}).

Due to property (2), the transformations, generated by classical variables
and quantum operators, have the same algebra.

For example, the quantization of energy $E$ gives:
\begin{equation*}
E\rightarrow \hat{E}=\mathrm{i}\hbar \,\partial _{t}.
\end{equation*}%
The relations between operators must be similar to the relations between the
relevant physical quantities observed in classical mechanics.

For example, the quantization of the classical equation $E=H$, where
\begin{equation*}
H=H(p_{i},q^{i})=T+U
\end{equation*}%
denotes the Hamilton's function of the total system energy (the sum of the
kinetic energy $T$ and potential energy $U$), gives the Schr\"{o}dinger
equation of motion of the state ket--vector $|\psi (t)\rangle $ in the
Hilbert state--space $\mathcal{H}$
\begin{equation*}
\mathrm{i}\hbar \,\partial _{t}|\psi (t)\rangle \,=\hat{H}\,|\psi (t)\rangle
.
\end{equation*}

In the simplest case of a single particle in the potential field $U$, the
operator of the total system energy -- Hamiltonian is given by:
\begin{equation*}
\hat{H}=-\frac{{\hbar ^{2}}}{2m}\nabla ^{2}+U,
\end{equation*}%
where $m$ denotes the mass of the particle and $\nabla $ is the classical
gradient operator. So the first term on the r.h.s denotes the kinetic energy
of the system, and therefore the momentum operator must be given by:%
\begin{equation*}
\hat{p}=-\mathrm{i}\hbar \nabla .
\end{equation*}

Now, for each pair of states $|\varphi \rangle ,|\psi \rangle $ their scalar
product $\langle \varphi |\psi \rangle $ is introduced, which is \cite%
{QuLeap}:

\begin{enumerate}
\item Linear (for right multiplier):
\begin{equation*}
\langle \varphi |\alpha _{1}\psi _{1}+\alpha _{2}\psi _{2}\rangle =\alpha
_{1}\langle \varphi |\psi _{1}\rangle +\alpha _{2}\langle \varphi |\psi
_{2}\rangle ;
\end{equation*}

\item In transposition transforms to complex conjugated:
\begin{equation*}
\langle \varphi |\psi \rangle =\overline{\langle \psi |\varphi \rangle };
\end{equation*}
this implies that it is `anti--linear' for left multiplier:
\begin{equation*}
\langle \alpha _{1}\varphi _{1}+\alpha _{2}\varphi _{2}\rangle =\bar{\alpha}%
_{1}\langle \varphi _{1}|\psi \rangle +\bar{\alpha}_{2}\langle \varphi
_{2}|\psi \rangle );
\end{equation*}

\item Additionally it is often required, that the scalar product should be
positively defined:
\begin{equation*}
\text{for all }|\psi \rangle ,\quad \langle \psi |\psi \rangle \geq 0\qquad
\text{and}\qquad \langle \psi |\psi \rangle =0\quad \ \ \text{iff \ \ }|\psi
\rangle =0.
\end{equation*}
\end{enumerate}

Complex conjugation of classical variables is represented as Hermitian
conjugation of operators.\footnote{%
Two operators $\hat{f},\hat{f}^{+}$ are called Hermitian conjugated (or
adjoint), if
\begin{equation*}
\langle \varphi |\hat{f}\psi \rangle =\langle \hat{f}^{+}\varphi |\psi
\rangle \qquad (\text{for all }\varphi ,\psi ).
\end{equation*}%
\par
This scalar product is also denoted by $\langle \varphi |\hat{f}|\psi
\rangle $ and called a matrix element of an operator.
\par
-- operator is Hermitian (self--adjoint) if $\hat{f}^{+}=\hat{f}$ and
anti--Hermitian if $\hat{f}^{+}=-\hat{f};$%
\par
-- operator is unitary, if $\hat{U}^{+}=\hat{U}^{-1}$; such operators
preserve the scalar product:
\begin{equation*}
\langle \hat{U}\varphi |\hat{U}\psi \rangle =\langle \varphi |\hat{U}^{+}%
\hat{U}|\psi \rangle =\langle \varphi |\psi \rangle .
\end{equation*}%
\par
Real classical variables should be represented by Hermitian operators;
complex conjugated classical variables $(a,\bar{a})$ correspond to Hermitian
conjugated operators $(\hat{a},\hat{a}^{+})$.
\par
Multiplication of a state by complex numbers does not change the state
physically.
\par
Any Hermitian operator in Hilbert space has only real eigenvalues:
\begin{equation*}
\hat{f}|\psi _{i}\rangle =f_{i}|\psi _{i}\rangle ,\qquad (\text{for all}%
~~f_{i}\in \mathbb{R}).
\end{equation*}%
\par
Eigenvectors $|\psi _{i}\rangle $ form complete orthonormal basis
(eigenvectors with different eigenvalues are automatically orthogonal; in
the case of multiple eigenvalues one can form orthogonal combinations; then
they can be normalized).}

If the two Hermitian operators $\hat{f}$ and $\hat{g}$ commute, i.e., $[\hat{%
f},\hat{g}]=0$ (see Heisenberg picture below), than the corresponding
quantities can simultaneously have definite values. If the two operators do
not commute, i.e., $[\hat{f},\hat{g}]\neq 0$, the quantities corresponding
to these operators cannot have definite values simultaneously, i.e., the
general \textit{Heisenberg uncertainty relation} is valid:
\begin{equation*}
(\Delta \hat{f})^{2}\cdot (\Delta \hat{g})^{2}\geq \frac{\hbar }{4}[\hat{f},%
\hat{g}]^{2},
\end{equation*}%
where $\Delta $ denotes the deviation of an individual measurement from the
mean value of the distribution. The well-known particular cases are ordinary
uncertainty relations for coordinate--momentum ($q-p$), and energy--time ($%
E-t$):
\begin{equation*}
\Delta q\cdot \Delta p_{q}\geq \frac{\hbar }{2},\text{ \ \ \ and \ \ \ }%
\Delta E\cdot \Delta t\geq \frac{\hbar }{2}.
\end{equation*}

For example, the rules of commutation, analogous to the classical ones
written by the Poisson's brackets, are postulated for canonically--conjugate
coordinate and momentum operators:
\begin{equation*}
\lbrack \hat{q}^{i},\hat{q}^{j}]=0,\qquad \lbrack \hat{p}_{i},\hat{p}%
_{j}]=0,\qquad \lbrack \hat{q}^{i},\hat{p}_{j}]=\mathrm{i}\hbar \delta
_{j}^{i}\hat{I},
\end{equation*}%
where $\delta _{j}^{i}$ is the Kronecker's symbol. By applying the
commutation rules to the system Hamiltonian $\hat{H}=\hat{H}(\hat{p}_{i},%
\hat{q}^{i})$, the \textit{quantum Hamilton's equations} are obtained:%
\begin{equation*}
\hat{\dot{p}}_{i}=-{\partial }_{\hat{q}^{i}}\hat{H}\qquad \text{and\qquad }%
\hat{\dot{q}}^{i}={\partial }_{\hat{p}_{i}}\hat{H}.
\end{equation*}%
\qquad

A quantum state can be observed either in the \emph{coordinate $q-$%
representation}, or in the \emph{momentum $p-$representation}. In the $q-$%
representation, operators of coordinate and momentum have respective forms: $%
\hat{q}=q$, and $\hat{p}_{q}=-\mathrm{i}\hbar \frac{\partial }{{\partial q}}$%
, while in the $p$--representation, they have respective forms: $\hat{q}=%
\mathrm{i}\hbar \frac{\partial }{{\partial }p_{q}}$, and $\hat{p}_{q}=p_{q}$%
. The forms of the state vector $|\psi (t)\rangle $ in these two
representations are mathematically related by a \emph{Fourier--transform pair%
}.

\subsubsection{Three quantum pictures}

In the $q-$representation, there are three main pictures (see e.g., \cite%
{QuLeap}):

\begin{enumerate}
\item \textit{Schr\"{o}dinger picture},

\item \textit{Heisenberg picture}, and

\item \textit{Dirac interaction picture}.
\end{enumerate}

These three pictures mutually differ in the time--dependence, i.e.,
time--evolution of the state--vector wavefunction $|\psi (t)\rangle $ and
the Hilbert coordinate basis $(q^{i})$ together with the system operators.

1. In the \emph{Schr\"{o}dinger} (S) \emph{picture}, under the action of the
\emph{evolution operator} $\hat{S}(t)$ the state--vector $|\psi (t)\rangle $
rotates:
\begin{equation*}
|\psi (t)\rangle \,=\hat{S}(t)\,|\psi (0)\rangle ,
\end{equation*}%
and the coordinate basis $(q^{i})$ is fixed, so the operators are constant
in time:
\begin{equation*}
\hat{F}(t)=\hat{F}(0)=\hat{F},
\end{equation*}%
and the system evolution is determined by the Schr\"{o}dinger wave equation:
\begin{equation*}
\mathrm{i}\hbar \,\partial _{t}|\psi ^{S}(t)\rangle ={\hat{H}}^{S}\,|\psi
^{S}(t)\rangle .
\end{equation*}%
If the Hamiltonian does not explicitly depend on time, $\hat{H}(t)=\hat{H}$,
which is the case with the absence of variables of macroscopic fields, the
state vector $|\psi (t)\rangle $ can be presented in the form:
\begin{equation*}
|\psi (t)\rangle \,=\exp \left( -\mathrm{i}\frac{E}{\hbar }t\right) |\psi
\rangle ,
\end{equation*}%
satisfying the time--independent Schr\"{o}dinger equation
\begin{equation*}
\hat{H}\,|\psi \rangle \,=E\,|\psi \rangle ,
\end{equation*}%
which gives the eigenvalues $E_{m}$ and eigenfunctions $|\psi _{m}\rangle $
of the Hamiltonian $\hat{H}$.

2. In the \emph{Heisenberg} (H) \emph{picture}, under the action of the
evolution operator $\hat{S}(t)$, the coordinate basis $(q^{i})$ rotates, so
the operators of physical variables evolve in time by the similarity
transformation:
\begin{equation*}
\hat{F}(t)=\hat{S}^{-1}(t)\,\hat{F}(0)\,\hat{S}(t),
\end{equation*}%
while the state vector $|\psi (t)\rangle $ is constant in time:
\begin{equation*}
|\psi (t)\rangle \,=|\psi (0)\rangle \,=|\psi \rangle ,
\end{equation*}%
and the system evolution is determined by the \emph{Heisenberg equation of
motion}:
\begin{equation*}
\mathrm{i}\hbar \,\partial _{t}\hat{F}^{H}(t)=[\hat{F}^{H}(t),\hat{H}%
^{H}(t)],
\end{equation*}%
where $\hat{F}(t)$ denotes arbitrary Hermitian operator of the system, while
the commutator, i.e., Poisson quantum bracket, is given by:
\begin{equation*}
\lbrack \hat{F}(t),\hat{H}(t)]=\hat{F}(t)\,\hat{H}(t)-\hat{H}(t)\,\hat{F}(t)=%
\hat{\imath}K.
\end{equation*}%
In both Schr\"{o}dinger and Heisenberg picture the evolution operator $\hat{S%
}(t)$ itself is determined by the Schr\"{o}dinger--like equation:
\begin{equation*}
\mathrm{i}\hbar \,\partial _{t}\hat{S}(t)=\hat{H}\,\hat{S}(t),
\end{equation*}%
with the initial condition $\hat{S}(0)=\hat{I}$. It determines the Lie group
of transformations of the Hilbert space $L^{2}(\psi )$ in itself, the
Hamiltonian of the system being the generator of the group.

3. In the \emph{Dirac interaction} (I) \emph{picture} both the state vector $%
|\psi (t)\rangle $ and coordinate basis $(q^{i})$ rotate; therefore the
system evolution is determined by both the Schr\"{o}dinger wave equation and
the Heisenberg equation of motion:
\begin{equation*}
\mathrm{i}\hbar \,\partial _{t}|\psi ^{I}(t)\rangle =\hat{H}^{I}\,|\psi
^{I}(t)\rangle ,\text{ \ \ \ and \ \ \ }\mathrm{i}\hbar \,\partial _{t}\hat{F%
}^{I}(t)=[\hat{F}^{I}(t),\hat{H}^{O}(t)].
\end{equation*}%
Here, $\hat{H}=\hat{H}^{0}+\hat{H}^{I}$, where $\hat{H}^{0}$ corresponds to
the Hamiltonian of the free fields and $\hat{H}^{I}$ corresponds to the
Hamiltonian of the interaction.

In particular, the stationary (time-independent) Schr\"{o}dinger equation,
\begin{equation*}
\hat{H}\,\psi =\hat{E}\,\psi ,
\end{equation*}%
can be obtained from the condition for the minimum of the \emph{quantum
action}:
\begin{equation*}
\delta S[\psi ]=0.
\end{equation*}%
The quantum action is usually defined by the integral:
\begin{equation*}
S[\psi ]=\,\langle \psi (t)|\,\hat{H}\,|\psi (t)\rangle \,=\int \psi ^{\ast }%
\hat{H}\psi \,\,dV,
\end{equation*}%
with the additional normalization condition for the unit--probability of the
psi--function:
\begin{equation*}
\langle \psi (t)|\psi (t)\rangle \,=\int \psi ^{\ast }\psi \,\,dV=1.
\end{equation*}%
When the functions $\psi $ and $\psi ^{\ast }$ are considered to be formally
independent and only one of them, say $\psi ^{\ast }$ is varied, we can
write the condition for an extreme of the action:
\begin{equation*}
\delta S[\psi ]=\int \delta \psi ^{\ast }\hat{H}\psi \,\,dV-E\int \delta
\psi ^{\ast }\psi \,\,dV=\int \delta \psi ^{\ast }(\hat{H}\psi -E\psi
)\,dV=0,
\end{equation*}%
where $E$ is a Lagrangian multiplier. Owing to the arbitrariness of $\delta
\psi ^{\ast }$, the Schr\"{o}dinger equation $\hat{H}\psi -\hat{E}\psi =0$
must hold.

\subsubsection{Dirac's probability amplitude and time--dependent perturbation}

Most quantum--mechanical problems cannot be solved exactly. For such
problems we can use \textit{Dirac's perturbation method}, which consists in
splitting up the time--dependent Hamiltonian $H=H(t)$ into two parts:
\[
H(t)=H_{0}+\epsilon H_{1}(t),
\]
in which $H_{0}=E$ must be simple, non-autonomous, energy function that can
be dealt with exactly, while $\epsilon H_{1}(t)=V(t)$ is small
time--dependent perturbation, which can be expanded as a power series in a
small numerical factor $\epsilon $. The first part, $H_{0},$ may then be
considered as the Hamiltonian of a simplified, or unperturbed system that
can be exactly solved, while the addition of the second part $\epsilon
H_{1}(t)$\ will require small corrections, of the nature of a power--series
expanded perturbation in the solution for the unperturbed system. Provided
the perturbation series in $\epsilon $ converges, the perturbation method
will give the answer to our problem with any desired accuracy; even when the
series does not converge, the first approximation obtained by means of it is
usually fairly accurate \cite{DiracBook}.

Therefore, we do not consider any modification to be made in the states of
the unperturbed system $E=H_{0}$, but we suppose that the perturbed system $%
H(t),$ instead of remaining permanently in one of its states, is continually
changing from one state to another (or, making transmissions), under the
influence of the perturbation $V(t)=\epsilon H_{1}(t)$.

We will work in the \textit{Heisenberg's representation} for the unperturbed
system $E,$ assuming that we have a general set of linear Hermitian
operators $\alpha $'s to label the representatives. Let us suppose that at
initial time $t_{0}$ the system is in a state for which the $\alpha $'s
certainly have the values $\alpha ^{\prime }$, so that the basic ket $\left|
\alpha ^{\prime }\right\rangle $ would correspond to this state. This state
would be stationary if there were no perturbation, i.e., if $H(t)=E.$ The
perturbation $V(t)$ cause the $E$ to change. At time $t$ the ket
corresponding to the state $\left| \alpha ^{\prime }\right\rangle $ in the
\textit{Schr\"{o}dinger's picture} will be $T\left| \alpha ^{\prime
}\right\rangle ,$ according to equation
\begin{eqnarray}
\left| At\right\rangle =T\left| At_{0}\right\rangle ,\quad  &&\text{as well
as}  \nonumber \\
\mathrm{i}\frac{dT}{dt}=H(t)T\quad  &&\text{and}~~\mathrm{\qquad i}\frac{%
d\left| At\right\rangle }{dt}=H(t)\left| At\right\rangle ,  \label{sc1}
\end{eqnarray}
where $T$ is a linear Hermitian operator independent of the ket $\left|
At\right\rangle $ and depending only on time ($t_{0}$ and $t$). The
probability of the $\alpha $'s having the values $\alpha ^{\prime \prime }$
is given by the absolute square of the \textit{probability amplitude} (or,
\textit{transition amplitude}) $\left\langle \alpha ^{\prime \prime }\right|
T\left| \alpha ^{\prime }\right\rangle \ ($for the system's transition from
the state $\left| \alpha ^{\prime }\right\rangle $ to the state $\left|
\alpha ^{\prime \prime }\right\rangle ),$
\begin{equation}
P(\alpha ^{\prime },\alpha ^{\prime \prime })=|\left\langle \alpha ^{\prime
\prime }\right| T\left| \alpha ^{\prime }\right\rangle |^{2}.  \label{pam}
\end{equation}
For $\alpha ^{\prime }\neq \alpha ^{\prime \prime },$ $P(\alpha ^{\prime
},\alpha ^{\prime \prime })$ is the probability of a transition taking place
from state $\alpha ^{\prime }$ to state $\alpha ^{\prime \prime }$ during
the time interval $[t_{0},t];$ $P(\alpha ^{\prime },\alpha ^{\prime })$ is
the probability of no transition taking place at all, while the sum of $%
P(\alpha ^{\prime },\alpha ^{\prime \prime })$ for all $\alpha ^{\prime
\prime }$ is unity.

Let us now suppose more generally that initially the system is in one of the
various states $\alpha ^{\prime }$ with the probability $P_{\alpha ^{\prime
}}$ for each. To deal effectively with this problem, we introduce the
\textit{von Neumann's quantum density function} $\rho $, a
quantum--mechanical analogue to the \textit{Gibbs statistical density
function} $\rho =\rho (t)$\ of a \textit{Gibbs ensemble} with the classical
Hamiltonian $H(q,p),$ which evolves within the ensemble's $n-$dimensional
phase--space $\mathcal{P}=\{(q^{i},p_{i})\,|\,i=1,...,n\}$ according to the
\textit{Poisson equation}
\begin{eqnarray*}
\partial _{t}\rho &=&-[\rho ,H(q,p)],\qquad \\
\text{with the normalizing condition} &:&\text{{}}\iint_{\mathcal{P}}\rho
\,dq^{i}dp_{i}=1.
\end{eqnarray*}
The von Neumann's quantum density function $\rho $ corresponding to the
initial probability distribution $P_{\alpha ^{\prime }}$ is given by
\[
\rho _{0}=\sum_{\alpha ^{\prime }}\left| \alpha ^{\prime }\right\rangle
P_{\alpha ^{\prime }}\left\langle \alpha ^{\prime }\right| .
\]
At time $t,$ each ket $\left| \alpha ^{\prime }\right\rangle $ will have
changed to $T\left| \alpha ^{\prime }\right\rangle $ and each bra $%
\left\langle \alpha ^{\prime }\right| $\ will change to $\left\langle \alpha
^{\prime }\right| \bar{T}$ (where $\bar{T}$ is complex--conjugate to $T),$
so $\rho _{0}$ will have changed to
\[
\rho (t)=\sum_{\alpha ^{\prime }}T\left| \alpha ^{\prime }\right\rangle
P_{\alpha ^{\prime }}\left\langle \alpha ^{\prime }\right| \bar{T}.
\]
The \textit{probability amplitude} of $\alpha $'s then having the values of $%
\alpha ^{\prime \prime }$ will be (using (\ref{pam}))
\[
\left\langle \alpha ^{\prime \prime }\right| \rho (t)\left| \alpha ^{\prime
}\right\rangle =\sum_{\alpha ^{\prime }}\left\langle \alpha ^{\prime \prime
}\right| T\left| \alpha ^{\prime }\right\rangle P_{\alpha ^{\prime
}}\left\langle \alpha ^{\prime }\right| \bar{T}\left| \alpha ^{\prime \prime
}\right\rangle =P_{\alpha ^{\prime }}P(\alpha ^{\prime },\alpha ^{\prime
\prime }).
\]
This result expresses that the probability of the system being in the state $%
\alpha ^{\prime \prime }$ at time $t$ equals the sum of the probabilities of
the system being initially in any state $\alpha ^{\prime }\neq \alpha
^{\prime \prime },$ and making a transition from state $\alpha ^{\prime }$
to the final state $\alpha ^{\prime \prime }.$ Thus, the various transition
probabilities act independently of one another, according to the ordinary
laws of probability.

The whole problem of calculating transitions thus reduces to the
determination of the probability amplitudes $\left\langle \alpha ^{\prime
\prime }\right| T\left| \alpha ^{\prime }\right\rangle \ $\cite{DiracBook}.
These can be worked out from (\ref{sc1}), or
\begin{equation}
\mathrm{i}\dot{T}=[H_{0}+\epsilon H_{1}(t)]T=(E+V)T\qquad \text{(where \ \ }%
\dot{T}=dT/dt\text{)}.\quad  \label{dt}
\end{equation}
This calculation can be simplified if instead of $T$ and $V$ operators, we
are working with
\begin{equation}
T^{\ast }=\exp [\mathrm{i}E(t-t_{0})]T\qquad \text{and\qquad }V^{\ast }=\exp
[\mathrm{i}E(t-t_{0})]V\exp [-\mathrm{i}E(t-t_{0})],  \label{tv}
\end{equation}
where $V^{\ast }$ is the result of applying a \textit{unitary transformation} to $V.$
Using (\ref{tv}) we obtain
\begin{equation}
\mathrm{i}\dot{T}^{\ast }=\exp [\mathrm{i}E(t-t_{0})]VT=V^{\ast }T^{\ast },
\label{t*}
\end{equation}
which is more convenient then (\ref{dt}) as it makes the time evolution of $%
T^{\ast }$depend only on the (unitary transformed) perturbation $V^{\ast }$
and not on the unperturbed state $E.$ From (\ref{tv}) we get the probability
amplitude
\begin{eqnarray*}
\left\langle \alpha ^{\prime \prime }\right| T^{\ast }\left| \alpha ^{\prime
}\right\rangle &=&\exp [\mathrm{i}E(t-t_{0})]\left\langle \alpha ^{\prime
\prime }\right| T\left| \alpha ^{\prime }\right\rangle ,\qquad \text{so that}
\\
P(\alpha ^{\prime },\alpha ^{\prime \prime }) &=&|\left\langle \alpha
^{\prime \prime }\right| T^{\ast }\left| \alpha ^{\prime }\right\rangle
|^{2},
\end{eqnarray*}
which shows that $T$ and $T^{\ast }$ are equally good for determining
transition probabilities.

So far, our work in this subsection has been exact. Now we assume the
perturbation $V(t)=\epsilon H_{1}(t)$ is a small quantity of the first order
in $\epsilon $ and express $T^{\ast }$ in the form
\begin{equation}
T^{\ast }=1+T_{1}^{\ast }+T_{2}^{\ast }+...,  \label{tex}
\end{equation}
where $T_{1}^{\ast }=T_{1}^{\ast }(\epsilon ),$ $T_{2}^{\ast }=T_{2}^{\ast
}(\epsilon ^{2}),$ etc. Substituting (\ref{tex}) into (\ref{t*}) we get the
expansion

\begin{eqnarray*}
\mathrm{i}\dot{T}_{1}^{\ast } &=&V^{\ast }, \\
\mathrm{i}\dot{T}_{2}^{\ast } &=&V^{\ast }T_{1}^{\ast }, \\
\mathrm{i}\dot{T}_{3}^{\ast } &=&V^{\ast }T_{2}^{\ast }, \\
&&...
\end{eqnarray*}
From the first of these equations we obtain
\[
T_{1}^{\ast }=-\mathrm{i}\int_{t_{0}}^{t}V^{\ast }(t^{\prime })\,dt^{\prime
},
\]
from the second we get
\[
T_{2}^{\ast }=-\int_{t_{0}}^{t}V^{\ast }(t^{\prime })\,dt^{\prime
}\int_{t_{0}}^{t}V^{\ast }(t^{\prime \prime })\,dt^{\prime \prime },
\]
and so on.

Now, the perturbation form of the transition probability $P(\alpha ^{\prime
},\alpha ^{\prime \prime })=|\left\langle \alpha ^{\prime \prime }\right|
T^{\ast }\left| \alpha ^{\prime }\right\rangle |^{2}$ is, if we retain only
the first--order term $T_{1}^{\ast }$ (which is sufficiently accurate for
many practical problems), given by
\[
P(\alpha ^{\prime },\alpha ^{\prime \prime })=\left|
\int_{t_{0}}^{t}\left\langle \alpha ^{\prime \prime }\right| V^{\ast
}(t^{\prime })\left| \alpha ^{\prime }\right\rangle \,dt^{\prime }\right|
^{2}.
\]
If we retain the first two terms, $T_{1}^{\ast }$ and $T_{2}^{\ast },$ the
transition probability $P(\alpha ^{\prime },\alpha ^{\prime \prime })$ is
given by
\begin{eqnarray*}
P(\alpha ^{\prime },\alpha ^{\prime \prime }) &=&\left|
\int_{t_{0}}^{t}\left\langle \alpha ^{\prime \prime }\right| V^{\ast
}(t^{\prime })\left| \alpha ^{\prime }\right\rangle \,dt^{\prime }\right.  \\
&-&\mathrm{i}\sum_{\alpha ^{\prime \prime \prime }\neq \alpha ^{\prime
\prime },\alpha ^{\prime }}\left. \int_{t_{0}}^{t}\left\langle \alpha
^{\prime \prime }\right| V^{\ast }(t^{\prime })\left| \alpha ^{\prime \prime
\prime }\right\rangle \,dt^{\prime }\int_{t_{0}}^{t}\left\langle \alpha
^{\prime \prime \prime }\right| V^{\ast }(t^{\prime \prime })\left| \alpha
^{\prime }\right\rangle \,dt^{\prime \prime }\right| ^{2},
\end{eqnarray*}
where $\alpha ^{\prime \prime \prime }$ is the so--called intermediate state
(between $\alpha ^{\prime }$ and $\alpha ^{\prime \prime }$). This shows the
perturbative calculation of the transition probability $P(\alpha ^{\prime
},\alpha ^{\prime \prime })$: we fist calculate the perturbative expansion
of the transition amplitude $\left\langle \alpha ^{\prime \prime }\right|
T^{\ast }\left| \alpha ^{\prime }\right\rangle ,$ and then take its absolute
square to obtain the overall transition probability. For more technical
details, see \cite{DiracBook}.

Both Dirac's concepts introduced in this subsection, namely transition amplitude and time--dependent perturbation, will prove essential later in the development of the \textit{Feynman path integral}, as well as the \textit{Feynman diagrams} approach to \textit{quantum field theory} (QFT).

\subsubsection{State--space for $n$ non-relativistic quantum particles}

Classical state--space for the system of $n$ particles is its $6N$D
phase--space $\mathcal{P}$, including all position and momentum vectors, $%
\mathbf{r}_{i}=(x,y,z)_{i}$ and $\mathbf{p}_{i}=(p_{x},p_{y},p_{z})_{i}$
respectively (for $i=1,...,n)$. The \emph{quantization} is performed as a
\emph{linear representation} of the real Lie algebra $\mathcal{L}_{P}$ of
the phase--space $\mathcal{P}$, defined by the Poisson bracket $\{A,B\}$ of
classical variables $A,B$ -- into the corresponding real Lie algebra $%
\mathcal{L}_{H}$ of the Hilbert space $\mathcal{H}$, defined by the
commutator $[\hat{A},\hat{B}]$ of skew--Hermitian operators $\hat{A},\hat{B}$
\cite{QuLeap}.

We start with the \emph{Hilbert space} $\mathcal{H}_{x}$ for a single 1D
quantum particle, which is composed of all vectors $|\psi _{x}\rangle $ of
the form
\begin{equation*}
|\psi _{x}\rangle =\int_{-\infty }^{+\infty }\psi \left( x\right)
\,|x\rangle \,\,dx,
\end{equation*}%
where $\psi \left( x\right) =\langle x|\psi \rangle $ are square integrable
Fourier coefficients,
\begin{equation*}
\int_{-\infty }^{+\infty }\left\vert \psi \left( x\right) \right\vert
\,^{2}\,dx<+\infty .
\end{equation*}%
The position and momentum Hermitian operators, $\hat{x}$ and $\hat{p}$,
respectively, act on the vectors $|\psi _{x}\rangle \in $ $\mathcal{H}_{x}$
in the following way:
\begin{eqnarray*}
\hat{x}|\psi _{x}\rangle &=&\int_{-\infty }^{+\infty }\hat{x}\,\psi \left(
x\right) \,|x\rangle \,\,dx,\qquad \int_{-\infty }^{+\infty }\left\vert
x\,\psi \left( x\right) \right\vert \,^{2}\,dx<+\infty , \\
\hat{p}|\psi _{x}\rangle &=&\int_{-\infty }^{+\infty }-\mathrm{i}\hbar
\partial _{\hat{x}}\psi \left( x\right) \,|x\rangle \,\,dx,\qquad
\int_{-\infty }^{+\infty }\left\vert -\mathrm{i}\hbar \partial _{x}\psi
\left( x\right) \right\vert ^{2}\,dx<+\infty .
\end{eqnarray*}

The \textit{orbit Hilbert space} $\mathcal{H}_{1}^{o}$ for a single 3D
quantum particle with the full set of compatible observable $\mathbf{\hat{r}=%
}(\hat{x},\hat{y},\hat{z}),\mathbf{\hat{p}}=(\hat{p}_{x},\hat{p}_{y},\hat{p}%
_{z}),$ is defined as
\begin{equation*}
\mathcal{H}_{1}^{o}=\mathcal{H}_{x}\otimes \mathcal{H}_{y}\otimes \mathcal{H}%
_{z},
\end{equation*}%
where $\mathbf{\hat{r}}$ has the common generalized eigenvectors of the form
\begin{equation*}
|\mathbf{\hat{r}}\rangle =|x\rangle \mathcal{\times }|y\rangle \mathcal{%
\times }|z\rangle \,.
\end{equation*}%
$\mathcal{H}_{1}^{o}$ is composed of all vectors $|\psi _{r}\rangle $ of the
form
\begin{equation*}
|\psi _{r}\rangle =\int_{\mathcal{H}^{o}}\psi \left( \mathbf{r}\right) \,|%
\mathbf{r}\rangle \,\,d\mathbf{r=}\int_{-\infty }^{+\infty }\int_{-\infty
}^{+\infty }\int_{-\infty }^{+\infty }\psi \left( x,y,z\right) \,|x\rangle
\mathcal{\times }|y\rangle \mathcal{\times }|z\rangle \,\,dxdydz,
\end{equation*}%
where $\psi \left( \mathbf{r}\right) =\langle \mathbf{r}|\psi _{r}\rangle $
are square integrable Fourier coefficients,
\begin{equation*}
\int_{-\infty }^{+\infty }\left\vert \psi \left( \mathbf{r}\right)
\right\vert \,^{2}\,d\mathbf{r}<+\infty .
\end{equation*}%
The position and momentum operators, $\mathbf{\hat{r}}$ and $\mathbf{\hat{p}}
$, respectively, act on the vectors $|\psi _{r}\rangle \in $ $\mathcal{H}%
_{1}^{o}$ in the following way:
\begin{eqnarray*}
\mathbf{\hat{r}}|\psi _{r}\rangle &=&\int_{\mathcal{H}_{1}^{o}}\mathbf{\hat{r%
}}\,\psi \left( \mathbf{r}\right) \,|\mathbf{r}\rangle \,\,d\mathbf{r}%
,\qquad \int_{\mathcal{H}_{1}^{o}}\left\vert \mathbf{r\,}\psi \left( \mathbf{%
r}\right) \right\vert \,^{2}\,d\mathbf{r}<+\infty , \\
\mathbf{\hat{p}}|\psi _{r}\rangle &=&\int_{\mathcal{H}_{1}^{o}}-\mathrm{i}%
\hbar \partial _{\mathbf{\hat{r}}}\psi \left( \mathbf{r}\right) \,|\mathbf{r}%
\rangle \,\,d\mathbf{r},\qquad \int_{\mathcal{H}_{1}^{o}}\left\vert -\mathrm{%
i}\hbar \partial _{\mathbf{r}}\psi \left( \mathbf{r}\right) \right\vert
^{2}\,d\mathbf{r}<+\infty .
\end{eqnarray*}

Now, if we have a system of $n$ 3D particles, let $\mathcal{H}_{i}^{o}$
denote the orbit Hilbert space of the $i$th particle. Then the composite
orbit state--space $\mathcal{H}_{n}^{o}$ of the whole system is defined as a
direct product
\begin{equation*}
\mathcal{H}_{n}^{o}=\mathcal{H}_{1}^{o}\otimes \mathcal{H}_{2}^{o}\otimes
...\otimes \mathcal{H}_{n}^{o}.
\end{equation*}%
$\mathcal{H}_{n}^{o}$ is composed of all vectors
\begin{equation*}
|\psi _{r}^{n}\rangle =\int_{\mathcal{H}_{n}^{o}}\psi \left( \mathbf{r}_{1},%
\mathbf{r}_{2},...,\mathbf{r}_{n}\right) \,|\mathbf{r}_{1}\rangle \mathcal{%
\times }|\mathbf{r}_{2}\rangle \mathcal{\times }...\mathcal{\times }|\mathbf{%
r}_{n}\rangle \,\,d\mathbf{r}_{1}d\mathbf{r}_{2}...d\mathbf{r}_{n}
\end{equation*}%
where $\psi \left( \mathbf{r}_{1},\mathbf{r}_{2},...,\mathbf{r}_{n}\right)
=\langle \mathbf{r}_{1},\mathbf{r}_{2},...,\mathbf{r}_{n}|\psi
_{r}^{n}\rangle $ are square integrable Fourier coefficients
\begin{equation*}
\int_{\mathcal{H}_{n}^{o}}\left\vert \psi \left( \mathbf{r}_{1},\mathbf{r}%
_{2},...,\mathbf{r}_{n}\right) \right\vert ^{2}\,d\mathbf{r}_{1}d\mathbf{r}%
_{2}...d\mathbf{r}_{n}<+\infty .
\end{equation*}%
The position and momentum operators $\mathbf{\hat{r}}_{i}$ and $\mathbf{\hat{%
p}}_{i}$ act on the vectors $|\psi _{r}^{n}\rangle \in $ $\mathcal{H}_{n}^{o}
$ in the following way:
\begin{eqnarray*}
\mathbf{\hat{r}}_{i}|\psi _{r}^{n}\rangle  &=&\int_{\mathcal{H}%
_{n}^{o}}\left\{ \mathbf{\hat{r}}_{i}\right\} \psi \left( \mathbf{r}_{1},%
\mathbf{r}_{2},...,\mathbf{r}_{n}\right) \,|\mathbf{r}_{1}\rangle \mathcal{%
\times }|\mathbf{r}_{2}\rangle \mathcal{\times }...\mathcal{\times }|\mathbf{%
r}_{n}\rangle \,\,d\mathbf{r}_{1}d\mathbf{r}_{2}...d\mathbf{r}_{n}, \\
\mathbf{\hat{p}}_{i}|\psi _{r}^{n}\rangle  &=&\int_{\mathcal{H}%
_{n}^{o}}\left\{ -\mathrm{i}\hbar \partial _{\mathbf{\hat{r}}_{i}}\right\}
\psi \left( \mathbf{r}_{1},\mathbf{r}_{2},...,\mathbf{r}_{n}\right) \,|%
\mathbf{r}_{1}\rangle \mathcal{\times }|\mathbf{r}_{2}\rangle \mathcal{%
\times }...\mathcal{\times }|\mathbf{r}_{n}\rangle \,\,d\mathbf{r}_{1}d%
\mathbf{r}_{2}...d\mathbf{r}_{n},
\end{eqnarray*}%
with the square integrable Fourier coefficients
\begin{eqnarray*}
\int_{\mathcal{H}_{n}^{o}}\left\vert \left\{ \mathbf{\hat{r}}_{i}\right\}
\psi \left( \mathbf{r}_{1},\mathbf{r}_{2},...,\mathbf{r}_{n}\right)
\right\vert ^{2}\,d\mathbf{r}_{1}d\mathbf{r}_{2}...d\mathbf{r}_{n}
&<&+\infty , \\
\int_{\mathcal{H}_{n}^{o}}\left\vert \left\{ -\mathrm{i}\hbar \partial _{%
\mathbf{r}_{i}}\right\} \psi \left( \mathbf{r}_{1},\mathbf{r}_{2},...,%
\mathbf{r}_{n}\right) \right\vert ^{2}\,d\mathbf{r}_{1}d\mathbf{r}_{2}...d%
\mathbf{r}_{n} &<&+\infty ,
\end{eqnarray*}%
respectively. In general, any set of vector Hermitian operators $\{\mathbf{%
\hat{A}}_{i}\}$ corresponding to all the particles, act on the vectors $%
|\psi _{r}^{n}\rangle \in $ $\mathcal{H}_{n}^{o}$ in the following way:
\begin{equation*}
\mathbf{\hat{A}}_{i}|\psi _{r}^{n}\rangle =\int_{\mathcal{H}_{n}^{o}}\{%
\mathbf{\hat{A}}_{i}\}\psi \left( \mathbf{r}_{1},\mathbf{r}_{2},...,\mathbf{r%
}_{n}\right) \,|\mathbf{r}_{1}\rangle \mathcal{\times }|\mathbf{r}%
_{2}\rangle \mathcal{\times }...\mathcal{\times }|\mathbf{r}_{n}\rangle \,\,d%
\mathbf{r}_{1}d\mathbf{r}_{2}...d\mathbf{r}_{n},
\end{equation*}%
with the square integrable Fourier coefficients
\begin{equation*}
\int_{\mathcal{H}_{n}^{o}}\left\vert \left\{ \mathbf{\hat{A}}_{i}\right\}
\psi \left( \mathbf{r}_{1},\mathbf{r}_{2},...,\mathbf{r}_{n}\right)
\right\vert ^{2}\,d\mathbf{r}_{1}d\mathbf{r}_{2}...d\mathbf{r}_{n}<+\infty .
\end{equation*}

\subsection{Transition to quantum fields}

\subsubsection{Amplitude, relativistic invariance and causality}

We will see later that in QFT, the fundamental quantity is not any more Schr\"odinger's \emph{wavefunction} but the rather the closely related, yet different, Dirac--Feynman's \textit{amplitude}. To introduce the amplitude concept within the non-relativistic quantum mechanics, suppose that $\mathbf{x=}\left( x,y,z\right)$ and consider the amplitude for a
free particle to propagate in time $t$ from $\mathbf{x}_{0}$ to $\mathbf{x}=\mathbf{x}(t)
$, which is given by
\[
U(t)=\left\langle \mathbf{x}|\,\mathrm{e}^{-{\rm i}Ht}|\mathbf{x}_{0}\right\rangle
.
\]
As the kinetic energy of a free particle is $E=\mathbf{p}^{2}/2m$, we have
\begin{eqnarray*}
U(t) &=&\left\langle \mathbf{x}|\,\mathrm{e}^{-\mathrm{i}(\mathbf{p}%
^{2}/2m)t}|\mathbf{x}_{0}\right\rangle ~=~\int d^{3}p\left\langle \mathbf{x}|\,\mathrm{e}^{-\mathrm{i}(\mathbf{p}%
^{2}/2m)t}|\mathbf{p}\right\rangle \left\langle \mathbf{p}|\mathbf{x}%
_{0}\right\rangle  \\
&=&\frac{1}{(2\pi )^{2}}\int d^{3}p\,\mathrm{e}^{-\mathrm{i}(\mathbf{p}%
^{2}/2m)t}\cdot \,\mathrm{e}^{\mathrm{i}\mathbf{p}\cdot (\mathbf{x-x}_{0})}
~=~\left( \frac{m}{2\pi \mathrm{i}t}\right) ^{3/2}\,\mathrm{e}^{\mathrm{i}m(%
\mathbf{x-x}_{0})^{2}/2t}.
\end{eqnarray*}
Later we will deal with the amplitude in the relativistic framework.

As we have seen in the previous section, in non-relativistic quantum
mechanics observables are represented by self-adjoint operators that in the
Heisenberg picture depend on time, so that for any \textit{quantum observable%
} $\mathcal{O}$, the Heisenberg equation of motion holds (in normal units):
\begin{equation}
\mathrm{i}\partial _{t}\mathcal{O}=[\mathcal{O},H].  \label{HE}
\end{equation}
Therefore measurements are localized in time but are global in space. The
situation is radically different in the relativistic case. Because no signal
can propagate faster than the speed of light, measurements have to be
localized both in time and space. Causality demands then that two
measurements carried out in causally-disconnected regions of space--time
cannot interfere with each other. In mathematical terms this means that if $%
\mathcal{O}_{R_{1}}$ and $\mathcal{O}_{R_{2}}$ are the observables
associated with two measurements localized in two causally-disconnected
regions $R_{1}$, $R_{2}$, they satisfy the \textit{commutator} relation \cite
{peskin}
\begin{equation}
\lbrack \mathcal{O}_{R_{1}},\mathcal{O}_{R_{2}}]=0,\hspace*{1cm}%
\mbox{if
$(x_{1}-x_{2})^2<0$, for all $x_1\in R_{1}$, $x_{2}\in R_{2}$}.
\label{commutator}
\end{equation}

Hence, in a relativistic theory, the basic operators in the Heisenberg
picture must depend on the space-time position $x^{\mu }$. Unlike the case
in non-relativistic quantum mechanics, here the position ${x}$ \textsl{is not%
} an observable, but just a label, similarly to the case of time in ordinary
quantum mechanics. Causality is then imposed microscopically by requiring
\begin{equation}
\lbrack \mathcal{O}(x),\mathcal{O}(y)]=0,\hspace*{1cm}\mbox{if $(x-y)^2<0$}.
\label{microcausality}
\end{equation}
A smeared operator $\mathcal{O}_{R}$ over a space-time region $R$ can then
be defined as
\begin{equation*}
\mathcal{O}_{R}=\int d^{4}x\,\mathcal{O}(x)\,f_{R}(x),
\end{equation*}
where $f_{R}(x)$ is the \emph{characteristic function} associated with $R$,
\begin{equation*}
f_{R}(x)=\left\{
\begin{array}{lll}
1, & \mathrm{for} & x\in R, \\
0, & \mathrm{for} & x\notin R.
\end{array}
\right.
\end{equation*}
Equation (\ref{commutator}) follows now from the micro--causality condition (%
\ref{microcausality}).

Therefore, relativistic invariance forces the introduction of quantum
fields. It is only when we insist in keeping a single-particle
interpretation that we crash against causality violations. To illustrate the
point, let us consider a single particle wave function $\psi (t,{x})$ that
initially is localized in the position ${x}=0$
\begin{equation*}
\psi (0,{x})=\delta ({x}).
\end{equation*}

Evolving this wave function using the Hamiltonian $H=\sqrt{-\nabla ^{2}+m^{2}%
}$ we find that the wave function can be written as
\begin{equation*}
\psi (t,{x})=\mathrm{e}^{-\mathrm{i}t\sqrt{-\nabla ^{2}+m^{2}}}\delta ({x}%
)=\int {\frac{d^{3}p}{(2\pi )^{3}}}\mathrm{e}^{\mathrm{i}{p}\cdot {x}-%
\mathrm{i}t\sqrt{p^{2}+m^{2}}}.
\end{equation*}
Integrating over the angular variables, the wave function can be recast in
the form
\begin{equation*}
\psi (t,{x})={\frac{1}{2\pi ^{2}|{x}|}}\int_{-\infty }^{\infty }p\,dk\,%
\mathrm{e}^{\mathrm{i}p|{x}|}\,\mathrm{e}^{-\mathrm{i}t\sqrt{p^{2}+m^{2}}}.
\end{equation*}
The resulting integral can be evaluated using the complex integration
contour $C$. The result is that, for any $t>0$, one finds that $\psi (t,{x}%
)\neq 0$ for any ${x}$. If we insist in interpreting the wave function $\psi
(t,{x})$ as the probability density of finding the particle at the location $%
{x}$ in the time $t$ we find that the probability leaks out of the light
cone, thus violating causality.

In the Heisenberg picture, the amplitude for a particle to propagate from
point $y$ to point $x$ in the \textit{field} $\phi $\ is defined as
\begin{equation*}
D(x-y)=\left\langle 0|\phi (x)\phi (y)|0\right\rangle =\int \frac{d^{3}p}{%
(2\pi )^{3}}\frac{1}{2E}\,\mathrm{e}^{-\mathrm{i}p\cdot (x-y)}.
\end{equation*}
In this picture we can make time-dependent the \textit{field operator} $\phi
=\phi (x)$ and its cannonically--conjugate \textit{momentum operator} $\pi
=\pi (x),$ as
\begin{equation*}
\phi (x)=\phi (\mathbf{x},t)=\,\mathrm{e}^{\mathrm{i}Ht}\phi (\mathbf{x})\,%
\mathrm{e}^{-\mathrm{i}Ht},\qquad \pi (x)=\pi (\mathbf{x},t)=\,\mathrm{e}^{%
\mathrm{i}Ht}\pi (\mathbf{x})\,\mathrm{e}^{-\mathrm{i}Ht}.
\end{equation*}
Using (\ref{HE})\ we can compute the time dependence of $\phi $ and $\pi $
as
\begin{equation*}
\mathrm{i}\dot{\phi}(\mathbf{x},t)=\mathrm{i}\pi (\mathbf{x},t),\qquad
\mathrm{i}\dot{\pi}(\mathbf{x},t)=-\mathrm{i}(-\nabla ^{2}+m^{2})\phi (%
\mathbf{x},t).
\end{equation*}
Combining the two results we get the \emph{Klein--Gordon equation}
\begin{equation*}
\ddot{\phi}=(\nabla ^{2}-m^{2})\phi .
\end{equation*}

\subsubsection{Gauge theories}

Recall that a \textit{gauge theory} is a theory that admits a symmetry with
a local parameter. For example, in every quantum theory the global phase of
the wave $\psi -$function is arbitrary and does not represent something
physical. Consequently, the theory is invariant under a global change of
phases (adding a constant to the phase of all wave functions, everywhere);
this is a global symmetry. In quantum electrodynamics, the theory is also
invariant under a local change of phase, that is, one may shift the phase of
all wave functions so that the shift may be different at every point in
space-time. This is a local symmetry. However, in order for a well--defined
derivative operator to exist, one must introduce a new field, the \textit{%
gauge field}, which also transforms in order for the local change of
variables (the phase in our example) not to affect the derivative. In
quantum electrodynamics this gauge field is the electromagnetic potential
1--form $A$ (see Appendix), in components within the $n$D coframe $\{dx^{\mu
}\}$ on a smooth manifold $M$ (dual to the frame, i.e., basis of tangent
vectors $\{\partial _{\mu }=\partial /\partial x^{\mu }\}$, given by
\begin{equation*}
A=A_{\mu }dx^{\mu },\quad \text{such that}\quad A_{\mathrm{new}}=A_{\mathrm{%
old}}+df,\qquad \text{($f$ is any scalar function)}
\end{equation*}%
-- leaves the electromagnetic field 2--form $F=dA$ unchanged. This change $%
df $ of local gauge of variable $A$ is termed \textit{gauge transformation}.
In quantum field theory the excitations of fields represent particles. The
particle associated with excitations of the gauge field is the \emph{gauge
boson}. All the fundamental interactions in nature are described by gauge
theories. In particular, in quantum electrodynamics, whose gauge
transformation is a local change of phase, the gauge group is the circle
group $U(1)$ (consisting of all complex numbers with absolute value $1$),
and the gauge boson is the photon (see e.g., \cite{Frampton}).

The \textit{gauge field} of classical electrodynamics, given in local
coframe $\{dx^{\mu }\}$ on $M$ as an electromagnetic potential 1--form
\begin{equation*}
A=A_{\mu }dx^{\mu }=A_{\mu }dx^{\mu }+df,\qquad (f=~\text{arbitrary scalar
field}),
\end{equation*}%
is globally a \textit{connection} on a $U(1)-$bundle of $M$.\footnote{%
Recall that in the 19th Century, Maxwell unified Faraday's electric and
magnetic fields. Maxwell's theory led to Einstein's special relativity where
this unification becomes a spin-off of the unification of space end time in
the form of the \textit{Faraday tensor} \cite{MTW}
\begin{equation*}
F=E\wedge dt+B,
\end{equation*}%
where $F$ is electromagnetic $2-$form on space-time, $E$ is electric $1-$%
form on space, and $B$ is magnetic $2-$form on space. Gauge theory considers
$F$ as secondary object to a connection--potential $1-$form $A$. This makes
half of the Maxwell equations into tautologies \cite{BaezGauge}, i.e.,
\begin{equation*}
F=dA\quad \Longrightarrow \quad dF=0\quad :\quad \text{Bianchi identity},
\end{equation*}%
but does not imply the second half of Maxwell's equations,
\begin{equation*}
\delta F=-4\pi J\quad :\quad \text{dual Bianchi identity}.
\end{equation*}%
To understand the deeper meaning of the connection--potential $1-$form $A$,
we can integrate it along a path $\gamma $ in space-time, ~$x\cone{\gamma}y$%
.~ Classically, the integral $\int_{\gamma }A$ represents an \emph{action}
for a charged point particle to move along the path $\gamma $.
Quantum--mechanically, $\exp \left( \mathrm{i}\int_{\gamma }A\right) $
represents a \emph{phase} (within the unitary Lie group $U(1)$) by which the
particle's wave--function changes as it moves along the path $\gamma $, so $%
A $ is a $U(1)-$connection.
\par
In other words, Maxwell's equations can be formulated using complex line
bundles, or principal bundles with fibre $U(1)$. The connection $\nabla $ on
the line bundle has a curvature $F=\nabla ^{2}$ which is a 2--form that
automatically satisfies $dF=0$\ and can be interpreted as a field--strength.
If the line bundle is trivial with flat reference connection $d,$ we can
write $\nabla =d+A$ \ and $F=dA$ with $A$ the 1--form composed of the
electric potential and the magnetic vector potential.} The corresponding
electromagnetic field, locally the 2--form on $M,$
\begin{eqnarray*}
F &=&dA,\qquad \text{in components given by} \\
F &=&\frac{1}{2}F_{\mu \nu }\,dx^{\mu }\wedge dx^{\nu },\qquad \text{with \
\ \ }F_{\mu \nu }=\partial _{\nu }A_{\mu }-\partial _{\mu }A_{\nu }
\end{eqnarray*}%
is globally the \textit{curvature} of the connection $A$\footnote{%
The only thing that matters here is the \emph{difference} $\alpha $ \emph{%
between two paths} $\gamma _{1}$ and $\gamma _{2}$ \emph{in the action} $%
\int_{\gamma }A$ \cite{BaezGauge}, which is a 2--morphism (see e.g., \cite%
{GaneshSprBig,GaneshADG})
\begin{equation*}
x\ctwodbl{\gamma_1}{\gamma_2}{\alpha}y
\end{equation*}%
} under the gauge--covariant derivative,
\begin{equation}
D_{\mu }=\partial _{\mu }-\mathrm{i}eA_{\mu },  \label{covd1}
\end{equation}%
where $e$ is the charge coupling constant.%
\footnote{%
If a gauge transformation is given by%
\begin{equation*}
\psi \mapsto \mathrm{e}^{i\Lambda }\psi
\end{equation*}%
and for the gauge potential%
\begin{equation*}
A_{\mu }\mapsto A_{\mu }+\frac{1}{e}(\partial _{\mu }\Lambda ),
\end{equation*}%
then the gauge--covariant derivative,
\begin{equation*}
D_{\mu }=\partial _{\mu }-ieA_{\mu }
\end{equation*}%
transforms as%
\begin{equation*}
D_{\mu }\mapsto \partial _{\mu }-\mathrm{i}eA_{\mu }-i(\partial _{\mu
}\Lambda )
\end{equation*}%
and $D_{\mu }\psi $ transforms as%
\begin{equation*}
D_{\mu }\mapsto \partial _{\mu }-ieA_{\mu }-i(\partial _{\mu }\Lambda ).
\end{equation*}%
} In particular, in 4D space-time electrodynamics, the 1--form \textit{%
electric current density} $J$ has the components $J_{\mu }=(\rho ,\mathbf{j}%
)=(\rho ,j_{x},j_{y},j_{z})$ (where $\rho $ is the charge density), the
2--form \textit{Faraday} $F$ is given in components of electric field $%
\mathbf{E}$ and magnetic field $\mathbf{B}$ by
\begin{equation*}
F_{\mu \nu }=\left(
\begin{array}{cccc}
0 & E_{x} & E_{y} & E_{z} \\
-E_{x} & 0 & -B_{z} & B_{y} \\
-E_{y} & B_{z} & 0 & -B_{x} \\
-E_{z} & -B_{y} & B_{x} & 0%
\end{array}%
\right) ,\qquad \text{with\qquad }F_{\nu \mu }=-F_{\mu \nu },
\end{equation*}%
while its dual 2--form \textit{Maxwell} $\star F$ has the following
components
\begin{equation*}
\star F_{\mu \nu }=\left(
\begin{array}{cccc}
0 & -B_{x} & -B_{y} & -B_{z} \\
B_{x} & 0 & -E_{z} & E_{y} \\
B_{y} & E_{z} & 0 & -E_{x} \\
B_{z} & -E_{y} & B_{x} & 0%
\end{array}%
\right) ,\qquad \text{with\qquad }\star F_{\nu \mu }=-\star F_{\mu \nu },
\end{equation*}%
so that classical electrodynamics is governed by the \emph{Maxwell equations}%
, which in modern exterior formulation read
\begin{eqnarray*}
dF &=&0,\qquad \delta F=-4\pi J,\qquad \text{or in components,} \\
F_{[\mu \nu ,\eta ]} &=&0,\qquad F_{\mu \nu },^{\mu }=-4\pi J_{\mu },
\end{eqnarray*}%
where $\star $ is the Hodge star operator and $\delta $ is the Hodge
codiferential (see section \ref{Hodge} below), comma denotes the partial
derivative and the 1--form of electric current $J=J_{\mu }dx^{\mu }$ is
conserved, by the electrical \textit{continuity equation},
\begin{equation*}
\delta J=0,\qquad \text{or in components,\qquad }J_{\mu },^{\mu }=0.
\end{equation*}

The first, sourceless Maxwell equation, $dF=0$, gives vector magnetostatics
and magnetodynamics,
\begin{eqnarray*}
\text{Magnetic Gauss' law} &:&\func{div}\mathbf{B}=0,\qquad \\
\text{Faraday's law} &\text{:}&\partial _{t}\mathbf{B}+\func{curl}\mathbf{E}%
=0.
\end{eqnarray*}%
The second Maxwell equation with source, $\delta F=J$, gives vector
electrostatics and electrodynamics,
\begin{eqnarray*}
\text{Electric Gauss' law} &:&\func{div}\mathbf{E}=4\pi \rho ,\qquad \\
\text{Amp\`{e}re's law} &:&\partial _{t}\mathbf{E}-\func{curl}\mathbf{B}%
=-4\pi \mathbf{j}.
\end{eqnarray*}

The standard \textit{Lagrangian} for the free electromagnetic field, $F=dA$,
is given by \cite{GaneshSprBig,GaneshADG,QuLeap}
\begin{equation*}
\mathcal{L}(A)=\frac{1}{2}(F\wedge \star \,F),
\end{equation*}%
with the corresponding \emph{action functional}
\begin{equation*}
S(A)=\frac{1}{2}\int F\wedge \star \,F.
\end{equation*}

Maxwell's equations are generally applied to macroscopic averages of the
fields, which vary wildly on a microscopic scale in the vicinity of
individual atoms, where they undergo quantum effects as well (see below).

\subsubsection{Free and interacting field theories}

A generic \emph{gauge--covariant derivative} with Lorentz index $\mu$ is
denoted by $D_{\mu }$. For a Maxwell field, $D_{\mu }$ is given by (\ref%
{covd1}).
\begin{eqnarray*}
\text{Dirac slash notation} &:&\partial \!\!\!\!/\ \overset{\mathrm{def}}{=}%
\ \gamma ^{\mu }\partial _{\mu },\qquad D\!\!\!\!/\ \overset{\mathrm{def}}{=}%
\ \gamma ^{\mu }D_{\mu }, \\
\text{Dirac algebra} &:&\{\gamma ^{\mu },\gamma ^{\nu }\}=\gamma ^{\mu
}\gamma ^{\nu }+\gamma ^{\nu }\gamma ^{\mu }=2g^{\mu \nu }\times \mathbf{1}%
_{n\times n}.
\end{eqnarray*}

Standard free theor\textrm{i}es are Klein--Gordon and Dirac fields:
\begin{eqnarray*}
\text{Klein--Gordon equation} &:&(\partial ^{2}+m^{2})\psi =0, \\
\text{Dirac equation} &:&(\mathrm{i}\partial \!\!\!/-m)\psi =0.
\end{eqnarray*}

Two main examples of interacting theor\textrm{i}es are $\phi ^{4}-$theory
and QED:

\begin{enumerate}
\item $\phi ^{4}-$theory:
\begin{eqnarray*}
\text{Lagrangian} &:&\mathcal{L}=\frac{1}{2}(\partial _{\mu }\phi )^{2}\,-%
\frac{1}{2}m^{2}\phi ^{2}-\frac{\lambda }{4!}\phi ^{4}, \\
\text{Equation of motion} &:&(\partial ^{2}+m^{2})\phi =-\frac{\lambda }{3!}%
\phi ^{3}.
\end{eqnarray*}

\item QED:
\begin{eqnarray*}
\text{Lagrangian} &:&\mathcal{L}=\mathcal{L}_{\mathrm{Maxwell}}+\mathcal{L}_{%
\mathrm{Dirac}}+\mathcal{L}_{\mathrm{int}} \\
&=&-\frac{1}{4}(F_{\mu \nu })^{2}+\bar{\psi}(\mathrm{i}D\!\!\!\!/-m)\psi . \\
\text{Gauge invariance} &:&\psi (x)\rightarrow \,\mathrm{e}^{\mathrm{i}%
\alpha (x)}\psi (x)\Longrightarrow A_{\mu }\rightarrow A_{\mu }-\frac{1}{e}%
\partial _{\mu }\alpha (x). \\
\text{Equation of motion} &:&(\mathrm{i}D\!\!\!\!/-m)\psi =0.
\end{eqnarray*}
\end{enumerate}

\subsubsection{Dirac QED}

The \textit{Dirac equation} for a particle with mass $m$ (in natural units)
reads (see, e.g., \cite{QuLeap})
\begin{equation}
(\mathrm{i}\gamma ^{\mu }\partial _{\mu }-m)\psi =0,\qquad (\mu =0,1,2,3)
\label{DiracEq}
\end{equation}%
where $\psi (x)$ is a 4--component spinor\footnote{%
The most convenient definitions for the 2--spinors, like the Dirac spinor,
are:
\par
$\phi ^{1}=%
\begin{bmatrix}
1 \\
0%
\end{bmatrix}%
,\,\phi ^{2}=%
\begin{bmatrix}
0 \\
1%
\end{bmatrix}%
\,$ and $\chi ^{1}=%
\begin{bmatrix}
0 \\
1%
\end{bmatrix}%
,\,\chi ^{2}=%
\begin{bmatrix}
1 \\
0%
\end{bmatrix}%
\,.$} wave--function, the so--called Dirac spinor, while $\gamma ^{\mu }$
are $4\times 4$ \textit{Dirac }$\gamma -$\textit{matrices},
\begin{eqnarray*}
&\gamma ^{0}=%
\begin{pmatrix}
1 & 0 & 0 & 0 \\
0 & 1 & 0 & 0 \\
0 & 0 & -1 & 0 \\
0 & 0 & 0 & -1%
\end{pmatrix}%
,\qquad &\gamma ^{1}\!=\!%
\begin{pmatrix}
0 & 0 & 0 & 1 \\
0 & 0 & 1 & 0 \\
0 & -1 & 0 & 0 \\
-1 & 0 & 0 & 0%
\end{pmatrix}%
, \\
&\gamma ^{2}\!=\!%
\begin{pmatrix}
0 & 0 & 0 & -{\rm i} \\
0 & 0 & {\rm i} & 0 \\
0 & {\rm i} & 0 & 0 \\
-{\rm i} & 0 & 0 & 0%
\end{pmatrix}%
,\qquad &\gamma ^{3}\!=\!%
\begin{pmatrix}
0 & 0 & 1 & 0 \\
0 & 0 & 0 & -1 \\
-1 & 0 & 0 & 0 \\
0 & 1 & 0 & 0%
\end{pmatrix}%
.
\end{eqnarray*}%
They obey the \textit{anticommutation relations}
\begin{equation*}
\{\gamma ^{\mu },\gamma ^{\nu }\}=\gamma ^{\mu }\gamma ^{\nu }+\gamma ^{\nu
}\gamma ^{\mu }=2g^{\mu \nu },
\end{equation*}%
where $g_{\mu \nu }$ is the metric tensor.

Dirac's $\gamma -$matrices are conventionally derived as%
\begin{equation*}
\gamma ^{k}=%
\begin{pmatrix}
0 & \sigma ^{k} \\
-\sigma ^{k} & 0%
\end{pmatrix}%
,\qquad (k=1,2,3)
\end{equation*}%
where $\sigma ^{k}$ are \textit{Pauli }$\sigma -$\textit{matrices}\footnote{%
In quantum mechanics, each Pauli matrix represents an observable describing
the spin of a spin ${\frac12} $ particle in the three spatial directions.
Also, $i\sigma _{j}$ are the generators of rotation acting on
non-relativistic particles with spin ${\frac12} $. The state of the
particles are represented as two--component spinors.
\par
In \emph{quantum information}, single--qubit quantum gates are $2\times 2$
unitary matrices. The Pauli matrices are some of the most important
single--qubit operations.} (a set of $2\times 2$ complex Hermitian and
unitary matrices), defined as%
\begin{equation*}
\sigma _{1}=\sigma _{x}=%
\begin{pmatrix}
0 & 1 \\
1 & 0%
\end{pmatrix}%
,\qquad \sigma _{2}=\sigma _{y}=%
\begin{pmatrix}
0 & -{\rm i} \\
{\rm i} & 0%
\end{pmatrix}%
,\qquad \sigma _{3}=\sigma _{z}=%
\begin{pmatrix}
1 & 0 \\
0 & -1%
\end{pmatrix}%
,
\end{equation*}%
obeying both the commutation and anticommutation relations%
\begin{equation*}
\lbrack \sigma _{i},\sigma _{j}] = 2{\rm i}\,\varepsilon _{ijk}\,\sigma
_{k},\qquad \{\sigma _{i},\sigma _{j}\} =2\delta _{ij}\cdot I,
\end{equation*}
where $\varepsilon _{ijk}$\ is the Levi--Civita symbol, $\delta _{ij}$ is
the Kronecker delta, and $I$ is the identity matrix.

Now, the Lorentz--invariant form of the Dirac equation (\ref{DiracEq}) for
an electron with a charge $e$ and mass $m_{\mathrm{e}}$, moving with a
4--momentum 1--form $p=p_{\mu }dx^{\mu }$ in a classical electromagnetic
field defined by 1--form $A=A_{\mu }dx^{\mu }$, reads (see, e.g., \cite{QuLeap}):
\begin{equation}
\left\{\mathrm{i}\gamma ^{\mu }\left[ p_{\mu }-eA_{\mu }\right] -m_{\mathrm{e%
}}\right\} \psi (x)=0,  \label{DiracCov}
\end{equation}%
and is called the \textit{covariant Dirac equation}.

The formal QED Lagrangian (density) includes three terms,%
\begin{equation}
\mathcal{L}(x)=\mathcal{L}_{\mathrm{em}}(x)+\mathcal{L}_{\mathrm{int}}(x)+%
\mathcal{L}_{\mathrm{e-p}}(x),  \label{QEDlagrangian}
\end{equation}%
related respectively to the free electromagnetic field 2--form ~$F=F_{\mu \nu
}dx^{\mu }\wedge dx^{\nu }$,~ the electron--positron field (in the presence
of the external vector potential 1--form $A_{\mu }^{\mathrm{ext}})$, and the
interaction field (dependent on the charge--current 1--form ~$J=J_{\mu
}dx^{\mu }$). The free electromagnetic field Lagrangian in (\ref%
{QEDlagrangian}) has the standard electrodynamic form%
\begin{equation*}
\mathcal{L}_{\mathrm{em}}(x)=-\frac{1}{4}F^{\mu \nu }F_{\mu \nu },
\end{equation*}%
where the electromagnetic fields are expressible in terms of components of
the potential 1--form ~
$A=A_{\mu }dx^{\mu }$~ by%
\begin{equation*}
F_{\mu \nu }=\partial _{\mu }A_{\nu }^{\mathrm{tot}}-\partial _{\nu }A_{\mu
}^{\mathrm{tot}},\quad \text{with\quad }A_{\mu }^{\mathrm{tot}}=A_{\mu }^{%
\mathrm{ext}}+A_{\mu }.
\end{equation*}

The electron-positron field Lagrangian is given by Dirac's equation (\ref%
{DiracCov}) as
\begin{equation*}
\mathcal{L}_{\mathrm{e-p}}(x)=\bar{\psi}(x)\left\{\mathrm{i}\gamma ^{\mu }%
\left[ p_{\mu }-eA_{\mu }^{\mathrm{ext}}\right] -m_{\mathrm{e}}\right\} \psi
(x),
\end{equation*}%
where $\bar{\psi}(x)$ is the Dirac adjoint spinor wave function.

The interaction field Lagrangian%
\begin{equation*}
\mathcal{L}_{\mathrm{int}}(x)=-J^{\mu }A_{\mu },
\end{equation*}%
accounts for the interaction between the uncoupled electrons and the
radiation field.

The field equations deduced from (\ref{QEDlagrangian}) read%
\begin{eqnarray}
\left\{ \mathrm{i}\gamma ^{\mu }\left[ p_{\mu }-eA_{\mu }^{\mathrm{ext}}%
\right] -m_{\mathrm{e}}\right\} \psi (x) &=&\gamma ^{\mu }\psi (x)A_{\mu },
\notag \\
\partial ^{\mu }F_{\mu \nu } &=&J_{\nu }.  \label{QEDformal}
\end{eqnarray}%
The formal QED requires the solution of the system (\ref{QEDformal}) when $%
A^{\mu }(x),$ $\psi (x)$ and $\bar{\psi}(x)$ are quantized fields.

\subsubsection{Abelian Higgs Model}

The Abelian\footnote{An \emph{Abelian} (or, commutative) \emph{group} (even better, Lie group, see Appendix), is such a group $G$ that satisfies the condition:
~$a \cdot b = b \cdot a$ for all $a, b \in G$. In other words, its \emph{commutator}, ~$[a, b] := a^{-1}b^{-1}ab$~
equals the identity element.} Higgs model is an example of gauge theory used in particle and condensed matter physics. Besides the electromagnetic field
it contains a self-interacting scalar field, the so-called \emph{Higgs field}%
, minimally coupled to electromagnetism. From the conceptual point of view,
it is advantageous to consider this field theory in ($2+1$)D space-time and
to extend it subsequently to ($3+1$)D for applications. The Abelian Higgs
Lagrangian reads \cite{Lenz}
\[
\mathcal{L}=-\frac{1}{4}F_{\mu \nu }F^{\mu \nu }+(D_{\mu }\phi )^{\ast
}(D^{\mu }\phi )-V(\phi ),
\]
which contains the complex (charged), self-interacting scalar field $\phi $.
The Higgs potential is the \emph{Mexican hat} function of the real and imaginary part of the
Higgs field,
\[
V(\phi )=\frac{1}{4}\lambda (|\phi |^{2}-a^{2})^{2}.
\]
By construction, this Higgs potential is minimal along a circle $|\phi |=a$
in the complex $\phi $ plane. The constant $\lambda $ controls the strength
of the self--interaction of the Higgs field and, for stability reasons, is
assumed to be positive, \ $\lambda \geq 0\,.$ The Higgs field is minimally
coupled to the radiation field $A_{\mu }$, i.e.,~the partial derivative $%
\partial _{\mu }$ is replaced by the covariant derivative, \ $D_{\mu
}=\partial _{\mu }+\mathrm{i}eA_{\mu }.$ Gauge fields and field strengths
are related by
\[
F_{\mu \nu }=\partial _{\mu }A_{\nu }-\partial _{\nu }A_{\mu }=\frac{1}{%
\mathrm{i}e}\left[ D_{\mu },D_{\nu }\right] .
\]

The inhomogeneous Maxwell equations are obtained from the  least action
principle,
\[
\delta S=0,\qquad \text{with \ \ \ }S=\int \mathcal{L}d^{4}x=0,
\]
by variation of the action $S$ with respect to the gauge fields $A_{\mu }$
(and their derivatives $\partial _{\mu }A_{\mu }$). With
\begin{eqnarray*}
\frac{\delta \mathcal{L}}{\delta \partial _{\mu }A_{\nu }} &=&-F^{\mu \nu
},\qquad \frac{\delta \mathcal{L}}{\delta A_{\nu }}=-j^{\nu }\,,\qquad \text{%
we get} \\
\partial _{\mu }F^{\mu \nu } &=&j^{\nu },\qquad j_{\nu }=\mathrm{i}e(\phi
^{\star }\partial _{\nu }\phi -\phi \partial _{\nu }\phi ^{\star
})-2e^{2}\phi ^{\ast }\phi A_{\nu }.
\end{eqnarray*}

We remark here that the homogeneous Maxwell equations are not dynamical
equations of motion. They are integrability conditions and guarantee that
the field strength can be expressed in terms of the gauge fields. The
homogeneous equations follow from the \textit{Jacobi identit}y of the
covariant derivative
\[
\left[ D_{\mu },\left[ D_{\nu },D_{\sigma }\right] \right] +\left[ D_{\sigma
},\left[ D_{\mu },D_{\nu }\right] \right] +\left[ D_{\nu },\left[ D_{\sigma
},D_{\mu }\right] \right] =0.
\]
Multiplication with the totally antisymmetric 4--tensor $\epsilon ^{\mu \nu
\rho \sigma },$ yields the homogeneous equations for the dual field strength
$\tilde{F}^{\mu \nu }$
\[
\left[ D_{\mu },\tilde{F}^{\mu \nu }\right] =0,\qquad \tilde{F}^{\mu \nu }=%
\frac{1}{2}\epsilon ^{\mu \nu \rho \sigma }F_{\rho \sigma }.
\]
The transition: \ $F\longrightarrow \tilde{F}$ \ corresponds to the
following duality relation of electric and magnetic fields, \ $\mathbf{E}%
\longrightarrow \mathbf{B},$ $\mathbf{B}\longrightarrow -\mathbf{E}.$

Variation with respect to the charged matter field yields the equation of
motion:
\[
D_{\mu }D^{\mu }\phi +\frac{\delta V}{\delta \phi ^{\ast }}=0.
\]

Gauge theories contain redundant variables. This redundancy manifests itself
in the presence of local symmetry transformations, or gauge transformations,
~$U(x)=e^{\mathrm{i}e\alpha (x)}, $~ which rotate the phase of the matter
field and shift the value of the gauge field in a space-time dependent
manner
\begin{equation}
\phi \longrightarrow \phi ^{\,[U]}=U(x)\phi (x)\,,\qquad A_{\mu
}\longrightarrow A_{\mu }^{\,[U]}=A_{\mu }+U(x)\,\frac{1}{\mathrm{i}e}%
\,\partial _{\mu }\,U^{\dagger }(x)\,.  \label{gt}
\end{equation}
The covariant derivative {\ $D_{\mu }$} has been defined such that {\ $%
D_{\mu }\phi $} transforms covariantly, i.e.,~like the matter field {\ $\phi
$} itself.
\[
D_{\mu }\phi (x)\longrightarrow U(x)\,D_{\mu }\phi (x).
\]
This transformation property together with the invariance of $F_{\mu \nu }$
guarantees invariance of $\mathcal{L}$ and of the equations of motion. A
gauge field which is gauge equivalent to $A_{\mu }=0$ is called a pure
gauge. According to (\ref{gt}) a pure gauge satisfies
\[
A_{\mu }^{pg}(x)=U(x)\frac{1}{\mathrm{i}e}\,\partial _{\mu }\,U^{\dagger
}(x)=-\partial _{\mu }\,\alpha (x),
\]
and the corresponding field strength vanishes.

Note that the non-Abelian Higgs model has the action:
$$S(\phi , A) = {1\over 4}\int {\rm Tr}(F^{\mu\nu}F_{\mu\nu}) + |D\phi|^2 + V(|\phi|),$$ where now the non-Abelian field $A$ is contained both in the covariant derivative $D$ and in the components $F^{\mu\nu}$ and $F_{\mu\nu}$ (see Yang--Mills theory below).

\section{Feynman Path Integral}

\subsection{The action--amplitude formalism}

The `driving engine' of quantum field theory is the Feynman path integral.
Very briefly, there are three basic forms of the path integral (see, e.g.,
\cite{Complexity,QuLeap}):\newline

1. \emph{Sum--over--histories,} developed in Feynman's version of quantum
mechanics (QM)\footnote{%
Feynman's \emph{amplitude} is a space-time version of the Schr\"{o}dinger's
\emph{wavefunction} $\psi $, which describes how the (non-relativistic)
quantum state of a physical system changes in space and time, i.e.,
\begin{equation*}
\langle \mathrm{Out}_{t_{fin}}|\mathrm{In}_{t_{ini}}\rangle =\psi (\mathbf{x}%
,t),\qquad (\text{for~~}\mathbf{x}\in \lbrack \mathrm{In},\mathrm{Out}%
],~t\in \lbrack t_{ini},t_{fin}]).
\end{equation*}%
In particular, quantum wavefunction $\psi $ is a complex--valued function of
real space variables $\mathbf{x}=(x_{1},x_{2},...,x_{n})\in \mathbb{R}^{n}$,
which means that its domain is in $\mathbb{R}^{n}$ and its range is in the
complex plane, formally $\psi (\mathbf{x}):\mathbb{R}^{n}\rightarrow \mathbb{%
C}.$ For example, the one--dimensional \emph{stationary plane wave} with
wave number $k$ is defined as
\begin{equation*}
\psi (x)=\mathrm{e}^{\mathrm{i}kx},\qquad (\mathrm{for~~}x\in \mathbb{R}),
\end{equation*}%
where the real number $k$ describes the wavelength, $\lambda =2\pi /k.$ In $%
n $ dimensions, this becomes
\begin{equation*}
\psi (x)=\mathrm{e}^{\mathrm{i}\mathbf{p\cdot x}},
\end{equation*}%
where the momentum vector $\mathbf{p=k}$ is the vector of the wave numbers $%
\mathbf{k}$ in natural units (in which $\hbar =m=1$).
\par
More generally, quantum wavefunction is also time dependent, $\psi =\psi (%
\mathbf{x,t}).$ The time--dependent plane wave is defined by%
\begin{equation}
\psi (\mathbf{x,t})=\mathrm{e}^{\mathrm{i}\mathbf{p\cdot x-}\mathrm{i}%
p^{2}t/2}.  \label{plW}
\end{equation}%
\par
In general, $\psi (\mathbf{x,t})$ is governed by the Schr\"{o}dinger
equation \cite{VQM,QuLeap} (in natural units $\hbar =m=0$)
\begin{equation}
\mathrm{i}\frac{\partial }{\partial t}\psi (\mathbf{x},t)=-\frac{1}{2}\Delta
\psi (\mathbf{x},t),  \label{schgen}
\end{equation}%
where $\Delta $ is the $n-$dimensional Laplacian. The solution of (\ref%
{schgen}) is given by the integral of the time--dependent plane wave (\ref%
{plW}),%
\begin{equation*}
\psi (\mathbf{x},t)=\frac{1}{(2\pi )^{n/2}}\int_{\mathbb{R}^{n}}\mathrm{e}^{%
\mathrm{i}\mathbf{p\cdot x-}\mathrm{i}p^{2}t/2}\hat{\psi}_{0}(\mathbf{p}%
)d^{n}p,
\end{equation*}%
which means that $\psi (\mathbf{x},t)$ is the inverse Fourier transform of
the function%
\begin{equation*}
\hat{\psi}(\mathbf{p},t)=\mathrm{e}^{\mathbf{-}\mathrm{i}p^{2}t/2}\hat{\psi}%
_{0}(\mathbf{p}),
\end{equation*}%
where $\hat{\psi}_{0}(\mathbf{p})$ has to be calculated for each initial
wavefunction. For example, if initial wavefunction is Gaussian,
\begin{equation*}
f(x)=\exp (-a\frac{x^{2}}{2}),\qquad \text{with the Fourier transform}\qquad
\hat{f}(p)=\frac{1}{\sqrt{a}}\exp (-\frac{p^{2}}{2a}).  \label{gaus2}
\end{equation*}%
$\hspace{2cm}\text{ then}\qquad \hat{\psi}_{0}(p)=\frac{1}{\sqrt{a}}\exp (-%
\frac{p^{2}}{2a}).$} \cite{FeynQM};

2. \emph{Sum--over--fields,} started in Feynman's version of quantum
electrodynamics (QED) \cite{FeynQED}\newline
and later improved by Fadeev--Popov \cite{Faddeev};

3. \emph{Sum--over--geometries/topologies} in quantum gravity (QG),
initiated by S. Hawking and properly developed in the form of causal
dynamical triangulations (see \cite{Ambjorn}; for a `softer' review, see
\cite{Loll}).

In all three versions, Feynman's \emph{action--amplitude formalism} includes
two components:

1. A real--valued, classical, \emph{Hamilton's action functional,}
\begin{equation}
S[\Phi ]~:=~\int_{t_{ini}}^{t_{fin}}{L}[\Phi ]\,dt,  \label{act}
\end{equation}%
with the Lagrangian energy function defined over the Lagrangian density $%
\mathcal{L}$,
\begin{equation*}
{L}[\Phi ]=\int d^{n}x\,\mathcal{L}(\Phi ,\partial _{\mu }\Phi ),\qquad
(\partial _{\mu }\equiv \partial /\partial x^{\mu }),
\end{equation*}%
while $\Phi $ is a common symbol denoting all three things to be summed upon
(histories, fields and geometries). The action functional $S[\Phi ]$ obeys
the \emph{Hamilton's least action principle,} ~$\delta S[\Phi ]=0,$~ and
gives, using standard variational methods,\footnote{%
In Lagrangian field theory, the fundamental quantity is the action
\begin{equation*}
S[\Phi ]=\int_{t_{in}}^{t_{out}}L\,dt=\int_{\mathbb{R}^{4}}d^{n}x\,\mathcal{L%
}(\Phi ,\partial _{\mu }\Phi )\,,
\end{equation*}%
so that the least action principle, $\delta S[\Phi ]=0,$ gives
\begin{eqnarray*}
0 &=&\int_{\mathbb{R}^{4}}d^{n}x\,\left\{ \frac{\partial \mathcal{L}}{%
\partial \Phi }\delta \Phi +\frac{\partial \mathcal{L}}{\partial (\partial
_{\mu }\Phi )}\delta (\partial _{\mu }\Phi )\right\}  \\
&=&\,\int_{\mathbb{R}^{4}}d^{n}x\,\left\{ \frac{\partial \mathcal{L}}{%
\partial \Phi }\delta \Phi -\partial _{\mu }\left( \frac{\partial \mathcal{L}%
}{\partial (\partial _{\mu }\Phi )}\right) \delta \Phi +\partial _{\mu
}\left( \frac{\partial \mathcal{L}}{\partial (\partial _{\mu }\Phi )}\delta
\Phi \right) \right\} .
\end{eqnarray*}%
The last term can be turned into a surface integral over the boundary of the
$\mathbb{R}^{4}$ (4D space-time region of integration). Since the initial
and final field configurations are assumed given, $\delta \Phi =0$ at the
temporal beginning $t_{in}$ and end $t_{out}$\ of this region, which implies
that the surface term is zero. Factoring out the $\delta \Phi $ from the
first two terms, and since the integral must vanish for arbitrary $\delta
\Phi $, we arrive at the Euler-lagrange equation of motion for a field,
\begin{equation*}
\partial _{\mu }\left( \frac{\partial \mathcal{L}}{\partial (\partial _{\mu
}\Phi )}\right) -\frac{\partial \mathcal{L}}{\partial \Phi }=0.
\end{equation*}%
If the Lagrangian (density) $\mathcal{L}$\ contains more fields, there is
one such equation for each. The momentum density $\pi (x)$ of a field,
conjugate to $\Phi (x)$ is defined as: ~ $\pi (x)=\frac{\partial \mathcal{L}%
}{\partial _{\mu }\Phi (x)}.$%
\par
For example, the standard electromagnetic action
\begin{equation*}
S=-\frac{1}{4}\int_{\mathbb{R}^{4}}d^{4}x\,F_{\mu \nu }F^{\mu \nu },\qquad
\text{where}\qquad F_{\mu \nu }=\partial _{\mu }A_{\nu }-\partial _{\nu
}A_{\mu },
\end{equation*}%
gives the sourceless Maxwell's equations: ~
\begin{equation*}
\partial _{\mu }F^{\mu \nu }=0,\qquad \epsilon ^{\mu \nu \sigma \eta
}\partial _{\nu }F_{\sigma \eta }=0,
\end{equation*}%
where the field strength tensor $F_{\mu \nu }$ and the Maxwell equations are
invariant under the \emph{gauge transformations,}
\begin{equation*}
A_{\mu }\longrightarrow A_{\mu }+\partial _{\mu }\epsilon .
\end{equation*}%
\par
The equations of motion of charged particles are given by the Lorentz--force
equation,
\begin{equation*}
m{\frac{du^{\mu }}{d\tau }}=eF^{\mu \nu }u_{\nu },
\end{equation*}%
where $e$ is the charge of the particle and $u^{\mu }(\tau )$ its
four-velocity as a function of the proper time.} the Euler--Lagrangian
equations, which define the shortest path, the extreme field, and the
geometry of minimal curvature (and without holes).

2. A complex--valued, quantum \emph{transition amplitude},\footnote{%
The transition amplitude is closely related to \emph{partition function} $Z,$
which is a quantity that encodes the statistical properties of a system in
thermodynamic equilibrium. It is a function of temperature and other
parameters, such as the volume enclosing a gas. Other thermodynamic
variables of the system, such as the total energy, free energy, entropy, and
pressure, can be expressed in terms of the partition function or its
derivatives. In particular, the partition function of a \emph{canonical
ensemble}~is defined as a sum ~~$Z(\beta )=\sum_{j}\mathrm{e}^{-\beta
E_{j}}, $~ where $\beta =1/(k_{B}T)$ is the `inverse temperature', where $T$
is an ordinary temperature and $k_{B}$ is the Boltzmann's constant. However,
as the position $x^{i}$ and momentum $p_{i}$ variables of an $i$th particle
in a system can vary continuously, the set of microstates is actually
uncountable. In this case, some form of \textit{coarse--graining} procedure
must be carried out, which essentially amounts to treating two mechanical
states as the same microstate if the differences in their position and
momentum variables are `small enough'. The partition function then takes the
form of an integral. For instance, the partition function of a gas
consisting of $N$ molecules is proportional to the $6N-$dimensional
phase--space integral,
\begin{equation*}
Z(\beta )\sim \int_{\mathbb{R}^{6N}}\,d^{3}p_{i}\,d^{3}x^{i}\exp [-\beta
H(p_{i},x^{i})],
\end{equation*}%
where $H=H(p_{i},x^{i}),$ ($i=1,...,N$) \ is the classical Hamiltonian
(total energy) function.
\par
Given a set of random variables $X_{i}$ taking on values $x^{i}$, and purely
potential Hamiltonian function $H(x^{i})$, the partition function is defined
as
\begin{equation*}
Z(\beta )=\sum_{x^{i}}\exp \left[ -\beta H(x^{i})\right] .
\end{equation*}%
The function $H$ is understood to be a real-valued function on the space of
states $\{X_{1},X_{2},\cdots \}$ while $\beta $ is a real-valued free
parameter (conventionally, the inverse temperature). The sum over the $x^{i}$
is understood to be a sum over all possible values that the random variable $%
X_{i}$ may take. Thus, the sum is to be replaced by an integral when the $%
X_{i}$ are continuous, rather than discrete. Thus, one writes
\begin{equation*}
Z(\beta )=\int dx^{i}\exp \left[ -\beta H(x^{i})\right] ,
\end{equation*}%
for the case of continuously-varying random variables $X_{i}$.
\par
Now, the number of variables $X_{i}$ need not be countable, in which case
the set of coordinates $\{x^{i}\}$ becomes a field\ $\phi =\phi (x),$ so\
the sum is to be replaced by the \emph{Euclidean path integral} (that is a
Wick--rotated Feynman transition amplitude (\ref{Wick}) in imaginary time),
as
\begin{equation*}
Z(\phi )=\int \mathcal{D}[\phi ]\exp \left[ -H(\phi )\right] .
\end{equation*}%
\par
More generally, in quantum field theory, instead of the field Hamiltonian $%
H(\phi )$ we have the action $S(\phi )$ of the theory. Both Euclidean path
integral,
\begin{equation}
Z(\phi )=\int \mathcal{D}[\phi ]\exp \left[ -S(\phi )\right] ,\qquad \text{%
real path integral in imaginary time,}  \label{Eucl}
\end{equation}%
and Lorentzian one,
\begin{equation}
Z(\phi )=\int \mathcal{D}[\phi ]\exp \left[ \mathrm{i}S(\phi )\right]
,\qquad \text{complex path integral in real time,}  \label{Lor}
\end{equation}%
are usually called `partition functions'. While the Lorentzian path integral
(\ref{Lor}) represents a quantum-field theory-generalization of the Schr\"{o}%
dinger equation, the Euclidean path integral (\ref{Eucl}) represents a
statistical-field-theory generalization of the Fokker--Planck equation.}
\begin{equation}
\langle \mathrm{Out}_{t_{fin}}|\mathrm{In}_{t_{ini}}\rangle ~:=\int_{\mathrm{%
\Omega }}\mathcal{D}[\Phi ]\,\mathrm{e}^{{\rm i}S[\Phi ]},  \label{pathInt1}
\end{equation}%
where $\mathcal{D}[\Phi ]$ is `an appropriate' Lebesgue--type measure,
\begin{equation*}
\mathcal{D}[\Phi ]=\lim_{N\rightarrow \infty }\prod_{s=1}^{N}\Phi
_{s}^{i},\qquad (i=1,...,n),
\end{equation*}%
so that we can `safely integrate over a continuous spectrum and sum over a
discrete spectrum of our problem domain $\Omega $', of which the absolute
square is the real--valued probability density function,
\begin{equation*}
P~:=~|\langle \mathrm{Out}_{t_{fin}}|\mathrm{In}_{t_{ini}}\rangle \rangle
|^{2}.
\end{equation*}

This procedure can be redefined in a mathematically cleaner way if we
Wick--rotate the time variable $t$ to imaginary values, $t\mapsto \tau={} t$%
, thereby making all integrals real:
\begin{equation}
\int \mathcal{D}[\Phi]\, \mathrm{e}^{{\rm i} S[\Phi]}~\cone{Wick}\quad \int
\mathcal{D}[\Phi]\, \mathrm{e}^{-S[\Phi]}.  \label{Wick}
\end{equation}

For example, in non-relativistic quantum mechanics, the propagation
amplitude from $x_{a}$\ to $x_{b}$\ is given by the \emph{configuration path
integral}
\begin{equation*}
U(x_{a},x_{b};T) =\left\langle x_{b}|x_{a}\right\rangle =\left\langle
x_{b}|\,\mathrm{e}^{-\mathrm{i}HT}|x_{a}\right\rangle =\int \mathcal{D}
[x(t)]\,\mathrm{e}^{\mathrm{i}S[x(t)]},
\end{equation*}
which satisfies the Schr\"{o}dinger equation (in natural units)
\begin{equation*}
i\frac{\partial }{\partial T}U(x_{a},x_{b};T)=\hat{H} U(x_{a},x_{b};T),%
\qquad \text{where \ \ \ \ \ }\hat{H}=-\frac{1}{2}\frac{\partial ^{2}}{%
\partial x_{b}^{2}}+V(x_{b}).
\end{equation*}

The \emph{phase--space path integral} (without peculiar constants in the
functional measure) reads
\begin{equation*}
U(q_{a},q_{b};T)=\left( \prod_{i}\int \mathcal{D}[q(t)]\mathcal{D}%
[p(t)]\right) \exp \left[ \mathrm{i}\int_{0}^{T}\left( p_{i}\dot{q}%
^{i}-H(q,p)\right) \,dt\right] ,
\end{equation*}
where the functions $q(t)$\ (space coordinates) are constrained at the
endpoints, but the functions $p(t)$ (canonically--conjugated momenta) are
not. The functional measure is just the product of the standard integral
over phase space at each point in time
\begin{equation*}
\mathcal{D}[q(t)]\mathcal{D}[p(t)]=\prod_{i}\frac{1}{2\pi }\int dq^{i}dp_{i}.
\end{equation*}
Applied to a non-relativistic real scalar field $\phi(x,t)$, this path
integral becomes
\begin{equation*}
\left\langle \phi _{b}(x,t)|\,\mathrm{e}^{-\mathrm{i}HT}|\phi
_{a}(x,t)\right\rangle =\int \mathcal{D}[\phi ]\exp \left[ \mathrm{i}%
\int_{0}^{T}\mathcal{L}(\phi)\,d^{4}x\right] ,\quad \text{with \ \ }\mathcal{%
L}(\phi)=\frac{1}{2}(\partial _{\mu }\phi )^{2}-V(\phi ).
\end{equation*}

\subsection{Correlation functions and generating functional}

If we have two fields in the interacting theory, the corresponding
two--point correlation function, or two--point Green's function, is denoted
by ~$\left\langle \Omega |T\{\phi (x)\phi (y)\}|\Omega \right\rangle , $~
where the notation $\left\vert \Omega \right\rangle $ is introduced to
denote the ground state of the interacting theory, which is generally
different from $\left\vert 0\right\rangle ,$ the ground state of the free
theory. The correlation function can be interpreted physically as the
amplitude for propagation of a particle or excitation between $y$ and $x.$
In the free theory, it is simply the Feynman propagator
\begin{equation*}
\left\langle 0|T\{\phi (x)\phi (y)\}|0\right\rangle _{\mathrm{free}%
}=D_{F}(x-y)=\int \frac{d^{4}p}{(2\pi )^{4}}\frac{\mathrm{i\,e}^{-ip\cdot
(x-y)}}{p^{2}-m^{2}+\mathrm{i}\epsilon }.
\end{equation*}
We would like to know how this expression changes in the interacting theory.
Once we have analyzed the two--point correlation functions, it will be easy
to generalize our results to higher correlation functions in which more than
two field operators appear.

In general we have:
\begin{equation*}
\left\langle \Omega |T\{\phi (x)\phi (y)\}|\Omega \right\rangle
=\lim_{T\rightarrow \infty (1-i\epsilon )}\frac{\left\langle 0|T\{\phi
_{I}(x)\phi _{I}(y)\exp [-\mathrm{i}\int_{-T}^{T}dt\,H_{I}(t)]\}|0\right%
\rangle }{\left\langle 0|T\{\exp [-\mathrm{i}\int_{-T}^{T}dt\,H_{I}(t)]\}|0%
\right\rangle },
\end{equation*}
\begin{equation*}
\left\langle 0|T\{\phi _{I}(x)\phi _{I}(y)\exp [-\mathrm{i}%
\int_{-T}^{T}dt\,H_{I}(t)]\}|0\right\rangle =\left(
\begin{array}{c}
\text{sum of all possible Feynman diagrams} \\
\text{with two external points}%
\end{array}
\right) ,
\end{equation*}
where each diagram is built out of Feynman propagators, vertices and
external points.

The virtue of considering the time--ordered product is clear: It allows us
to put everything inside one large $T-$operator. A similar formula holds for
higher correlation functions of arbitrarily many fields; for each extra
factor of $\phi $ on the left, put an extra factor of $\phi _{I}$ on the
right.

In the interacting theory, the corresponding two--point correlation function
is given by
\begin{equation*}
\left\langle \Omega |T\{\phi (x)\phi (y)\}|\Omega \right\rangle =\left(
\begin{array}{c}
\text{sum of all connected diagrams} \\
\text{with two external points}%
\end{array}
\right) .
\end{equation*}
This is generalized to higher correlation functions as
\begin{equation*}
\left\langle \Omega |T\{\phi (x_{1})...\phi (x_{n})\}|\Omega \right\rangle
=\left(
\begin{array}{c}
\text{sum of all connected diagrams} \\
\text{with }n\text{ external points}%
\end{array}
\right) .
\end{equation*}

In a scalar field theory, the generating functional of correlation functions
is defined as
\begin{equation*}
Z[J]=\int \mathcal{D}[\phi ]\exp \left[ \mathrm{i}\int d^{4}x\left[ \mathcal{%
L}+J(x)\phi (x)\right] \,\right] =\left\langle \Omega |\,\mathrm{e}^{-%
\mathrm{i}HT}|\Omega \right\rangle =\,\mathrm{e}^{-iE[J]};
\end{equation*}
this is a functional integral over $\phi (x)$ in which we have added a
source term $J(x)\phi (x)$ to $\mathcal{L}=\mathcal{L}(\phi)$.

For example, the generating functional of the free Klein--Gordon theory is
simply
\begin{equation*}
Z[J]=Z_{0}\exp \left[ -\frac{1}{2}\int d^{4}xd^{4}yJ(x)D_{F}(x-y)J(y)\right]
.
\end{equation*}

\subsection{Quantization of the electromagnetic field}

Consider the path integral
\begin{eqnarray*}
Z[A] &=&\int \mathcal{D}[A]\,\mathrm{e}^{\mathrm{i}S[A]},\qquad \text{where
the action for the free e.-m. field is} \\
S[A] &=&\int d^{4}x\left[ -\frac{1}{4}(F_{\mu \nu })^{2}\right] =\frac{1}{2}%
\int d^{4}x\,A_{\mu }(x)\left( \partial ^{2}g^{\mu \nu }-\partial ^{\mu
}\partial ^{\nu }\right) A_{\nu }(x).
\end{eqnarray*}

$Z[A]$ is the path integral over each of the four spacetime components:
\begin{equation*}
\mathcal{D}[A]=\mathcal{D}[A]^{0}\mathcal{D}[A]^{1}\mathcal{D}[A]^{2}%
\mathcal{D}[A]^{3}.
\end{equation*}
This functional integral is badly divergent, due to gauge invariance. Recall
that $F_{\mu \nu },$ and hence $L,$ is invariant under a general gauge
transformation of the form
\begin{equation*}
A_{\mu }(x)\rightarrow A_{\mu }(x)+\frac{1}{e}\partial _{\mu }\alpha (x).
\end{equation*}
The troublesome modes are those for which \ $A_{\mu }(x)=\partial _{\mu
}\alpha (x),$ that is, those that are gauge--equivalent to \ $A_{\mu }(x)=0.$
The path integral is badly defined because we are redundantly integrating
over a continuous infinity of physically equivalent field configurations. To
fix this problem, we would like to isolate the interesting part of the path
integral, which counts each physical configuration only once. This can be
accomplished using the \emph{Faddeev--Popov trick}, which effectively adds a
term to the system Lagrangian and after which we get
\begin{equation*}
Z[A]=\int \mathcal{D}[A]\exp \left[ \mathrm{i}\int_{-T}^{T}d^{4}x\,\left[
\mathcal{L}-\frac{1}{2\xi }(\partial ^{\mu }A_{\mu })^{2}\right] \right],
\end{equation*}
where $\xi $ is any finite constant.

This procedure needs also to be applied to the formula for the two--point
correlation function
\begin{equation*}
\left\langle \Omega |T\,\mathcal{O}(A)|\Omega \right\rangle
=\lim_{T\rightarrow \infty (1-i\epsilon )}\frac{\int \mathcal{D}[A]\,%
\mathcal{O}(A)\exp \left[ \mathrm{i}\int_{-T}^{T}d^{4}x\,\mathcal{L}\right]
\mathcal{\,}}{\int \mathcal{D}[A]\exp \left[ \mathrm{i}\int_{-T}^{T}d^{4}x\,%
\mathcal{L}\right] },
\end{equation*}%
which after Faddeev--Popov procedure becomes
\begin{equation*}
\left\langle \Omega |T\,\mathcal{O}(A)|\Omega \right\rangle
=\lim_{T\rightarrow \infty (1-i\epsilon )}\frac{\int \mathcal{D}[A]\,%
\mathcal{O}(A)\exp \left[ \mathrm{i}\int_{-T}^{T}d^{4}x\,\left[ \mathcal{L}-%
\frac{1}{2\xi }(\partial ^{\mu }A_{\mu })^{2}\right] \right] \mathcal{\,}}{%
\int \mathcal{D}[A]\exp \left[ \mathrm{i}\int_{-T}^{T}d^{4}x\,\left[
\mathcal{L}-\frac{1}{2\xi }(\partial ^{\mu }A_{\mu })^{2}\right] \right] }.
\end{equation*}

\section{Path--Integral TQFT}

\subsection{Schwarz--type and Witten--type theories}

Consider a set of fields $\{\phi _{i}\}$ on a Riemannian $n-$manifold $M$
(with a metric $g_{\mu \nu }$) and real functional of these fields, $S[\phi
_{i}]$, which is the action of the theory. Also consider a set of operators $%
\mathcal{O}_{\alpha }(\phi _{i})$ (labeled by some set of indices $\alpha $%
), which are arbitrary functionals of the fields $\{\phi _{i}\}$. The \emph{%
vacuum expectation value} (VEV) of a product of these operators is defined
as the path integral (see \cite{Labastida})
\[
\langle \mathcal{O}_{\alpha _{1}}\mathcal{O}_{\alpha _{2}}\cdots \mathcal{O}%
_{\alpha _{p}}\rangle =\int \mathcal{D}[\phi_{i}]\mathcal{O}_{\alpha _{1}}(\phi _{i})%
\mathcal{O}_{\alpha _{2}}(\phi _{i})\cdots \mathcal{O}_{\alpha _{p}}(\phi
_{i})\exp \left( -S[\phi _{i}]\right) .
\]
A quantum field theory is considered \textit{topological} if it possesses
the following property:
\begin{equation}
\frac{\delta }{\delta g^{\mu \nu }}\langle \mathcal{O}_{\alpha _{1}}\mathcal{%
O}_{\alpha _{2}}\cdots \mathcal{O}_{\alpha _{p}}\rangle =0,  \label{reme}
\end{equation}
i.e., if the VEVs of some set of selected operators remain invariant under
variations of the metric $g_{\mu \nu }$ on $M$. In this case, the operators $%
\mathcal{O}_{\alpha }(\phi _{i})$ are called \textit{observables}.

There are two ways to formally guarantee that condition (\ref{reme}) is
satisfied. The first one corresponds to the situation in which both, the
action, S, as well as the operators $\mathcal{O}_{\alpha }$, are metric
independent. These TQFTs are called \textit{Schwarz--type}. In the case of
Schwarz--type theories one must first construct an action which is
independent of the metric $g_{\mu \nu }$. The method is best illustrated by
considering an example. Let us take into consideration the most interesting
case of this type of theories: \emph{Chern--Simons gauge theory}. The data
in Chern--Simons gauge theory are the following: a differentiable compact
3--manifold $M$, a gauge group $G$, which will be taken simple and compact,
and an integer parameter $k$. The action is the integral of the
\emph{Chern--Simons form} associated to a \emph{gauge connection} $A$ corresponding to the
group $G$,
\begin{equation}
S_{\mathrm{CS}}[A]=\int_{M}\mathrm{Tr}(A\wedge dA+\frac{2}{3}A\wedge A\wedge
A).  \label{valery}
\end{equation}

Observables are constructed out of operators which do not contain the metric
$g_{\mu \nu }$. In gauge invariant theories, as it is the case, one must
also demand for these operators invariance under gauge transformations. An
important set of observables in Chern--Simons gauge theory is constituted by
the trace of the holonomy of the gauge connection $A$ in some representation
$R$ along a 1--cycle $\gamma $, that is the \emph{Wilson loop},\footnote{A \emph{holonomy} on a smooth manifold is a general geometrical consequence of the curvature of the manifold connection, measuring the extent to which parallel transport around closed loops fails to preserve the geometrical data being transported. Related to holonomy is a \emph{Wilson loop}, which is a gauge--invariant observable obtained from the holonomy of the gauge connection around a given loop. More precisely, a Wilson loop is a quantity defined by the trace of a path--ordered exponential of a gauge field $A_\mu$ transported along a closed curve (loop) $\gamma$,
~$W_\gamma = \mathrm{Tr}\,(\, {P}\exp [{\rm i} \oint_\gamma A_\mu dx^\mu ]\,),$~ where $P$ is the path-ordering operator.}
\begin{equation}
\mathrm{Tr}_{R}\left( {\mathrm{Hol}}_{\gamma }(A)\right) =\mathrm{Tr}_{R}{%
\mathrm{P}}\exp \int_{\gamma }A.  \label{silvie}
\end{equation}
The VEVs are labeled by representations $R_{i}$ and embeddings $\gamma
_{i}$ of $S^{1}$ into $M$ \cite{Labastida}
\begin{equation}
\langle \mathrm{Tr}_{R_{1}}{\mathrm{P}}\mathrm{e}^{\int_{\gamma _{1}}A}\dots
\mathrm{Tr}_{R_{n}}{\mathrm{P}}\mathrm{e}^{\int_{\gamma _{n}}A}\rangle
\nonumber \\
=\int [DA]\mathrm{Tr}_{R_{1}}{\mathrm{P}}\mathrm{e}^{\int_{\gamma
_{1}}A}\dots \mathrm{Tr}_{R_{n}}{\mathrm{P}}\mathrm{e}^{\int_{\gamma _{n}}A}%
\mathrm{e}^{\frac{\mathrm{i}k}{4\pi }S_{\mathrm{CS}}(A)}.  \label{encarna}
\end{equation}
A non-perturbative analysis of Chern--Simons gauge theory shows that the
invariants associated to the observables $\mathcal{O}_{\alpha }(\phi _{i})$
are knot and link invariants of polynomial type as the Jones polynomial and
its generalizations. The perturbative analysis has also led to this result
and has shown to provide a very useful framework to study Vassiliev
invariants.

The second way to guarantee (\ref{reme}) corresponds to the case in which
there exists a symmetry, whose infinitesimal form is denoted by $\delta $,
satisfying the following properties:
\begin{equation}
\delta \mathcal{O}_{\alpha }(\phi _{i})=0,\;\;\;\;\;T_{\mu \nu }(\phi
_{i})=\delta G_{\mu \nu }(\phi _{i}),  \label{angela}
\end{equation}
where $T_{\mu \nu }(\phi _{i})$ is the energy--momentum tensor of the
theory, given by
\begin{equation}
T_{\mu \nu }(\phi _{i})=\frac{\delta }{\delta g^{\mu \nu }}S[\phi _{i}],
\label{rosi}
\end{equation}
while $G_{\mu \nu }(\phi _{i})$ is some tensor.

The fact that $\delta $ in (\ref{angela}) is a symmetry of the theory means
that the transformations $\delta \phi _{i}$ of the fields are such that
$\ \delta S[\phi _{i}]=0$~ and $\ \delta \mathcal{O}_{\alpha }(\phi
_{i})=0$. Conditions (\ref{angela}) lead formally to the following relation
for VEVs:
\begin{eqnarray}
&&\frac{\delta }{\delta g^{\mu \nu }}\langle \mathcal{O}_{\alpha _{1}}%
\mathcal{O}_{\alpha _{2}}\cdots \mathcal{O}_{\alpha _{p}}\rangle =  \nonumber
\\
&&-\int \mathcal{D}[\phi_{i}]\mathcal{O}_{\alpha _{1}}(\phi _{i})\mathcal{O}_{\alpha
_{2}}(\phi _{i})\cdots \mathcal{O}_{\alpha _{p}}(\phi _{i})T_{\mu \nu }\exp
\left( -S[\phi _{i}]\right)   \label{rosina} \\
&=&-\int \mathcal{D}[\phi_{i}]\delta \left( \mathcal{O}_{\alpha _{1}}(\phi _{i})%
\mathcal{O}_{\alpha _{2}}(\phi _{i})\cdots \mathcal{O}_{\alpha _{p}}(\phi
_{i})G_{\mu \nu }\exp \left( -S[\phi _{i}]\right) \right) =0,  \nonumber
\end{eqnarray}
which implies that the quantum field theory can be regarded as topological.
In (\ref{rosina}) it has been assumed that the action and the measure $%
\mathcal{D}[\phi_{i}]$ are invariant under the symmetry $\delta $. We have assumed
also in (\ref{rosina}) that the observables are metric--independent. This is
a common situation in this type of theories, but it does not have to be
necessarily so. In fact, in view of (\ref{rosina}), it would be possible to
consider a wider class of operators satisfying:
\begin{equation}
{\frac{\delta }{{\delta g_{\mu \nu }}}}\mathcal{O}_{\alpha }(\phi
_{i})=\delta O_{\alpha }^{\mu \nu }(\phi _{i}),  \label{queso}
\end{equation}
where $O_{\alpha }^{\mu \nu }(\phi _{i})$ is a certain functional of the
fields of the theory.

This second type of TQFTs are called \textit{cohomological} of \textit{%
Witten--type}. One of its main representatives is \textit{Donaldson--Witten
theory}, which can be regarded as a certain \textit{twisted} version of $N=2$
supersymmetric Yang--Mills theory. It is important to remark that the
symmetry $\delta $ must be a scalar symmetry. The reason is that, being a
global symmetry, the corresponding parameter must be covariantly constant
and for arbitrary manifolds this property, if it is satisfied at all,
implies strong restrictions unless the parameter is a scalar.

Most of the TQFTs of cohomological type satisfy the relation:
~$
S[\phi _{i}]=\delta \Lambda (\phi _{i}),
$~
for some functional $\Lambda (\phi _{i})$. This means that the topological observables of the theory
(in particular the partition function itself) are independent of the value
of the coupling constant. For example, consider the VEV \cite
{Labastida}
\begin{equation}
\langle \mathcal{O}_{\alpha _{1}}\mathcal{O}_{\alpha _{2}}\cdots \mathcal{O}%
_{\alpha _{p}}\rangle =\int \mathcal{D}[\phi_{i}]\mathcal{O}_{\alpha _{1}}(\phi _{i})%
\mathcal{O}_{\alpha _{2}}(\phi _{i})\cdots \mathcal{O}_{\alpha _{p}}(\phi
_{i})\exp \left( -{\frac{1}{{g^{2}}}}S[\phi _{i}]\right) .  \label{sansed}
\end{equation}
Under a change in the coupling constant, $1/g^{2}\rightarrow 1/g^{2}-\Delta $%
, one has (assuming that the observables do not depend on the coupling), up
to first order in $\Delta $:
\begin{eqnarray*}
&&\langle \mathcal{O}_{\alpha _{1}}\mathcal{O}_{\alpha _{2}}\cdots \mathcal{O%
}_{\alpha _{p}}\rangle \longrightarrow \langle \mathcal{O}_{\alpha _{1}}%
\mathcal{O}_{\alpha _{2}}\cdots \mathcal{O}_{\alpha _{p}}\rangle  \nonumber
\\
&&\Delta \int \mathcal{D}[\phi_{i}]\delta \left[ \mathcal{O}_{\alpha _{1}}(\phi _{i})%
\mathcal{O}_{\alpha _{2}}(\phi _{i})\cdots \mathcal{O}_{\alpha _{p}}(\phi
_{i})\Lambda (\phi _{i})\exp \left( -{\frac{1}{{g^{2}}}}S[\phi _{i}]\right) %
\right]  \nonumber \\
&&=\langle \mathcal{O}_{\alpha _{1}}\mathcal{O}_{\alpha _{2}}\cdots \mathcal{%
O}_{\alpha _{p}}\rangle .
\end{eqnarray*}
Hence, observables can be computed either in the weak coupling limit,
$g\rightarrow 0$, or in the strong coupling limit, $g\rightarrow \infty $.

\subsection{Hodge decomposition theorem}
\label{Hodge}

The \textit{Hodge star} operator $\star :\Omega ^{p}(M)\rightarrow \Omega
^{n-p}(M)$, which maps any exterior $p-$form $\alpha \in \Omega ^{p}(M)$
into its \emph{dual} $(n-p)-$form $\star \,\alpha \in \Omega ^{n-p}(M)$ on a
smooth $n-$manifold $M$, is defined as (see, e.g. \cite{de
Rham,Hodge})
\begin{equation*}
\alpha \wedge \star \,\beta =\beta \wedge \star \,\alpha =\left\langle
\alpha ,\beta \right\rangle \mu ,\qquad \star \star \alpha
=(-1)^{p(n-p)}\alpha ,\qquad (\text{for }\alpha ,\beta \in \Omega ^{p}(M)),
\end{equation*}%
The $\star $ operator depends on the Riemannian metric $g=g_{ij}$ on $M$ and
also on the orientation (reversing orientation will change the sign) \cite%
{GaneshSprBig,GaneshADG}. Using the star operator, for any two $p-$forms $%
\alpha ,\beta \in \Omega ^{p}(M)$ with compact support on $M$ we define
bilinear and positive--definite Hodge $L^{2}-$inner product as
\begin{equation}
\langle \alpha ,\beta \rangle :=\int_{M}\alpha \wedge \star \,\beta .
\label{2.3}
\end{equation}%
where $\alpha \wedge \star \,\beta $ is an $n-$form.

Given the exterior derivative $d:\Omega ^{p}(M)\rightarrow \Omega ^{p+1}(M)$
on a smooth manifold $M$ (see Appendix), its Hodge dual (or, formal adjoint)
is the \emph{codifferential} $\delta $, a linear map $\delta :\Omega
^{p}(M)\rightarrow \Omega ^{p-1}(M)$, which is a generalization of the
divergence, defined by \cite{de Rham,Hodge}
\begin{equation*}
\delta =(-1)^{n(p+1)+1}\star d\star \qquad \text{so that\qquad }%
d=(-1)^{np}\star \delta \star .
\end{equation*}%
That is, if the dimension $n$ of the manifold $M$ is even, then $\delta
=-\star d\,\star $.

Applied to any $p-$form $\omega \in \Omega ^{p}(M)$, the codifferential $%
\delta $ gives
\begin{equation*}
\delta \omega =(-1)^{n(p+1)+1}\star d\star \omega ,\qquad \delta d\omega
=(-1)^{np+1}\star d\star d\omega .
\end{equation*}
If $\omega =f$ is a $0-$form, or function, then $\delta f=0$. If a $p-$form $%
\alpha$ is a codifferential of a $(p+1)-$form $\beta$, that is $\alpha
=\delta \beta$, then $\beta$ is called the \emph{coexact} form. A $p-$form $%
\alpha$ is \emph{coclosed} if $\delta\alpha=0$; then $\star\,\alpha$ is
closed (i.e., $d\star\alpha=0$) and conversely.

The Hodge codifferential $\delta $ satisfies the following set of rules:
\begin{itemize}
\item $\delta \delta =\delta ^{2}=0,$ \ the same as $\ dd=d^{2}=0;$

\item $\delta \star =(-1)^{p+1}\star d$; \ $\star \,\delta =(-1)^{p}\star d$;

\item $d\delta \star =\star \,\delta d$; \ $\star \,d\delta =\delta d\star $.
\end{itemize}

The codifferential $\delta $ can be coupled with the exterior derivative $d$
to construct the \emph{Hodge Laplacian} $\Delta :$ $\Omega
^{p}(M)\rightarrow \Omega ^{p}(M),$ a harmonic generalization of the
Laplace--Beltrami differential operator, given by
\begin{equation*}
\Delta =\delta d+d\delta =(d+\delta )^{2}.
\end{equation*}
$\Delta $ satisfies the following set of rules:
\begin{equation*}
\delta \,\Delta =\Delta \,\delta =\delta d\delta ;\qquad d\,\Delta =\Delta
\,d=d\delta d;\qquad \star \,\Delta =\Delta \star .
\end{equation*}

A $p-$form $\alpha $ is called \textit{harmonic} iff \
\begin{equation*}
\Delta \alpha =0~\Longleftrightarrow ~d\alpha =\delta \alpha =0.
\end{equation*}
Thus, $\alpha $ is harmonic in a compact domain $D\subset M$ iff it is both
closed and coclosed in $D$. Informally, every harmonic form is both closed
and coclosed. As a proof, we have:
\begin{equation*}
0=\left\langle \alpha ,~\Delta \alpha \right\rangle =\left\langle \alpha
,d\delta \alpha \right\rangle +\left\langle \alpha ,\delta d\alpha
\right\rangle =\left\langle \delta \alpha ,\delta \alpha \right\rangle
+\left\langle d\alpha ,d\alpha \right\rangle .
\end{equation*}
Since $\left\langle \beta ,~\beta \right\rangle \geq 0$ for any form $\beta $%
, $\left\langle \delta \alpha ,\delta \alpha \right\rangle $ and $%
\left\langle d\alpha ,d\alpha \right\rangle $ must vanish separately. Thus, $%
d\alpha =0$ and $\delta \alpha =0.$ All harmonic $p-$forms on a smooth
manifold $M$ form the vector space $H_{\Delta }^{p}(M)$.

Now, the celebrated \emph{Hodge decomposition theorem} (HDT) states that, on
a compact orientable smooth $n-$manifold $M$ (with $n\geq p$), any exterior $%
p-$form can be written as a unique sum of an \emph{exact} form, a \emph{%
coexact} form, and a \emph{harmonic} form. More precisely, for any form $%
\omega \in \Omega ^{p}(M)$ there are unique forms $\alpha \in \Omega
^{p-1}(M),$ $\beta \in \Omega ^{p+1}(M)$ and a harmonic form $\gamma \in
\Omega ^{p}(M),$ such that \bigbreak%
\centerline{\fbox{\parbox{11cm}{
{\large\begin{equation*} {\rm HDT:}\qquad\stackrel{\rm
any\,form}{\omega }\quad =\quad
\stackrel{\rm exact}{d\alpha }\quad + \quad\stackrel{\rm coexact}{\delta\beta }\quad + \quad\stackrel{\rm harmonic}{\gamma }
\end{equation*}}}}}\bigbreak\noindent For the proof, see \cite{de Rham,Hodge}.

In physics community, the exact form $d\alpha$ is called \emph{longitudinal}, while the coexact form $\delta\beta$ is called \emph{transversal}, so that
they are mutually orthogonal. Thus any form can be orthogonally decomposed
into a harmonic, a longitudinal and transversal form. For example, in fluid
dynamics, any vector-field $v$ can be decomposed into the sum of two
vector-fields, one of which is divergence--free, and the other is curl--free.

Since $\gamma $ is harmonic, $d\gamma =0.$ Also, by Poincar\'{e} lemma, $%
d(d\alpha )=0.$ In case $\omega $ is a closed $p-$form, $d\omega =0,$ then
the term $\delta \beta $ in HDT is absent, so we have the \emph{short Hodge
decomposition},
\begin{equation}
\omega =d\alpha +\gamma ,  \label{sHd}
\end{equation}%
thus $\omega $ and $\gamma $ differ by $d\alpha$. In topological
terminology, $\omega $ and $\gamma$ belong to the same \textit{cohomology
class} $[\omega ]\in H^{p}(M)$. Now, by the de Rham theorems it follows that
if $C$ is any $p-$cycle, then
\begin{equation*}
\int_{C}\omega =\int_{C}\gamma ,
\end{equation*}%
that is, $\gamma $ and $\omega $ have the same periods$.$ More precisely, if
$\omega $ is any closed $p-$form, then there exists a unique harmonic $p-$%
form $\gamma $ with the same periods as those of $\omega $ (see \cite{de
Rham,Flanders}).

The \emph{Hodge--Weyl theorem} \cite{de Rham,Hodge} states that every de
Rham cohomology class has a unique harmonic representative. In other words,
the space $H^p_\Delta(M)$ of harmonic $p-$forms on a smooth manifold $M$ is
isomorphic to the $p$th de Rham cohomology group,
\begin{equation}
H^{p}_{DR}(M):=\frac{Z^{p}(M)}{B^{p}{M}}=\frac{\limfunc{Ker}\left( d:\Omega
^{p}(M)\rightarrow \Omega ^{p+1}(M)\right) }{\limfunc{Im}\left( d:\Omega
^{p-1}(M)\rightarrow \Omega ^{p}(M)\right) },  \label{DR}
\end{equation}
or, ~$H^p_\Delta(M) \cong H^p_{DR}(M)$. That is, the harmonic part $\gamma$
of HDT depends only on the global structure, i.e., the topology of $M$.

For example, in $(2+1)$D electrodynamics, $p-$form Maxwell equations in the
Fourier domain $\Sigma $ are written as \cite{Teixeira}%
\begin{equation*}
\begin{array}{ll}
dE=\mathrm{i}\omega B, & \qquad dB=0, \\
dH=-\mathrm{i}\omega D+J, & \qquad dD=Q,%
\end{array}%
\end{equation*}%
where $H$ and $\omega$ are 0--forms (magnetizing field and field frequency),
$D$ (electric displacement field), $J$ (electric current density) and $E$
(electric field) are 1--forms, while $B$ (magnetic field) and $Q$ (electric
charge density) are 2--forms. From $d^{2}=0$ it follows that the $J$ and the
$Q$ satisfy the \textit{continuity equation}%
\begin{equation*}
dJ=\mathrm{i}\omega Q.
\end{equation*}
Constitutive equations, which include all metric information in this
framework, are written in terms of Hodge star operators (that fix an
isomorphism between $p$ forms and $(2-p)$ forms in the $(2+1)$ case)%
\begin{equation*}
D=\star \,E,\qquad B=\star \,H.
\end{equation*}

Applying HDT to the electric field intensity 1--form $E$, we get \cite{He}%
\begin{equation*}
E=d\phi +\delta A+\chi ,
\end{equation*}%
where $\phi $ is a 0--form (a scalar field) and $A$ is a 2--form; $d\phi $
represents the static field and $\delta A$ represents the dynamic field, and
$\chi $ represents the harmonic field component. If domain $\Sigma $ is
contractible, $\chi $ is identically zero and we have the short Hodge
decomposition,%
\begin{equation*}
E=d\phi +\delta A.
\end{equation*}

\subsection{Hodge decomposition and gauge path integral}

\subsubsection{Functional measure on the space of differential forms}

The Hodge inner product (\ref{2.3}) leads to a natural (metric--dependent)
functional measure $\mathcal{D}\mu \lbrack \omega ]$ on $\Omega ^{p}(M)$,
which normalizes the \textit{Gaussian functional integral}
\begin{equation}
\int \mathcal{D}\mu \lbrack \omega ]\,\mathrm{e}^{\mathrm{i}\langle \omega
|\omega \rangle }=1.  \label{2.4}
\end{equation}

One can use the invariance of (\ref{2.4}) to determine how the functional
measure transforms under the Hodge decomposition. Using HDT and its
orthogonality with respect to the inner product (\ref{2.3}), it was shown in
\cite{Gegenberg} that
\begin{equation}
\langle \omega ,\omega \rangle =\langle \gamma ,\gamma \rangle +\langle
d\alpha ,d\alpha \rangle +\langle \delta \beta ,\delta \beta \rangle
=\langle \gamma ,\gamma \rangle +\langle \alpha ,\delta d\alpha \rangle
+\langle \beta ,d\delta \beta \rangle\, ,  \label{2.5}
\end{equation}
where the following differential/conferential identities were used \cite%
{Choquet}
\begin{equation*}
\langle d\alpha ,d\alpha \rangle =\langle \alpha ,\delta d\alpha \rangle
\qquad \text{and\qquad }\langle \delta \beta ,\delta \beta \rangle =\langle
\beta ,d\delta \beta \rangle .
\end{equation*}
Since, for any linear operator $O$, one has
\begin{equation*}
\int \mathcal{D}\mu \lbrack \omega ]\exp \mathrm{i}\langle \omega |O\omega
\rangle ={\det }^{-1/2}(O),
\end{equation*}
(\ref{2.4}) and (\ref{2.5}) imply that
\begin{equation*}
\mathcal{D}\mu \lbrack \omega ]=\mathcal{D}\mu \lbrack \gamma ]\mathcal{D}%
\mu \lbrack \alpha ]\mathcal{D}\mu \lbrack \beta ]\,{\det }^{1/2}(\delta d){%
\det }^{1/2}(d\delta ).
\end{equation*}

\subsubsection{Abelian Chern--Simons theory}

Recall that the classical action for an Abelian Chern--Simons theory,
\begin{equation*}
S=\int_{M}A\wedge dA\,,
\end{equation*}%
is invariant (up to a total divergence) under the gauge transformation:
\begin{equation}
A\longmapsto A+d\varphi .  \label{2.10}
\end{equation}%
We wish to compute the \textit{partition function} for the theory
\begin{equation*}
Z:=\int \frac{1}{V_{G}}\mathcal{D}\mu \lbrack A]\,\mathrm{e}^{\mathrm{i}%
S[A]},
\end{equation*}%
where $V_{G}$ denotes the volume of the group of gauge transformations in (%
\ref{2.10}), which must be factored out of the partition function in order
to guarantee that the integration is performed only over physically distinct
gauge fields. We can handle this by using the Hodge decomposition to
parametrize the potential $A$ in terms of its gauge invariant, and gauge
dependent parts, so that the volume of the group of gauge transformations
can be explicitly factored out, leaving a functional integral over gauge
invariant modes only \cite{Gegenberg}.

We now transform the integration variables:
\begin{equation*}
A\longmapsto \alpha ,\beta ,\gamma ,
\end{equation*}
where $\alpha ,\beta ,\gamma $ parameterize respectively the exact, coexact,
and harmonic parts of the connection A. Using the Jacobian (\ref{2.5}) as
well as the following identity on 0--forms $\Delta =\delta d,$ we get \cite%
{Gegenberg}
\begin{equation*}
Z=\int \frac{1}{V_{G}}\mathcal{D}\mu \lbrack \alpha ]\mathcal{D}\mu \lbrack
\beta ]\mathcal{D}\mu \lbrack \gamma ]\,{\det }^{1/2}\left( \Delta \right) {%
\det }^{1/2}\left( d\delta \right) \mathrm{e}^{\mathrm{i}S},
\end{equation*}
from which it follows that
\begin{equation}
V_{G}=\int \mathcal{D}\mu \lbrack \alpha ],  \label{2.13}
\end{equation}
while the classical action functional becomes, after integrating by parts,
using the harmonic properties of $\gamma $ and the nilpotency of the
exterior derivative operators, and dropping surface terms:
\begin{equation*}
S=-\langle \beta ,\star \delta d\delta \beta \rangle \,\,.
\end{equation*}
Note that $S$ depends only the coexact (transverse) part of $A$. Using (\ref%
{2.13}) and integrating over $\beta $ yields:
\begin{equation*}
Z=\int \mathcal{D}\mu \lbrack \gamma ]{\det }^{-1/2}\left( \star \delta
d\delta \right) {\det }^{1/2}\left( \Delta \right) {\det }^{1/2}\left(
d\delta \right) .
\end{equation*}
Also, it was proven in \cite{Gegenberg} that
\begin{equation*}
{\det }(\star \delta d\delta )={\det }^{1/2}((d\delta d)(\delta d\delta ))={%
\det }^{\frac{3}{2}}(d\delta ).
\end{equation*}
As a consequence of Hodge duality we have the identity
\begin{equation*}
{\det }(\delta d)={\det }(d\delta ),
\end{equation*}
from which it follows that
\begin{equation*}
Z=\int \mathcal{D}\mu \lbrack \gamma ]\,{\det }^{-3/4}\left( \Delta
_{(1)}^{T}\right) {\det }^{1/2}\left( \Delta \right) {\det }^{1/2}\left(
\Delta _{(1)}^{T}\right) \,.
\end{equation*}
The operator $\Delta _{(1)}^{T}$ is the transverse part of the Hodge
Laplacian acting on $1-$forms:
\begin{equation*}
\Delta _{(1)}^{T}:=(\delta d)_{(1)}.
\end{equation*}
Applying identity for the Hodge Laplacian $\Delta _{(p)}$ \cite{Gegenberg}

\begin{equation*}
\det \left( \Delta _{(p)}\right) =\det \left( (\delta d)_{(p)}\right) \det
\left( (\delta d)_{(p-1)}\right) ,
\end{equation*}
we get
\begin{equation*}
{\det }\left( \Delta _{(1)}^{T}\right) ={\det }\left( \Delta _{(1)}\right) /{%
\det }\left( \Delta \right)
\end{equation*}
and hence
\begin{equation*}
Z=\int \mathcal{D}\mu \lbrack \gamma ]\,{\det }^{-1/4}\left( \Delta
_{(1)}\right) {\det }^{3/4}\left( \Delta \right).  \label{Zfin}
\end{equation*}
The space of harmonic forms $\gamma $\ (of any order) is a finite set.
Hence, the integration over harmonic forms (\ref{Zfin}) is a simple sum.

\section{Non-Abelian Gauge Theories}

\subsection{Intro to non-Abelian theories}

QED is the simplest example of a gauge theory coupled to matter based in the
Abelian gauge symmetry of local U(1) phase rotations. However, it is
possible also to construct gauge theories based on non-Abelian groups.
Actually, our knowledge of the strong and weak interactions is based on the
use of such non-Abelian generalizations of QED.

Let us consider a gauge group ${G}$ (see Appendix) with generators $T^{a}$, (%
$a=1,\ldots ,\mathrm{dim}{G}$) satisfying the Lie
algebra
\[
\lbrack T^{a},T^{b}]=\mathrm{i}f^{abc}T^{c}.
\]
A gauge field taking values on the Lie algebra of $\mathcal{G}$ can be
introduced $A_{\mu }\equiv A_{\mu }^{a}T^{a}$, which transforms under a
gauge transformations as
\[
A_{\mu }\longrightarrow {\frac{1}{\mathrm{i}g}}U\partial _{\mu
}U^{-1}+UA_{\mu }U^{-1},\hspace*{1cm}U=\mathrm{e}^{\mathrm{i}\chi
^{a}(x)T^{a}},
\]
where $g$ is the coupling constant. The associated field strength is defined
as
\[
F_{\mu \nu }^{a}=\partial _{\mu }A_{\nu }^{a}-\partial _{\nu }A_{\mu
}^{a}-gf^{abc}A_{\mu }^{b}A_{\nu }^{c}.
\]
Notice that this definition of the $F_{\mu \nu }^{a}$ reduces to the one
used in QED in the Abelian case when $f^{abc}=0$. In general, however,
unlike the case of QED the field strength is not gauge invariant. In terms
of $F_{\mu \nu }=F_{\mu \nu }^{a}T^{a}$ it transforms as
\[
F_{\mu \nu }\longrightarrow UF_{\mu \nu }U^{-1}.
\]

The coupling of matter to a non-Abelian gauge field is done by introducing
again the \textit{covariant derivative}. For a field $\Phi \longrightarrow
U\Phi $\ in a representation of $\mathcal{G}$, the covariant derivative is
given by
\[
D_{\mu }\Phi =\partial _{\mu }\Phi -\mathrm{i}gA_{\mu }^{a}T^{a}\Phi .
\]
With the help of this we can write a generic Lagrangian for a non-Abelian
gauge field coupled to scalars $\phi $ and spinors $\psi $ as
\[
\mathcal{L}=-{\frac{1}{4}}F_{\mu \nu }^{a}F^{\mu \nu \,a}+\mathrm{i}%
\overline{\psi }D\!\!\!\!/\psi +\overline{D_{\mu }\phi }D^{\mu }\phi -%
\overline{\psi }\left[ M_{1}(\phi )+\mathrm{i}\gamma _{5}M_{2}(\phi )\right]
\psi -V(\phi ).
\]

\subsection{Yang--Mills theory}

In non-Abelian gauge theories, {gauge fields} {are matrix-valued functions
of space-time. In SU(N) gauge theories they can be represented by the
generators of the corresponding Lie algebra, i.e.,~gauge fields and their
color components are related by
\begin{equation}
A_{\mu }(x)=A_{\mu }^{a}(x)\frac{\lambda ^{a}}{2},  \label{gfla}
\end{equation}
where the color sum runs over the {$N^{2}-1$} generators. The generators are
hermitian, traceless $N\times N$ matrices whose commutation relations are
specified by the structure constants {$f^{abc}$ \cite{Lenz}
\[
\left[ \frac{\lambda ^{a}}{2},\frac{\lambda ^{b}}{2}\right] =\mathrm{i}%
f^{abc}\frac{\lambda ^{c}}{2}\,.
\]
The normalization is chosen as
\[
\mathrm{Tr}\left( \frac{\lambda ^{a}}{2}\cdot \frac{\lambda ^{b}}{2}\right) =%
\frac{1}{2}\delta _{ab}.
\]
Most of our applications will be concerned with {$SU(2)$} gauge theories; in
this case the generators are the \emph{Pauli matrices},
\[
\lambda ^{a}=\tau ^{a},\qquad \text{{with structure constants \ \ \ \ }}%
f^{abc}=\epsilon ^{abc}.
\]
Covariant derivative, field strength tensor, and its color components are
respectively defined by
\begin{equation}
D_{\mu }=\partial _{\mu }+\mathrm{i}gA_{\mu },
\end{equation}
\begin{equation}
F^{\mu \nu }=\frac{1}{\mathrm{i}g}[D_{\mu },D_{\nu }],\qquad F_{\mu \nu
}^{a}=\partial _{\mu }A_{\nu }^{a}-\partial _{\nu }A_{\mu
}^{a}-gf^{abc}A_{\mu }^{b}A_{\nu }^{c}.  \label{fs}
\end{equation}
The definition of {\ electric and magnetic fields} in terms of the field
strength tensor is the same as in electrodynamics
\begin{equation}
E^{ia}\left( x\right) =-F^{0ia}\left( x\right) ,\qquad B^{ia}\left( x\right)
=-\frac{1}{2}\epsilon ^{ijk}F^{jka}\left( x\right) \ .  \label{eb}
\end{equation}
The dimensions of gauge field and field strength in {4}D space-time are
\[
\lbrack A]=\ell ^{-1},\qquad \lbrack F]=\ell ^{-2},
\]
and therefore in absence of a scale, ~ $A_{\mu }^{a}\sim M_{\mu \nu }^{a}%
\frac{x^{\nu }}{x^{2}}, $~ with arbitrary constants $M_{\mu \nu }^{a}$. In
general, the action associated with these fields exhibits infrared and
ultraviolet logarithmic divergencies. {\ In the following we will discuss }}}

\begin{itemize}
\item  \emph{Yang--Mills Theories:~} Only gauge fields are present. The
Yang--Mills Lagrangian is
\begin{equation}
\mathcal{L}_{YM}=-\frac{1}{4}F^{\mu \nu a}F_{\mu \nu }^{a}=-\frac{1}{2}%
\mathrm{Tr}\,\left( F^{\mu \nu }F_{\mu \nu }\right) =\frac{1}{2}(\mathbf{E}%
^{2}-\mathbf{B}^{2}).  \label{LYM}
\end{equation}

\item  \emph{Quantum Chromodynamics:~} QCD contains besides the gauge fields
(gluons), fermion fields (quarks). Quarks are in the fundamental
representation, i.e.,~in {SU(2)} they are represented by {2-component} color
spinors. The QCD Lagrangian is (flavor dependences suppressed)
\begin{equation}
\mathcal{L}_{QCD}=\mathcal{L}_{YM}+\mathcal{L}_{m},\qquad \mathcal{L}_{m}={%
\bar{\psi}}\left( \mathrm{i}\gamma ^{\mu }D_{\mu }-m\right) \psi ,
\label{matt}
\end{equation}
with the action of the covariant derivative on the quarks given by
\[
(D_{\mu }\psi )^{i}=(\partial _{\mu }\delta ^{ij}+\mathrm{i}gA_{\mu
}^{ij})\,\psi ^{j},\qquad (i,j=1\ldots N\,).
\]

\item  \emph{Georgi--Glashow Model:~} In the Georgi--Glashow model \cite
{GEGL72} (non-Abelian Higgs model), the gluons are coupled to a scalar,
self-interacting ($V(\phi )$) (Higgs) field in the adjoint representation.
The Higgs field has the same representation in terms of the generators as
the gauge field (\ref{gfla}) and can be thought of as a 3-component color
vector in $SU(2)$. Lagrangian and action of the covariant derivative are
respectively
\begin{equation}
\mathcal{L}_{GG}=\mathcal{L}_{YM}+\mathcal{L}_{m},\qquad \mathcal{L}_{m}=%
\frac{1}{2}D_{\mu }\phi D^{\mu }\phi -V(\phi ),  \label{GEGL}
\end{equation}
\begin{equation}
(D_{\mu }\phi )^{a}=\left[ D_{\mu },\phi \,\right] \,^{a}=(\partial _{\mu
}\delta ^{ac}-gf^{abc}A_{\mu }^{b})\phi ^{c}\,.  \label{dphi}
\end{equation}
\end{itemize}

\subsubsection{Yang--Mills action}

The general principle of least action,
\[
\delta S=0,\qquad\text{with}\quad S=\int \mathcal{L}\,d^{4}x,
\]
applied to the gauge fields,
\begin{eqnarray*}
\delta S_{YM} &=&-\int d^{4}x\,\mathrm{Tr}\left( F_{\mu \nu }\delta F^{\mu
\nu }\right) =-\int d^{4}x\,\mathrm{Tr}\left( F_{\mu \nu }\frac{2}{\mathrm{i}%
g}\left[ D^{\mu },\delta A^{\nu }\right] \right) \\
&=&2\int d^{4}x\,\mathrm{Tr}\left( \delta A^{\nu }\left[ D^{\mu },F_{\mu \nu
}\right] \right)
\end{eqnarray*}
gives the {inhomogeneous} field equations \cite{Lenz}
\begin{equation}
\left[ D_{\mu },F^{\mu \nu }\right] =j^{\nu },  \label{emga}
\end{equation}
with $j^{\nu }$ the {color current} associated with the matter fields
\begin{equation}
j^{a\nu }=\frac{\delta \mathcal{L}_{m}}{\delta A_{\nu }^{a}}.  \label{jmat}
\end{equation}
For QCD and the Georgi--Glashow model, these currents are given respectively
by
\begin{equation}
j^{a\nu }=g\bar{\psi}\gamma ^{\nu }\frac{\tau ^{a}}{2}\psi ,\qquad j^{a\nu
}=gf^{abc}\phi ^{b}(D^{\nu }\phi )^{c}\,.
\end{equation}
As in electrodynamics, the {homogeneous} field equations for the Yang--Mills
field strength
\[
\left[ D_{\mu },\tilde{F}^{\mu \nu }\right] =0,
\]
with the dual field strength tensor
\[
\tilde{F}^{\mu \nu }=\frac{1}{2}\,\epsilon ^{\mu \nu \sigma \rho }F_{\sigma
\rho },
\]
are obtained as the \textit{Jacobi identities} of the covariant derivative,
\[
\lbrack D_{\mu },[D_{\nu },D_{\rho }]]+[D_{\nu },[D_{\rho },D_{\mu
}]]+[D_{\rho },[D_{\nu },D_{\mu }]]=0.
\]

\subsubsection{Gauge transformations}

Gauge transformations change the color orientation of the matter fields
locally, i.e.,~in a space-time dependent manner, and are defined as
\[
U\left( x\right) =\exp \left\{ \mathrm{i}g\alpha \left( x\right) \right\}
=\exp \left\{ \mathrm{i}g\alpha ^{a}\left( x\right) \frac{\tau ^{a}}{2}%
\right\} ,
\]
with the arbitrary gauge function $\alpha ^{a}\left( x\right) $. {Matter
fields transform {covariantly} with} $U$
\begin{equation}
\psi vU\psi ,\qquad \phi \longrightarrow U\phi U^{\dagger }.  \label{matr}
\end{equation}
The transformation property of $A$ is chosen such that the covariant
derivatives of the matter fields $D_{\mu }\psi $ and $D_{\mu }\phi $ {\
transform as} the matter fields $\psi $ and $\phi $ respectively. As in
electrodynamics, this requirement makes the gauge fields transform {%
inhomogeneously} \cite{Lenz}
\begin{equation}
A_{\mu }\left( x\right) \longrightarrow U\left( x\right) \left( A_{\mu
}\left( x\right) +\frac{1}{\mathrm{i}g}\partial _{\mu }\right) U^{\dagger
}\left( x\right) =A_{\mu }^{\,\left[ \,U\right] }\left( x\right)
\label{FGT}
\end{equation}
resulting in a covariant transformation law for the field strength
\begin{equation}
F_{\mu \nu }\longrightarrow UF_{\mu \nu }U^{\dagger }.  \label{ufu}
\end{equation}
Under {infinitesimal} gauge transformations ($|g\alpha ^{a}\left( x\right)
|\ll 1$)
\begin{equation}
A_{\mu }^{a}\left( x\right) \longrightarrow A_{\mu }^{a}\left( x\right)
-\partial _{\mu }\alpha ^{a}\left( x\right) -gf^{abc}\alpha ^{b}\left(
x\right) A_{\mu }^{c}\left( x\right) .  \label{ifgt}
\end{equation}
As in electrodynamics, gauge fields which are gauge transforms of $A_{\mu }=0
$ are called pure gauges and are, according to (\ref{FGT}), given by
\begin{equation}
A_{\mu }^{pg}\left( x\right) =U\left( x\right) \frac{1}{\mathrm{i}g}%
\,\partial _{\mu }\,U^{\dagger }\left( x\right) \,.  \label{puga}
\end{equation}
Physical observables must be independent of the choice of gauge (coordinate
system in color space). Local quantities such as the Yang--Mills action
density $\mathrm{Tr}$\thinspace $\left( F^{\mu \nu }(x)F_{\mu \nu
}(x)\right) $ {or matter field bilinears like} $\bar{\psi}(x)\psi (x),\phi
^{a}(x)\phi ^{a}(x)$ {are {gauge invariant}, i.e.,~their value does not
change under local gauge transformations. One also introduces {non-local}
quantities which, in generalization of the transformation law (\ref{ufu})
for the field strength, change homogeneously under gauge transformations. In
this construction a basic building block is the} {\ path ordered integral}
\begin{equation}
\Omega \left( x,y,\mathcal{C}\right) =P\exp \left\{ -\mathrm{i}%
g\int_{s_{0}}^{s}d\sigma \frac{dx^{\mu }}{d\sigma }A_{\mu }\left( x(\sigma
)\right) \right\} =P\exp \left\{ -\mathrm{i}g\int_{\mathcal{C}}dx^{\mu
}A_{\mu }\right\} .
\end{equation}
It describes a {gauge string} between the space-time points $x=x(s_{0})$ and
$y=x(s)$. $\Omega $ satisfies the differential equation
\begin{equation}
\frac{d\Omega }{ds}=-\mathrm{i}g\frac{dx^{\mu }}{ds}A_{\mu }\Omega .
\label{dipo}
\end{equation}
Gauge transforming this differential equation yields the transformation
property of $\Omega $
\begin{equation}
\Omega \left( x,y,\mathcal{C}\right) \longrightarrow U\left( x\right) \Omega
\left( x,y,\mathcal{C}\right) U^{\dagger }\left( y\right) .  \label{gaom}
\end{equation}
With the help of $\Omega $, non-local, gauge invariant quantities like
\[
\mathrm{Tr}\left( F^{\mu \nu }(x)\Omega \left( x,y,\mathcal{C}\right) F_{\mu
\nu }(y)\right) ,\qquad \bar{\psi}(x)\Omega \left( x,y,\mathcal{C}\right)
\psi (y),
\]
or closed gauge strings, the following SU(N)--Wilson loops
\begin{equation}
W_{\mathcal{C}}=\frac{1}{N}\mathrm{Tr}\,\left( \Omega \left( x,x,\mathcal{C}%
\right) \right)   \label{wlop}
\end{equation}
can be constructed. For pure gauges (\ref{puga}), the differential equation (%
\ref{dipo}) is solved by
\begin{equation}
\Omega ^{pg}\left( x,y,\mathcal{C}\right) =U(x)\,U^{\dagger }(y).
\label{wlpg}
\end{equation}
{While $\bar{\psi}(x)\Omega \left( x,y,\mathcal{C}\right) \psi (y)$ is an
operator which connects the vacuum with {meson states} for {$SU(2)$} and {$%
SU(3)$}, {fermionic baryons} appear only in {$SU(3)$} in which gauge
invariant states containing an odd number of fermions can be constructed. In
{SU(3)} a point-like gauge invariant baryonic state is obtained by creating
three quarks in a color antisymmetric state at the same space-time point
\[
\psi (x)\sim \epsilon ^{abc}\psi ^{a}(x)\psi ^{b}(x)\psi ^{c}(x).
\]
Under gauge transformations},
\begin{eqnarray*}
\psi (x) &\longrightarrow &\epsilon ^{abc}U_{a\alpha }(x)\psi ^{\alpha
}(x)U_{b\beta }(x)\psi ^{\beta }(x)U_{c\gamma }(x)\psi ^{\gamma }(x) \\
&=&\det \left( U(x)\right) \epsilon ^{abc}\psi ^{a}(x)\psi ^{b}(x)\psi
^{c}(x)\,.
\end{eqnarray*}
Operators that create finite size baryonic states must contain appropriate
gauge strings as given by the following expression
\[
\psi (x,y,z)\sim \epsilon ^{abc}[\Omega (u,x,\mathcal{C}_{1})\psi
(x)]^{a}\,[\Omega (u,y,\mathcal{C}_{2})\psi (y)]^{b}\,[\Omega (u,z,\mathcal{C%
}_{3})\psi (z)]^{c}\,.
\]
The presence of these gauge strings makes $\psi $ gauge invariant as is
easily verified with the help of the transformation property (\ref{gaom}).
Thus, gauge invariance is enforced by color exchange processes taking place
between the quarks.

\subsection{Quantization of Yang--Mills theory}

{Gauge theories are formulated in terms of redundant variables. Only in this
way, a covariant, local representation of the dynamics of gauge degrees of
freedom is possible. For quantization of the theory both canonically or in
the path integral, redundant variables have to be eliminated. This procedure
is called gauge fixing. It is not unique and the implications of a
particular {choice} are generally not well understood. In the path integral
one performs a sum over all field configurations. In gauge theories this
procedure has to be modified by making use of the decomposition of the space
of gauge fields into equivalence classes, the gauge orbits. Instead of
summing in the path integral over formally different but physically
equivalent fields, the integration is performed over the {equivalence classes%
} of such fields, i.e.,~over the corresponding gauge orbits. The value of the
{action} is gauge invariant, i.e.,~the same for all members of a given gauge
orbit. Therefore, the action is seen to be a functional defined on classes (gauge
orbits) \cite{Lenz}. Also the integration measure
\[
\mathcal{D}\left[ A\right] =\prod_{x,\mu ,a}dA_{\mu }^{a}\left( x\right) .
\]
is gauge invariant since shifts and rotations of an integration variable do
not change the value of an integral. Therefore, in the naive path integral
\[
Z[A]=\int \mathcal{D}\left[ A\right] \mathrm{e}^{\mathrm{i}S\left[ A\right]
}\propto \int \prod_{x}dU\left( x\right) .
\]
a `volume' associated with the gauge transformations {\ $\prod_{x}dU\left(
x\right) $} can be factorized and thereby the integration be performed over
the gauge orbits. To turn this property into a working algorithm, redundant
variables are eliminated by imposing a \emph{gauge condition, }\ $f[A]=0,$
which is supposed to eliminate all gauge copies of a certain field
configuration {$A$}. In other words, the functional {$f$} has to be chosen
such that, for arbitrary field configurations, the equation, \ $f[A^{\,\left[
\,U\right] }\,]=0,$ determines uniquely the gauge transformation {$U$}. If
successful, the set of all gauge equivalent fields, the gauge orbit, is
represented by exactly {one representative}. In order to write down an
integral over gauge orbits, we insert into the integral the gauge--fixing {$%
\delta-$functional}
\[
\delta \left[ f\left( A\right) \right] =\prod_{x}\prod_{a=1}^{N^{2}-1}\delta %
\left[ f^{a}\left( A\left( x\right) \right) \right] .
\]
This modification of the integral however changes the value depending on the
representative chosen, as the following elementary identity shows }
\[
{\delta \left( g\left( x\right) \right) \ =\ }\frac{\delta \left( x-a\right)
}{|g^{\prime }\left( a\right) |}{,\qquad g\left( a\right) =0.}
\]
{\ This difficulty is circumvented with the help of the {\ Faddeev--Popov
determinant} {\ $\Delta _{f}\left[ A\right] $} defined implicitly by
\[
\Delta _{f}\left[ A\right] \int \mathcal{D}\left[ U\right] \delta \left[
f\left( A^{\left[ U\right] }\right) \right] =1.
\]
Multiplication of the path integral $Z[A]$ with the above {``1''} and taking
into account the gauge invariance of the various factors yields
\begin{eqnarray*}
Z[A] &=&\int \mathcal{D}\left[ U\right] \int \mathcal{D}\left[ A\right]
\mathrm{e}^{\mathrm{i}S\left[ A\right] }\Delta _{f}\left[ A\right] \delta %
\left[ f\left( A^{\left[ U\right] }\right) \right]  \\
&=&\int \mathcal{D}\left[ U\right] \int \mathcal{D}\left[ A\right] \mathrm{e}%
^{\mathrm{i}S\left[ A^{\left[ U\right] }\right] }\Delta _{f}\left[ A^{\left[
U\right] }\right] \delta \left[ f\left( A^{\left[ U\right] }\right) \right].
\end{eqnarray*}
{The gauge volume has been factorized and, being independent of the
dynamics, can be dropped. In summary, the final definition of the {%
generating functional} for gauge theories is given in terms of a \textit{sum
over gauge orbits,}
\[
Z\left[ J\right] =\int \mathcal{D}\left[ A\right] \Delta _{f}\left[ A\right]
\delta \left( f\left[ A\right] \,\right) \mathrm{e}^{\mathrm{i}S\left[ A%
\right] +\mathrm{i}\int d^{4}xJ^{\mu }A_{\mu }}.
\]
}

\subsubsection{Faddeev--Popov determinant}

{For the calculation of {$\Delta _{f}\left[ A\right] $}, we first consider
the change of the gauge condition {\ $f^{a}\left[ A\right] $} under
infinitesimal gauge transformations. Taylor expansion
\begin{eqnarray*}
f_{x}^{a}\left[ A^{\left[ U\right] }\right] &\approx &f_{x}^{a}\left[ A%
\right] +\int d^{4}y\sum_{b,\mu }\frac{\delta f_{x}^{a}\left[ A\right] }{%
\delta A_{\mu }^{b}\left( y\right) }\delta A_{\mu }^{b}\left( y\right) \\
&=&f_{x}^{a}\left[ A\right] +\int d^{4}y\sum_{b}M\left( x,y;a,b\right)
\alpha ^{b}\left( y\right),
\end{eqnarray*}
with $\delta A_{\mu }^{a}$ given by infinitesimal gauge transformations,
\begin{eqnarray*}
A_{\mu }^{a}\left( x\right) &\rightarrow &A_{\mu }^{a}\left( x\right)
-\partial _{\mu }\alpha ^{a}\left( x\right) -gf^{abc}\alpha ^{b}\left(
x\right) A_{\mu }^{c}\left( x\right) ,\qquad \text{{yields}} \\
M\left( x,y;a,b\right) &=&\left( \partial _{\mu }\delta
^{b,c}+gf^{bcd}A_{\mu }^{d}\left( y\right) \right) \frac{\delta f_{x}^{a}%
\left[ A\right] }{\delta A_{\mu }^{c}\left( y\right) }.
\end{eqnarray*}
In the second step, we compute the integral
\[
\Delta _{f}^{-1}\left[ A\right] =\int \mathcal{D}\left[ U\right] \delta %
\left[ f\left( A^{\left[ U\right] }\right) \right],
\]
by expressing the integration }$\mathcal{D}\left[ U\right] ${\ as an
integration over the gauge functions {$\alpha $}. We finally change to the
variables $\beta =M\alpha $,
\[
\Delta _{f}^{-1}\left[ A\right] =|\det M|^{-1}\int \mathcal{D}\left[ \beta %
\right] \delta \left[ f\left( A\right) -\beta \right],
\]
and arrive at the final expression for the Faddeev--Popov determinant \cite{Lenz}
\[
\Delta _{f}\left[ A\right] =|\det M|\,.
\]
Examples: }

\begin{itemize}
\item  Lorentz gauge
\begin{eqnarray*}
f_{x}^{a}\left( A\right)  &=&\partial ^{\mu }A_{\mu }^{a}\left( x\right)
-\chi ^{a}\left( x\right) , \\
M\left( x,y;a,b\right)  &=&-\left( \delta ^{ab}\Box -gf^{abc}A_{\mu
}^{c}\left( y\right) \partial _{y}^{\mu }\right) \delta ^{\left( 4\right)
}\left( x-y\right) .
\end{eqnarray*}

\item  Coulomb gauge
\begin{eqnarray*}
f_{x}^{a}\left( A\right)  &=&\mathrm{div}\mathbf{A}^{a}\left( x\right) -\chi
^{a}\left( x\right) , \\
M\left( x,y;a,b\right)  &=&\left( \delta ^{ab}\Delta +gf^{abc}\mathbf{A}%
^{c}\left( y\right) \nabla _{y}\right) \delta ^{\left( 4\right) }\left(
x-y\right) .
\end{eqnarray*}

\item  Axial gauge
\begin{eqnarray*}
f_{x}^{a}\left( A\right)  &=&n^{\mu }A_{\mu }^{a}\left( x\right) -\chi
^{a}\left( x\right) , \\
M\left( x,y;a,b\right)  &=&-\delta ^{ab}n_{\mu }\partial _{y}^{\mu }\delta
^{\left( 4\right) }\left( x-y\right) .
\end{eqnarray*}
\end{itemize}

\subsection{Basics of Conformal field theory}

A conformal field theory (CFT) is a quantum field theory (or, a statistical mechanics model at the critical point) that is invariant under \textit{conformal transformations}. Conformal field theory is often studied in 2D where there is an infinite-dimensional group of local conformal transformations, described by the holomorphic functions. CFT has important applications in string theory, statistical mechanics, and condensed matter physics.
For a good introduction to CFT see \cite{Belavin84,Senechal97}. We consider here only chiral CFTs in 2D (see \cite{FreedmanRMP}), where `chiral'\footnote{In general, a chiral field is a holomorphic field $W(z)$ which transforms as $$L_n W(z)=-z^{n+1} \frac{\partial}{\partial z} W(z) - (n+1)\Delta z^n W(z),\qquad \text{with}\qquad \bar L_n W(z)=0,$$ and similarly for an anti-chiral field. Here, $\Delta$ is the conformal weight of the chiral field $W$.} means that all of our fields will be functions of a complex number $z=x+\mathrm{i}y$
only and not functions of its conjugate $\bar{z}$.

To formally describe a 2D CFT we give its `conformal data', including a set of primary
fields, each with a conformal dimension $\Delta $, a table of fusion rules
of these fields and a central charge $c$. Data for three CFTs are given in Table %
\ref{tab:conformaldata}.

The \textit{operator product expansion} (OPE) describes what happens to two
fields when their positions approach each other. We write the OPE for two
arbitrary fields $\phi _{i}$ and $\phi _{j}$ as
\[
\lim_{z\rightarrow w}\phi _{i}(z)\phi
_{j}(w)=\sum_{k}C_{ij}^{k}(z-w)^{\Delta _{k}-\Delta _{i}-\Delta _{j}}\,\phi
_{k}(w),
\]%
where the \textit{structure constants} $C_{ij}^{k}$ are only nonzero as indicated by
the fusion table. Note that the OPE works \emph{inside} a correlator. For
example, in the $\mathbb{Z}_{3}$ para-fermion CFT (see Table \ref%
{tab:conformaldata}), since $\sigma _{1}\times \psi _{1}=\epsilon $, for
arbitrary fields $\phi _{i}$ we have \cite{FreedmanRMP}
\begin{eqnarray*}
&&\lim_{z\rightarrow w}\,\langle \,\phi _{1}(z_{1})\ldots \phi
_{M}(z_{M})\,\sigma _{1}(z)\psi _{1}(w)\,\rangle \sim \\
&&(z-w)^{2/5-1/15-2/3}\langle \,\phi _{1}(z_{1})\ldots \phi
_{M}(z_{M})\epsilon (w)\,\rangle .
\end{eqnarray*}

In addition to the OPE, there is also an important `neutrality' condition: a
correlator is zero unless all of the fields can fuse together to form the
identity field $\mathbf{1}$. For example, in the $\mathbb{Z}_{3}$
para-fermion field theory $\langle \psi _{2}\psi _{1}\rangle \neq 0$ since $%
\psi _{2}\times \psi _{1}=\mathbf{1}$, but $\langle \psi _{1}\psi
_{1}\rangle =0$ since $\psi _{1}\times \psi _{1}=\psi _{2}\neq \mathbf{1}$.

\begin{table}[tbph]
\begin{minipage}{3.5in}
\begin{minipage}{2in}
Chiral Bose Vertex: ($c=1$)

\begin{tabular}{|c||c|} \hline
$ $ & $\Delta$  \\
\hline \hline ${\rm e}^{{\rm i}\alpha\phi}$ & $\alpha^2/2$  \\ \hline
\end{tabular}
\hspace*{2pt}
\begin{tabular}{ |c||c||} \hline
$\times$ & ${\rm e}^{{\rm i} \alpha \phi}$  \\
\hline \hline ${\rm e}^{{\rm i} \beta \phi} $ & ${\rm e}^{{\rm i}(\alpha+\beta)\phi}$    \\ \hline
\end{tabular}
\end{minipage}\smallskip
\begin{minipage}{2in}
 Ising CFT:  ($c=1/2$)

\begin{tabular}{|c||c|} \hline
$ $ & $\Delta$  \\
\hline \hline $\psi$ & $1/2$  \\ \hline
 $\sigma$ & $1/16$
\\ \hline
\end{tabular}
\hspace*{2pt}
\begin{tabular}{ |c||c|c||} \hline
$\times$ & $\psi$ & $\sigma$  \\
\hline \hline $\psi $ & {\bf 1} &   \\ \hline
 $\sigma $ & $\sigma$ &  ${\bf 1} + \psi $ \\ \hline
\end{tabular}
\end{minipage}

\vspace*{5pt}

$\mathbb{Z}_3$ Parafermion CFT:  ($c=4/5$)

\begin{tabular}{|c||c|} \hline
$ $ & $\Delta$  \\
\hline \hline $\psi_{1}$ & $2/3$  \\ \hline $\psi_{2}$ & $2/3$  \\
\hline
 $\sigma_{1}$ & $1/15$ \\
 \hline
 $\sigma_{2}$ & $1/15$ \\
 \hline
 $\epsilon$ & 2/5
\\ \hline
\end{tabular}
\hspace*{5pt}
\begin{tabular}{ |c||c|c|c|c|c||} \hline
$\times$ & $\psi_1 $ & $\psi_2 $ & $\sigma_1$ & $\sigma_ 2 $& $\epsilon$  \\
\hline \hline $\psi_1 $ & $\psi_2$ &  & &    & \\
\hline $\psi_2 $ & ${\bf 1}$ &  $\psi_1$  & & &  \\
\hline $\sigma_1 $ & $\epsilon$ &  $\sigma_2$  &  $\sigma_2 + \psi_1$ & &  \\
\hline $\sigma_2 $ & $\sigma_1$ &  $\epsilon$  &  ${\bf 1} + \epsilon$  & $\sigma_1 + \psi_2$ &   \\
\hline $\epsilon $ & $\sigma_2$ &  $\sigma_1$  &  $\sigma_1 + \psi_2$  & $\sigma_2 + \psi_1$ &  ${\bf 1}+ \epsilon$ \\
\hline
\end{tabular}
\end{minipage} \vspace*{5pt}
\caption{Conformal data for three CFTs. Given is the list of primary fields
in the CFT with their conformal dimension $\Delta$, as well as the fusion
table. In addition, every CFT has an identity field $\mathbf{1}$ with
dimension $\Delta=0$ which fuses trivially with any field ($\mathbf{1}
\times \protect\phi_i = \protect\phi_i$ for any $\protect\phi_i$). Note that
fusion tables are symmetric so only the lower part is given. In the Ising
CFT the field $\psi$ is frequently notated as $\epsilon$. This
fusion table indicates the nonzero elements of the fusion matrix $N_{ab}^c$.
For example in the $\mathbb{Z}_3$ CFT, since ${\protect\sigma_1} \times {%
\protect\sigma_2} = \mathbf{1} + \protect\epsilon$, $N_{{\protect\sigma_1} {%
\protect\sigma_2}}^1 = N_{{\protect\sigma_1} {\protect\sigma_2}}^\protect%
\protect\epsilon = 1$ and $N_{{\protect\sigma_1} {\protect\sigma_2}}^c = 0$
for all $c$ not equal to $\mathbf{1}$ or $\protect\epsilon$. \protect%
\vspace*{-10pt} }
\label{tab:conformaldata}
\end{table}

Let us look at what happens when a fusion has more than one possible result.
For example, in the Ising CFT, $\sigma \times \sigma =\mathbf{1}+\psi $.
Using the OPE, we have
\begin{equation}
\lim_{w_{1}\rightarrow w_{2}}\!\sigma (w_{1})\sigma (w_{2})\!\sim \frac{%
\mathbf{1}}{(w_{1}-w_{2})^{1/8}}\!+(w_{1}-w_{2})^{3/8}\,{\psi ,}
\label{eq:sigmasigma}
\end{equation}%
where we have neglected the constants $C_{ij}^{k}$. If we consider $\langle
\sigma \sigma \rangle $, the neutrality condition picks out only the first
term in (\ref{eq:sigmasigma}) where the two $\sigma $'s fuse to form $\mathbf{1%
}$. Similarly, $\langle \sigma \sigma \psi \rangle $ results in the second
term of (\ref{eq:sigmasigma}) where the two $\sigma $'s fuse to form $\psi $
which then fuses with the additional $\psi $ to make $\mathbf{1}$.

Fields may also fuse to form the identity in more than one way. For example,
in the correlator $\langle \sigma (w_{1})\sigma (w_{2})\sigma (w_{3})\sigma
(w_{4})\rangle $ of the Ising CFT, the identity is obtained via two possible
fusion paths --- resulting in two different so-called `conformal blocks'. On
the one hand, one can fuse $\sigma (w_{1})$ and $\sigma (w_{2})$ to form $%
\mathbf{1}$ and similarly fuse $\sigma (w_{3})$ and $\sigma (w_{4})$ to form
$\mathbf{1}$. Alternately, one can fuse $\sigma (w_{1})$ and $\sigma (w_{2})$
to form $\psi $ and fuse $\sigma (w_{3})$ and $\sigma (w_{4})$ to form $\psi
$ then fuse the two resulting $\psi $ fields together to form $\mathbf{1}$.
The correlator generally gives a linear combination of the possible
resulting conformal blocks. We should thus think of such a correlator as
living in a vector space rather than having a single value. (If we instead
choose to fuse $1$ with $3$, and $2$ with $4$, we would obtain two blocks
which are linear combinations of the ones found by fusing 1 with 2 and 3
with 4. The resulting vectors space, however, is independent of the order of
fusion). Crucially, transporting the coordinates $w_{i}$ around each other
makes a rotation within this vector space.

To be clear about the notion of conformal blocks, let us look at the
explicit form of the Ising CFT correlator \cite{FreedmanRMP}
\begin{eqnarray*}
&&\lim_{w\rightarrow \infty }\langle \sigma (0)\sigma (z)\sigma (1)\sigma
(w)\rangle =a_{+}\,F_{+}+a_{-}\,F_{-}, \\
&&F_{\pm }(z)\sim (wz(1-z))^{-1/8}\sqrt{1\pm \sqrt{1-z}},
\end{eqnarray*}%
where $a_{+}$ and $a_{-}$ are arbitrary coefficients. When $z\rightarrow 0$
we have $F_{+}\sim z^{-1/8}$ whereas $F_{-}\sim z^{3/8}$. Comparing to (\ref%
{eq:sigmasigma}) we conclude that $F_{+}$ is the result of fusing $\sigma
(0)\times \sigma (z)\rightarrow \mathbf{1}$ whereas $F_{-}$ is the result of
fusing $\sigma (0)\times \sigma (z)\rightarrow \psi $. As $z$ is taken in a
clockwise circle around the point $z=1$, the inner square-root changes sign,
switching $F_{+}$ and $F_{-}$. Thus, this `braiding' (or `monodromy')
operation transforms
\[
\mbox{${a_+}\choose{a_-}$}\rightarrow \mathrm{e}^{2\pi \mathrm{i}/8}%
\mbox{${0 ~~ 1}\choose{1 ~~
    0}$}\mbox{${a_+}\choose{a_-}$}
\]%
Having a multiple valued correlator (I.e., multiple conformal blocks) is a
result of having such branch cuts. Braiding the coordinates ($w$'s) around
each other results in the correlator changing values within its allowable
vector space.

A useful technique for counting conformal blocks is the \textit{Bratteli diagram}.
In Figure \ref{BratteliDiag} we give the Bratteli diagram for the fusion of
multiple $\sigma $ fields in the Ising CFT. Starting with $\mathbf{1}$ at
the lower left, at each step moving from the left to the right, we fuse with
one more $\sigma $ field. At the first step, the arrow points from $\mathbf{1%
}$ to $\mathbf{\sigma }$ since $\mathbf{1}\times \sigma =\sigma $. At the
next step $\sigma $ fuses with $\sigma $ to produce either $\psi $ or $%
\mathbf{1}$ and so forth. Each conformal block is associated with a path
through the diagram. Thus to determine the number of blocks in $\langle
\sigma \sigma \sigma \sigma \rangle $ we count the number of paths of four
steps in the diagram starting at the lower left and ending at $\mathbf{1}$.
\begin{figure}[htb]
\label{epil1} \centerline{\includegraphics[width=6cm]{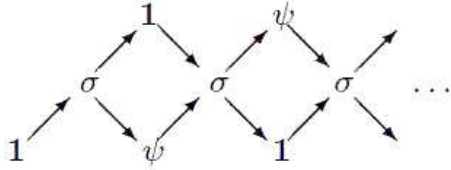}}
\caption{Bratteli diagram
for fusion of multiple $\sigma_1$ fields in the $\mathbb{Z}_3$ para-fermion CFT (modified and adapted from \cite{FreedmanRMP}).} \label{BratteliDiag}
\end{figure}

A particularly important CFT is obtained from a free Bose field theory in
(1+1)D by keeping only the left moving modes. The free
chiral Bose field $\phi (z)$, which is a sum of left moving creation and
annihilation operators, has a correlator $\langle \phi (z)\phi (z^{\prime
})\rangle =-\log (z-z^{\prime })$. We then define the normal ordered `chiral
vertex operator', $\mathrm{e}^{\mathrm{i}\alpha \phi (z)}$, which is a
conformal field. Since $\phi $ is a free field, Wick's theorem can be used
to obtain \cite{Senechal97}
\[
\left\langle \,\mathrm{e}^{\mathrm{i}{\alpha _{1}}\phi }(z_{1})\ldots
\mathrm{e}^{\mathrm{i}{\alpha _{N}}\phi }(z_{N})\,\right\rangle =\mathrm{e}%
^{-\sum_{i<j}\alpha _{i}\alpha _{j}\langle \phi (z_{i})\phi (z_{j})\rangle }=%
{\mbox{$\prod_{i<j}$}}\,\,\,(z_{i}-z_{j})^{\alpha _{i}\alpha _{j}}.
\]

\subsection{Chern--Simons theory and Jones polynomial}

\subsubsection{Chern--Simons theory and link invariants}

Consider $SU(2)_{k}$ non-Abelian Chern--Simons theory%
\begin{equation*}
S_{CS}[a]=\frac{k}{4\pi }\int_{M}\mathrm{tr}\left( a\wedge da+\frac{2}{3}%
a\wedge a\wedge a\right) .
\end{equation*}%
We modify the action by the addition of sources, $j^{\mu {\underline{a}}}$,
according to $\mathcal{L}\rightarrow \mathcal{L}+\mathrm{tr}\left( j\cdot
a\right) $. We take the sources to be a set of particles on prescribed
classical trajectories. The $i^{\text{th}}$ particle carries the spin $j_{i}$
representation of $SU(2)$. As we have seen above,
there are only $k+1$ allowed representations; later in this subsection, we
will see that if we give a particle a higher spin representation than $j=k/2$%
, then the amplitude will vanish identically. Therefore, $j_{i}$ must be in
allowed set of $k+1$ possibilities: $0,\frac{1}{2},\ldots ,\frac{k}{2}$. The
functional integral in the presence of these sources can be written in terms
of Wilson loops, ${W_{{\gamma _{i}},{j_{i}}}}[a]$, which are defined as
follows. The holonomy $U_{\gamma ,j}[a]$ is an $SU(2)$ matrix associated
with a curve $\gamma $. It is defined as the path--ordered exponential
integral of the gauge field along the path $\gamma $ \cite{FreedmanRMP}
\begin{multline*}
U_{\gamma ,j}[a]\equiv \mathcal{P}{\mathrm{e}^{\mathrm{i}{\oint_{\gamma }}{%
\mathbf{a}^{\underline{c}}T^{\underline{c}}\cdot d\mathbf{l}}}}={%
\sum_{n=0}^{\infty }}\mathrm{i}{^{n}}{\int_{0}^{2\pi }}\!\!{ds_{1}}{%
\int_{0}^{s_{1}}}\!\!{ds_{2}}\ldots \\
{\int_{0}^{s_{n-1}}}\!\!{ds_{n}}\left[ \dot{\mathbf{\gamma }}({s_{1}})\cdot {%
\mathbf{a}^{{\underline{a}}_{1}}}\left( \mathbf{\gamma }({s_{1}})\right) {T^{%
{\underline{a}}_{1}}}\ldots \,\dot{\mathbf{\gamma }}({s_{n}})\cdot {\mathbf{a%
}^{{\underline{a}}_{n}}}\left( \mathbf{\gamma }({s_{n}})\right) {T^{{%
\underline{a}}_{n}}}\right] ,
\end{multline*}%
where $\mathcal{P}$ is the path--ordering symbol. The Lie algebra generators
$T^{\underline{a}}$ are taken in the spin $j$ representation. $\vec{\gamma}%
(s)$, $s\in \lbrack 0,2\pi ]$ is a parametrization of $\gamma $; the \textit{%
holonomy} is clearly independent of the parametrization. The \textit{Wilson
loop} is the trace of the holonomy:
\begin{equation}
{W_{\gamma ,j}}[a]=\mathrm{tr}\left( U_{\gamma ,j}[a]\right) .
\label{eqn:Wilson-loop-def}
\end{equation}%
Let us consider the simplest case, in which the source is a
quasi-particle--quasi-hole pair of type $j$ which is created out of the ground
state, propagated for a period of time, and then annihilated, returning the
system to the ground state. The \textit{transition amplitude} for such a
process is given by the path integral:
\begin{equation*}
\left\langle 0|0\right\rangle _{\gamma ,j}=\int \mathcal{D}[a]\,\mathrm{e}^{%
\mathrm{i}S_{CS}[a]}\,{W_{\gamma ,j}}[a],
\end{equation*}%
where $\gamma $ is the spacetime loop formed by the trajectory of the
quasi-particle--quasi-hole pair. The Wilson loop was introduced as an order
parameter for confinement in a gauge theory because this amplitude roughly
measures the force between the quasi-particle and the quasi-hole. If they were
to interact with a confining force $V(r)\sim r$, then the logarithm of this
amplitude would be proportional to the the area of the loop; if they were to
have a short-ranged interaction, it would be proportional to the perimeter
of the loop. However, Chern--Simons theory is independent of a metric, so
the amplitude cannot depend on any length scales. It must simply be a
constant. For $j=1/2$, we will call this constant $d$. As the notation
implies, it is, in fact, the quantum dimension of a $j=1/2$ particle. As we
will see below, $d$ can be determined in terms of the level $k$, and the
quantum dimensions of higher spin particles can be expressed in terms of $d$.

We can also consider the amplitude for two pairs of quasi-particles to be
created out of the ground state, propagated for some time, and then
annihilated, returning the system to the ground state \cite{FreedmanRMP}
\begin{equation*}
\left\langle 0|0\right\rangle _{{\gamma _{1}},{j_{1}};{\gamma _{2}},{j_{2}}%
}=\int \mathcal{D}[a]\,\mathrm{e}^{\mathrm{i}S_{CS}[a]}\,{W_{{\gamma },{j}}}%
[a]\,{W_{{\gamma ^{\prime }},{j^{\prime }}}}[a].
\end{equation*}%
This amplitude can take different values depending on how $\gamma $ and $%
\gamma ^{\prime }$ are linked. If the curves are unlinked the integral must give $d^{2}$,
but when they are linked the value can be nontrivial. In a similar way, we
can formulate the amplitudes for an arbitrary number of sources.

It is useful to think about this history as a two
step process: from $t=-\infty $ to $t=0$ and from $t=0$ to $t=\infty $ (the
two pairs are created at some time $t<0$ and annihilated at some time $t>0$%
). At $t=0^{-}$, the system is in a four-quasi-particle state.\footnote{%
Quasiparticles and quasi-holes are topologically equivalent if $G=SU(2)$, so
we will use `quasi-particle' to refer to both.} Let us call this state $\psi $%
:
\begin{equation*}
\psi \lbrack A]=\int_{a(\mathbf{x},0)=A(\mathbf{x})}\!\!\!\!\mathcal{D}[a(\mathbf{x},t)]:W_{{\gamma _{-}^{{}}},j}[a]:W_{{\gamma _{-}^{\prime }}%
,j^{\prime }}[a]\,\mathrm{e}^{{\int_{-\infty }^{0}}dt\int {d^{2}}x:\mathcal{L%
}_{\mathrm{CS}}},
\end{equation*}%
where ${\gamma _{-}}$ and ${\gamma _{-}^{\prime }}$ are the arcs given by $%
\gamma (t)$ and $\gamma ^{\prime }(t)$ for $t<0$. $A(\mathbf{x})$ is the
value of the gauge field on the $t=0$ spatial slice; the wave-functional $%
\psi \lbrack A]$ assigns an amplitude to every spatial gauge field
configuration. For $G=SU(2)$ and $k>1$, there are actually two different
four-quasi-particle states: if particles $1$ and $2$ fuse to the identity
field $j=0$, then particles $3$ and $4$ must as well; if particles $1$ and $%
2 $ fuse to $j=1$, then particles $3$ and $4$ must as well. These are the
only possibilities.\footnote{For $k=1$, fusion to $j=1$ is not possible.} Which one
the system is in depends on how the trajectories of the four quasi-particles
are intertwined. Although quasi-particles $1$ and $2$ were created as a pair
from the vacuum, quasi-particle $2$ braided with quasi-particle $3$, so $1$
and $2$ may no longer fuse to the vacuum. In just a moment, we will see an
example of a different four-quasi-particle state.

We now interpret the $t=0$ to $t=\infty $ history as the conjugate of a $%
t=-\infty $ to $t=0$ history. In other words, it gives us a four
quasi-particle bra rather than a four quasi-particle ket \cite{FreedmanRMP}
\begin{equation*}
{\chi ^{\ast }}[A]=\int_{a(\mathbf{x},0)=A(\mathbf{x})}\!\!\!\!\mathcal{D}[a(\mathbf{x},t)]:W_{{\gamma _{+}},j}[a]:W_{{\gamma _{+}^{\prime }},j^{\prime
}}[a]\,\mathrm{e}^{{\int_{0}^{\infty }}dt\int {d^{2}}x:\mathcal{L}_{\mathrm{%
CS}}}.
\end{equation*}%
In the state $|\chi \rangle $, quasi-particles $1$ and $2$ fuse to form the
trivial quasi-particle, as do quasi-particles $3$ and $4$. Then we can
interpret the functional integral from $t=-\infty $ to $t=\infty $ as the
matrix element between the bra and the ket:
\begin{equation*}
\langle \chi |\psi \rangle =\int \mathcal{D}[a]\,\mathrm{e}^{\mathrm{i}%
S_{CS}[a]}\,{W_{{\gamma _{1}},{j_{1}}}}[a]\,{W_{{\gamma _{2}},{j_{2}}}}[a].
\end{equation*}

Now, observe that $|\psi \rangle $ is obtained from $|\chi \rangle $ by
taking quasi-particle $2$ around quasi-particle $3$, i.e., by exchanging
quasi-particles $2$ and $3$ twice, $|\psi \rangle =\rho \!\left( {\sigma
_{2}^{2}}\right) |\chi \rangle $. Hence,
\begin{equation}
\langle \chi |\rho \!\left( {\sigma _{2}^{2}}\right) |\chi \rangle =\int
\mathcal{D}[a]\,\mathrm{e}^{\mathrm{i}S_{CS}[a]}\,{W_{{\gamma _{1}},{j_{1}}}}%
[a]\,{W_{{\gamma _{2}},{j_{2}}}}[a].  \label{eqn:braiding-matrix-integral}
\end{equation}%
In this way, we can compute the entries of the braiding matrices $\rho
\!\left( {\sigma _{i}}\right) $ by computing functional integrals such as
the one on the right-hand-side of (\ref{eqn:braiding-matrix-integral}).

Consider, now, the state $\rho \!\left( {\sigma _{2}}\right) |\chi \rangle $%
, in which particles $2$ and $3$ are exchanged just once. From the figure,
we see that
\begin{equation*}
\langle \chi |\rho \!\left( {\sigma _{2}}\right) |\chi \rangle =d,\qquad
\langle \chi |\rho \!\left( {\sigma _{2}^{-1}}\right) |\chi \rangle =d,
\end{equation*}%
since both histories contain just a single unknotted loop. Meanwhile, ~
$
\langle \chi |\chi \rangle =d^{2}.
$

Since the four-quasi-particle Hilbert space is 2D, $\rho \!\left( {\sigma _{2}%
}\right) $ has two eigenvalues, $\lambda _{1}$, $\lambda _{2}$, so that
\begin{equation*}
\rho \!\left( {\sigma }\right) -\left( {\lambda _{1}}+{\lambda _{2}}\right) +%
{\lambda _{1}}{\lambda _{2}}\rho \!\left( {\sigma ^{-1}}\right) =0.
\end{equation*}%
Taking the expectation value in the state $|\chi \rangle $, we find:
\begin{equation*}
d-\left( {\lambda _{1}}+{\lambda _{2}}\right) {d^{2}}+{\lambda _{1}}{\lambda
_{2}}d =0,\qquad \text{so that}\qquad
d =\frac{1+{\lambda _{1}}{\lambda _{2}}}{{\lambda _{1}}+{\lambda _{2}}}.
\end{equation*}
Since the braiding matrix is unitary, $\lambda _{1}$ and $\lambda _{2}$ are
phases. The overall phase is unimportant for quantum computation, so we
really need only a single number. In fact, this number can be obtained from
self-consistency conditions \cite{Freedman04a}. However, the details of the
computation of $\lambda _{1}$, $\lambda _{2}$ within is technical and
requires a careful discussion of framing; the result is \cite{Witten89} that
$${\lambda _{1}}=-\mathrm{e}^{-3\pi \mathrm{i}/2(k+2)},\qquad {\lambda _{2}}=%
\mathrm{e}^{\pi \mathrm{i}/2(k+2)}.$$ These eigenvalues are simply ${R_{0}^{%
\frac{1}{2},\frac{1}{2}}}~=~{\lambda _{1}}$, ${R_{1}^{\frac{1}{2},\frac{1}{2}%
}}~=~{\lambda _{2}}$. Consequently,
\begin{equation}
d=2\,\cos \left( \frac{\pi }{k+2}\right) \qquad \text{and}
\label{eqn:d-value}
\end{equation}%
\begin{equation}
q^{-1/2}\rho ({\sigma _{i}})-q^{1/2}{\rho \!\left( \sigma _{i}^{-1}\right) }%
=q-q^{-1},  \label{eqn:Jones-skein}
\end{equation}%
where $q=-\mathrm{e}^{\pi \mathrm{i}/(k+2)}$. Since this operator equation
applies regardless of the state to which it is applied, we can apply it
locally to any given part of a knot diagram to relate the amplitude to the
amplitude for topologically simpler processes. This is an example of a \textit{skein relation}; in this case,
it is the skein relation which defines the Jones polynomial. In arriving at
this skein relation, we are retracing the connection between Wilson loops in
Chern--Simons theory and knot invariants which was made in the remarkable
paper \cite{Witten89}. In this paper, Witten showed that correlation
functions of Wilson loop operators in $SU(2)_{k}$ Chern--Simons theory are
equal to corresponding evaluations of the \textit{Jones polynomial},\footnote{A \textit{link} is a finite family of disjoint, smooth, oriented or un-oriented, closed curves in $\mathbb{R}^3$ or equivalently $S^3$. A \textit{knot} is a link with one component. The Jones polynomial $V_L(t)$ is a \textit{Laurent polynomial} in the variable $\sqrt{t}$, which
is defined for every oriented link $L$ but depends on that link only up to
orientation preserving diffeomorphism, or equivalently \textit{isotopy}, of $\mathbb{R}^3$. Links can be represented by diagrams in the plane. The Jones polynomial of a knot (and generally a link with an odd number
of components) is a Laurent polynomial in $t$ \cite{Jones85}.} which is a
\textit{topological invariant} of \textit{knot theory}\footnote{Knot theory is the area of topology that studies mathematical knots. Formally, a knot is an \textit{embedding} of a circle in 3D Euclidean space, $\mathbb{R}^3$. Two knots are equivalent if one can be transformed into the other via a deformation of $\mathbb{R}^3$ upon itself (known as an \textit{ambient isotopy}). These transformations correspond to manipulations of a knotted string that do not involve cutting the string or passing the string through itself.
Knots can be described in various ways. Given a method of description, however, there may be more than one description that represents the same knot. For example, a common method of describing a knot is a planar \textit{knot diagram}. However, any given knot can be drawn in many different ways using a knot diagram. Therefore, a fundamental problem in knot theory is determining when two descriptions represent the same knot. One way of distinguishing knots is by using a \textit{knot invariant}, a quantity which remains the same even with different descriptions of a knot. The concept of a knot has also been extended to higher dimensions by considering $m$D spheres $S^m$ in $\mathbb{R}^n$.} \cite{Jones85}:
\begin{equation}
\int \mathcal{D}[a]\,W_{{\gamma _{1}},\frac{1}{2}}[a]\ldots W_{{\gamma _{n}},%
\frac{1}{2}}[a]\,\mathrm{e}^{\mathrm{i}S_{CS}[a]}=V_{L}(q).
\label{eqn:Jones-CS}
\end{equation}%
$V_{L}(q)$ is the Jones polynomial associated with the link $L={\gamma _{1}}%
\cup \ldots \cup {\gamma _{n}}$, evaluated at $q=-\mathrm{e}^{\pi \mathrm{i}%
/(k+2)}$ using the skein relation (\ref{eqn:Jones-skein}). Note that we
assume here that all of the quasi-particles transform under the $j=\frac{1}{2}
$ representation of $SU(2)$. The other quasi-particle types can be obtained
through the fusion of several $j=1/2$ quasi-particles.

\section{Appendix}

\subsection{Manifolds and bundles}

Geometrically, a manifold is a nonlinear (i.e., curved) space which is locally homeomorphic (i.e., topologically equivalent) to a linear (i.e., flat) Euclidean space $\mathbb{R}^n$; e.g., in a magnifying glass, each local patch of the apple surface looks like a plane, although globally (as a whole) the apple surface is totally different from the plane. Physically, a configuration manifold is a set of all degrees of freedom of a dynamical system.

More precisely, consider Consider a set $M$ (see Figure \ref{Manifold1}) which is a
\emph{candidate} for a manifold. Any point $x\in M$\footnote{Note
that sometimes we will denote the point in a manifold $M$ by $m$,
and sometimes by $x$ (thus implicitly assuming the existence of
coordinates $x=(x^i)$).} has its \textit{Euclidean chart}, given
by a 1--1 and \emph{onto} map $\varphi _{i}:M\rightarrow
\Bbb{R}^{n}$, with its \textit{Euclidean image}
$V_{i}=\varphi_{i}(U_{i})$. More precisely, a chart $ \varphi_{i}$
is defined by
\[
\varphi _{i}:M\supset U_{i}\ni x\mapsto \varphi _{i}(x)\in V_{i}\subset \Bbb{%
R}^{n},
\]
where $U_{i}\subset M$ and $V_{i}\subset \Bbb{R}^{n}$ are open
sets.
\begin{figure}[h]
\centerline{\includegraphics[width=9cm]{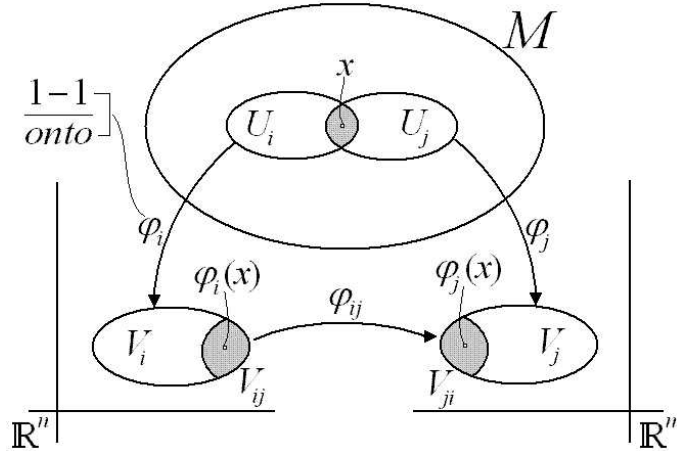}}
\caption{Geometric picture of the manifold concept.}
\label{Manifold1}
\end{figure}

Clearly, any point $x\in M$ can have several different charts (see
Figure \ref{Manifold1}). Consider a case of two charts, $\varphi
_{i},\varphi _{j}:M\rightarrow \Bbb{R}^{n}$,
having in their images two open sets, $V_{ij}=\varphi _{i}(U_{i}\cap U_{j})$ and $%
V_{ji}=\varphi _{j}(U_{i}\cap U_{j})$. Then we have
\textit{transition functions} $\varphi _{ij}$ between them,
\[
\varphi _{ij}=\varphi _{j}\circ \varphi
_{i}^{-1}:V_{ij}\rightarrow V_{ji},\qquad \text{locally given
by\qquad }\varphi _{ij}(x)=\varphi _{j}(\varphi _{i}^{-1}(x)).
\]
If transition functions $\varphi _{ij}$ exist, then we say that
two charts, $\varphi _{i}$ and $\varphi _{j}$ are
\emph{compatible}. Transition functions represent a general
(nonlinear) \emph{transformations of coordinates}, which are the
core of classical \emph{tensor calculus}.

A set of compatible charts $\varphi _{i}:M\rightarrow
\Bbb{R}^{n},$ such that each point $x\in M$ has its Euclidean
image in at least one chart, is called an \textit{atlas}. Two
atlases are \emph{equivalent} iff all their charts are compatible
(i.e., transition functions exist between them), so their union is
also an atlas. A \textit{manifold structure} is a class of
equivalent atlases.

Finally, as charts $\varphi _{i}:M\rightarrow \Bbb{R}^{n}$ were
supposed to be 1-1 and onto maps, they can be either
\emph{homeomorphism}\emph{s}, in which case we have a
\emph{topological} ($C^0$) manifold, or
\emph{diffeomorphism}\emph{s}, in which case we have a
\emph{smooth} ($C^{k}$) manifold.

On the other hand, tangent and cotangent bundles, $TM$ and
$T^{\ast }M$, respectively, of a smooth manifold $M$, are special
cases of a more general geometrical object called \emph{fibre
bundle}, where the word \emph{fiber} $V$ of a map $\pi
:Y\rightarrow X$ denotes the \emph{preimage} $\pi^{-1}(x)$ of an
element $x\in X$. It is a space which \emph{locally} looks like a
product of two spaces (similarly as a manifold locally looks like
Euclidean space), but may possess a different \emph{global}
structure. To get a visual intuition behind this fundamental
geometrical concept, we can say that a fibre bundle $Y$ is a
\emph{homeomorphic generalization} of a \emph{product space}
$X\times V$ (see Figure \ref{Fibre1}), where $X$ and $V$ are
called the \emph{base} and the \emph{fibre}, respectively. $\pi
:Y\rightarrow X$ is called the \emph{projection}, $Y_{x}=\pi
^{-1}(x)$ denotes a fibre over a point $x$ of the base $X$, while
the map $f=\pi ^{-1}:X\rightarrow Y$ defines the
\emph{cross--section}, producing the \textit{graph} $(x,f(x))$ in
the bundle $Y$ (e.g., in case of a tangent bundle, $f=\dot{x}$
represents a velocity vector--field, so that the graph in a the
bundle $Y$ reads $(x,\dot{x})$).
\begin{figure}[h]
 \centerline{\includegraphics[width=12cm]{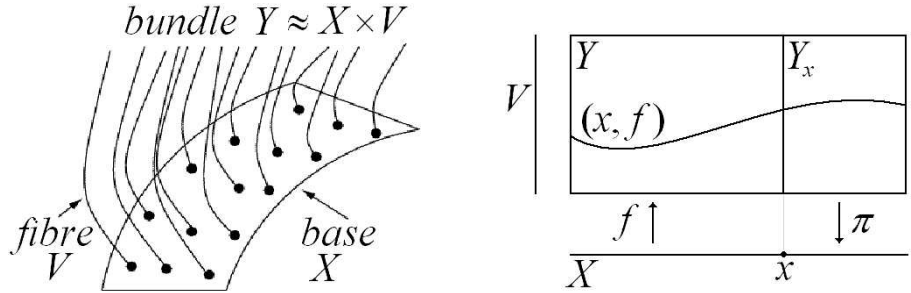}}
\caption{A sketch of a fibre bundle $Y\approx X\times V$ as a
generalization of a product space $X\times V$; left -- main
components; right -- a few details (see text for
explanation).}\label{Fibre1}
\end{figure}

A principal $G-$bundle is a bundle $\pi : Y \to X$ generated by a Lie group $G$ (see below) such that the group $G$ preserves the fibers of the bundle $Y$.

The main reason why we need to study fibre bundles is that
\emph{all dynamical objects} (including vectors, tensors,
differential forms and gauge potentials) are their
\emph{cross--sections}, representing \emph{generalizations of
graphs of continuous functions}. For more technical details, see \cite{GaneshSprBig,GaneshADG}.

\subsection{Lie groups}

A Lie group is a both a group and a manifold.
More precisely, a \textit{Lie group} is a smooth manifold $M$ that has at
the same time a group $G-$structure consistent with its manifold
$M-$structure in the sense that \textit{group
multiplication} ~$ \mu :G\times G\rightarrow G,~~ (g,h)\mapsto
gh$ and the\textit{group inversion} ~$ \nu :G\rightarrow
G,~~ g\mapsto g^{-1}$ are smooth functions. A point $e\in G$ is
called the \textit{group identity element}. For any Lie group $G$ in a neighborhood of its identity element $e$
it can be expressed in terms of a set of generators $T^{a}$ ($a=1,\ldots ,%
\mathrm{\dim \,}G$) as
\[
D(g)=\exp [-\mathrm{i}\alpha _{a}T^{a}]\equiv \sum_{n=0}^{\infty }{\frac{(-%
\mathrm{i})^{n}}{n!}}\alpha _{a_{1}}\ldots \alpha _{a_{n}}T^{a_{1}}\ldots
T^{a_{n}},
\]
where $\alpha _{a}\in \Bbb{C}$ are a set of coordinates of $M$ in a
neighborhood of $e$. Because of the general \textit{%
Baker--Campbell--Haussdorf formula}, the multiplication of two group elements
is encoded in the value of the commutator of two generators, that in general
has the form
\[
\lbrack T^{a},T^{b}]=\mathrm{i}f^{abc}T^{c},
\]
where $f^{abc}\in \Bbb{C}$ are called the structure constants. The set of
generators with the commutator operation form the Lie algebra associated
with the Lie group. Hence, given a representation of the Lie algebra of
generators we can construct a representation of the group by exponentiation
(at least locally near the identity).

In particular, for $SU(2)-$group, each group element is labeled by three real numbers $\alpha _{k},$
$(k=1,2,3)$. We have two basic representations: one is the fundamental
representation (or spin ${\frac{1}{2}}$) defined by
\[
D_{\frac{1}{2}}(\alpha _{k})=\mathrm{e}^{-{\frac{\mathrm{i}}{2}}\alpha
_{k}\sigma ^{k}},
\]
with $\sigma ^{\mathrm{i}}$ the Pauli matrices. The second one is the
adjoint (or spin 1) representation which can be written as
\[
D_{1}(\alpha _{k})=\mathrm{e}^{-\mathrm{i}\alpha _{k}J^{k}},\qquad \text{%
where}
\]
\[
J^{1}=\left(
\begin{array}{ccc}
0 & 0 & 0 \\
0 & 0 & 1 \\
0 & -1 & 0
\end{array}
\right) ,\hspace*{1cm}J^{2}=\left(
\begin{array}{ccc}
0 & 0 & -1 \\
0 & 0 & 0 \\
1 & 0 & 0
\end{array}
\right) ,\hspace*{1cm}J^{3}=\left(
\begin{array}{ccc}
0 & 1 & 0 \\
-1 & 0 & 0 \\
0 & 0 & 0
\end{array}
\right) .
\]
Actually, $J^{k}$ generate rotations around the $x$, $y$ and $z$ axis
respectively.

Let $M$ be a smooth manifold. An \textit{action of a Lie group}
$G$ (with the unit element $e$) on $M$ is a smooth map $\phi
:G\times M\rightarrow M,$ such that for all $x\in M$ and $g,h\in
G$, (i) $\phi (e,x)=x$ and (ii) $\phi \left( g,\phi (h,x)\right)
=\phi (gh,x).$ For more technical details, see \cite{GaneshSprBig,GaneshADG}.

\subsection{Differential forms and Stokes theorem}

Given the space of exterior differential $p-$forms $\Omega ^{p}(M)$ on a
smooth manifold $M$, we have the \emph{exterior derivative operator} $%
d:\Omega (M)\rightarrow \Omega ^{p+1}(M)$ which generalizes ordinary vector
differential operators (\textsl{grad}, \textsl{div} and \textsl{curl} see
\cite{de Rham,Flanders,GaneshSprBig}) and transforms $p-$forms $\omega $
into $(p+1)-$forms $d\omega ,$ with the main property: \ $dd=d^{2}=0.\,$%
Given a $p-$form $\alpha \in \Omega ^{p}(M)$ and a $q-$form $\beta \in
\Omega ^{q}(M),$ their exterior product is a $(p+q)-$form $\alpha \wedge
\beta \in \Omega ^{p+q}(M), $ where $\wedge $ is their anti-commutative
exterior (or, `wedge') product.

As differential forms are meant for integration, we have a generalization of
all integral theorems from vector calculus in the form of the Stokes
theorem: for the $p-$form $\omega $, in an oriented $n$D domain $C$, which
is a $p-$chain with a $(p-1)-$boundary $\partial C$,
\begin{equation}
\int_{\partial C}\omega =\int_{C}d\omega .  \label{Stok}
\end{equation}%
For any $p-$chain on a manifold $M,$ \emph{the boundary of a boundary is zero%
} \cite{MTW}, that is, $\partial \partial C=\partial ^{2}=0$.

A $p-$form $\beta $ is called \textit{closed} if its exterior derivative $%
d=\partial _{i}dx^{i}$ is equal to zero, \ $d\beta =0.$From this condition
one can see that the closed form (the \textit{kernel} of the exterior
derivative operator $d$) is conserved quantity. Therefore, closed $p-$forms
possess certain invariant properties, physically corresponding to the
conservation laws (see e.g., \cite{Abraham,GaneshADG}).

Also, a $p-$form $\beta $ that is an exterior derivative of some $(p-1)-$%
form $\alpha $, $\ \beta =d\alpha ,$is called \textit{exact} (the \textit{%
image} of the exterior derivative operator $d$). By Poincar\'{e} lemma,
exact forms prove to be closed automatically, \ $d\beta =d(d\alpha )=0.$

Since $d^{2}=0$, \emph{every exact form is closed.} The converse is only
partially true, by Poincar\'{e} lemma: every closed form is \textit{locally
exact}. In particular, there is a Poincar\'{e} lemma for contractible
manifolds: Any closed form on a smoothly contractible manifold is exact. The
Poincar\'{e} lemma is a generalization and unification of two well--known
facts in vector calculus:\newline
(i) If $\limfunc{curl}F=0$, then locally $F=\limfunc{grad}f$; ~ and ~ (ii)
If $\limfunc{div}F=0$, then locally $F=\limfunc{curl}G$.

A \textit{cycle} is a $p-$chain, (or, an oriented $p-$domain) $C\in \mathcal{%
C}_{p}(M)$ such that $\partial C=0$. A \textit{boundary} is a chain $C$ such
that $C=\partial B,$ for any other chain $B\in \mathcal{C}_{p}(M)$.
Similarly, a \textit{cocycle} (i.e., a \textit{closed form}) is a cochain $%
\omega $ such that $d\omega =0$. A \textit{coboundary} (i.e., an \textit{%
exact form}) is a cochain $\omega $ such that $\omega =d\theta ,$ for any
other cochain $\theta $. All exact forms are closed ($\omega =d\theta
\Rightarrow d\omega =0$) and all boundaries are cycles ($C=\partial
B\Rightarrow \partial C=0$). Converse is true only for smooth contractible
manifolds, by Poincar\'{e} lemma.

Integration on a smooth manifold $M$ should be thought of as a nondegenerate
bilinear pairing $\left( ,\right) $ between $p-$forms and $p-$chains
(spanning a finite domain on $M$). Duality of $p-$forms and $p-$chains on $M$
is based on the de Rham's `period', defined as \cite{{de Rham,Choquet}}
\begin{equation*}
\text{Period}:=\int_{C}\omega :=\left( C,\omega \right) ,
\end{equation*}%
where $C$ is a cycle, $\omega $ is a cocycle, while $\left\langle C,\omega
\right\rangle =\omega (C)$ is their inner product $\left( C,\omega \right)
:\Omega ^{p}(M)\times \mathcal{C}_{p}(M)\rightarrow \mathbb{R}$. From the
Poincar\'{e} lemma, a closed $p-$form $\omega $ is exact iff $\left(
C,\omega \right) =0$.

The fundamental topological duality is based on the Stokes theorem (\ref%
{Stok}), which can be re written as
\begin{equation*}
\left( \partial C,\omega \right) =\left( C,d\omega \right) ,
\end{equation*}%
where $\partial C$ is the boundary of the $p-$chain $C$ oriented coherently
with $C$ on $M$. While the \textit{boundary operator} $\partial $ is a
global operator, the coboundary operator $d$ is local, and thus more
suitable for applications. The main property of the exterior differential,
\begin{equation*}
d\circ d\equiv d^{2}=0\quad \Longrightarrow \quad \partial \circ \partial
\equiv \partial ^{2}=0,\qquad (\text{and converse}),
\end{equation*}%
\noindent can be easily proved using the Stokes' theorem as
\begin{equation*}
0=\left( \partial ^{2}C,\omega \right) =\left( \partial C,d\omega \right)
=\left( C,d^{2}\omega \right) =0.
\end{equation*}

\subsection{De Rham cohomology}

In the Euclidean 3D space $\mathbb{R}^{3}$ we have the following de Rham
\emph{cochain complex}
\begin{equation*}
0\rightarrow \Omega ^{0}(\mathbb{R}^{3})\underset{\mathrm{grad}}{\overset{d}{%
\longrightarrow }}\Omega ^{1}(\mathbb{R}^{3})\underset{\mathrm{curl}}{%
\overset{d}{\longrightarrow }}\Omega ^{2}(\mathbb{R}^{3})\underset{\mathrm{%
div}}{\overset{d}{\longrightarrow }}\Omega ^{3}(\mathbb{R}^{3})\rightarrow 0.
\end{equation*}%
Using the \textit{closure property} for the exterior differential in $%
\mathbb{R}^{3},~d\circ d\equiv d^{2}=0$, we get the standard identities from
vector calculus
\begin{equation*}
\limfunc{curl}\cdot \limfunc{grad}=0\text{ \ \ \ \ \ \ and \ \ \ \ \ \ }%
\limfunc{div}\cdot \limfunc{curl}=0.
\end{equation*}

As a duality, in $\mathbb{R}^{3}$ we have the following \emph{chain complex}
\begin{equation*}
0\leftarrow \mathcal{C}_{0}(\mathbb{R}^{3}){\overset{\partial }{%
\longleftarrow }}\mathcal{C}_{1}(\mathbb{R}^{3}){\overset{\partial }{%
\longleftarrow }}\mathcal{C}_{2}(\mathbb{R}^{3}){\overset{\partial }{%
\longleftarrow }}\mathcal{C}_{3}(\mathbb{R}^{3})\leftarrow 0,
\end{equation*}
(with the closure property $\partial\circ \partial\equiv \partial^{2}=0$)
which implies the following three boundaries:
\begin{equation*}
C_{1}\overset{\partial }{\mapsto }C_{0}=\partial (C_{1}),\qquad C_{2}\overset%
{\partial }{\mapsto }C_{1}=\partial (C_{2}),\qquad C_{3}\overset{\partial }{%
\mapsto }C_{2}=\partial (C_{3}),
\end{equation*}
where $C_0\in\mathcal{C}_0$ is a 0--boundary (or, a point), $C_1\in\mathcal{C%
}_1$ is a 1--boundary (or, a line), $C_2\in\mathcal{C}_2$ is a 2--boundary
(or, a surface), and $C_3\in\mathcal{C}_3$ is a 3--boundary (or, a
hypersurface). Similarly, the de Rham complex implies the following three
coboundaries:
\begin{equation*}
C^{0}\overset{d }{\mapsto }C^{1}=d (C^{0}),\qquad C^{1}\overset{d }{\mapsto }%
C^{2}=d (C^{1}),\qquad C^{2}\overset{d }{\mapsto }C^{3}=d (C^{2}),
\end{equation*}
where $C^0\in\Omega^0$ is 0--form (or, a function), $C^1\in\Omega^1$ is a
1--form, $C^2\in\Omega^2$ is a 2--form, and $C^3\in\Omega^3$ is a 3--form.

In general, on a smooth $n$D manifold $M$ we have the following de Rham
cochain complex \cite{de Rham}
\begin{equation*}
0\rightarrow \Omega ^{0}(M)\overset{d}{\longrightarrow }\Omega ^{1}(M)%
\overset{d}{\longrightarrow }\Omega ^{2}(M)\overset{d}{\longrightarrow }%
\Omega ^{3}(M)\overset{d}{\longrightarrow }\cdot \cdot \cdot \overset{d}{%
\longrightarrow }\Omega ^{n}(M)\rightarrow 0,
\end{equation*}%
satisfying the closure property on $M,~d\circ d\equiv d^{2}=0$.
\begin{figure}[tbh]
\centerline{\includegraphics[width=13cm]{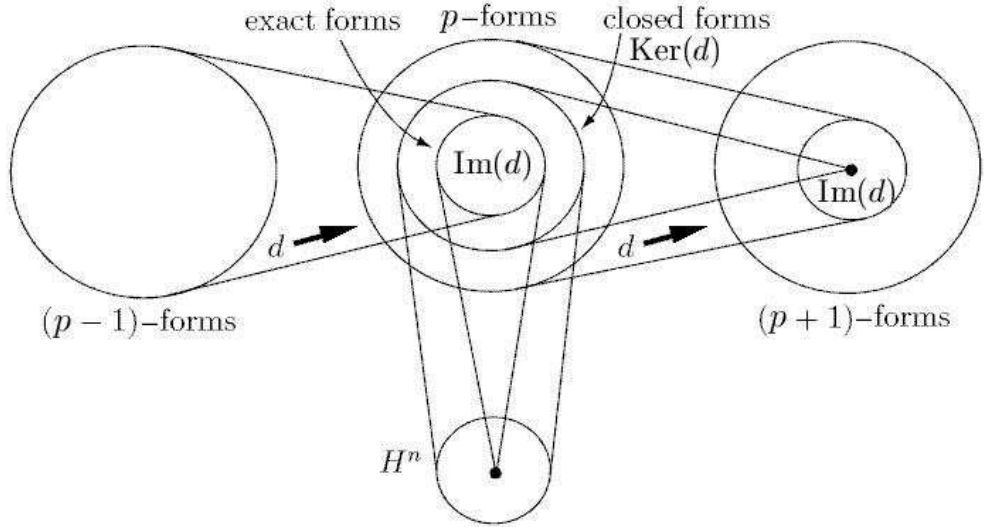}}
\caption{A small portion of the de Rham cochain complex, showing a
homomorphism of cohomology groups.}
\label{Cohomology}
\end{figure}

Informally, the de Rham cohomology is the (functional) space of closed
differential $p-$forms modulo exact ones on a smooth manifold.

More precisely, the subspace of all closed $p-$forms (cocycles) on a smooth
manifold $M$ is the kernel $\limfunc{Ker}(d)$ of the de Rham $d-$%
homomorphism (see Figure \ref{Cohomology}), denoted by $Z^{p}(M)\subset
\Omega ^{p}(M)$, and the sub-subspace of all exact $p-$forms (coboundaries)
on $M$ is the image $\limfunc{Im}(d)$ of the de Rham homomorphism denoted by
$B^{p}(M)\subset Z^{p}(M)$. The \textit{quotient space}
\begin{equation*}
H_{DR}^{p}(M):=\frac{Z^{p}(M)}{B^{p}{M}}=\frac{\limfunc{Ker}\left( d:\Omega
^{p}(M)\rightarrow \Omega ^{p+1}(M)\right) }{\limfunc{Im}\left( d:\Omega
^{p-1}(M)\rightarrow \Omega ^{p}(M)\right) },
\end{equation*}%
is called the $p$th de Rham \textit{cohomology group} of a manifold $M$. It
is a topological invariant of a manifold. Two $p-$cocycles $\alpha $,$\beta
\in \Omega ^{p}(M)$ are \emph{cohomologous}, or belong to the same \textit{%
cohomology class} $[\alpha ]\in H^{p}(M)$, if they differ by a $(p-1)-$%
coboundary $\alpha -\beta =d\theta \in \Omega ^{p-1}(M)$. The dimension $%
b_{p}=\dim H^{p}(M)$ of the de Rham cohomology group $H_{DR}^{p}(M)$ of the
manifold $M$ is called the Betti number $b_{p}$.

Similarly, the subspace of all $p-$cycles on a smooth manifold $M$ is the
kernel $\limfunc{Ker}(\partial )$ of the $\partial -$homomorphism, denoted
by $Z_{p}(M)\subset \mathcal{C}_{p}(M)$, and the sub-subspace of all $p-$%
boundaries on $M$ is the image $\limfunc{Im}(\partial )$ of the $\partial -$%
homomorphism, denoted by $B_{p}(M)\subset \mathcal{C}_{p}(M)$. Two $p-$%
cycles $C_{1}$,$C_{2}\in \mathcal{C}_{p}$ are \emph{homologous}, if they
differ by a $(p-1)-$boundary $C_{1}-C_{2}=\partial B\in \mathcal{C}_{p-1}(M)$%
. Then $C_{1}$ and $C_{2}$ belong to the same \textit{homology class} $%
[C]\in H_{p}(M)$, where $H_{p}(M)$ is the homology group of the manifold $M$%
, defined as
\begin{equation*}
H_{p}(M):=\frac{Z_{p}(M)}{B_{p}(M)}=\frac{\limfunc{Ker}(\partial :\mathcal{C}%
_{p}(M)\rightarrow \mathcal{C}_{p-1}(M))}{\limfunc{Im}(\partial :\mathcal{C}%
_{p+1}(M)\rightarrow \mathcal{C}_{p}(M))},
\end{equation*}%
where $Z_{p}$ is the vector space of cycles and $B_{p}\subset Z_{p}$ is the
vector space of boundaries on $M$. The dimension $b_{p}=\dim H_{p}(M)$ of
the homology group $H_{p}(M)$ is, by the de Rham theorem, the same Betti
number $b_{p}$.

If we know the Betti numbers for all (co)homology groups of the manifold $M$%
, we can calculate the \textit{Euler--Poincar\'{e} characteristic} of $M$ as
\begin{equation*}
\chi (M)=\sum_{p=1}^{n}(-1)^{p}b_{p}.
\end{equation*}

For example, consider a small portion of the de Rham cochain complex of
Figure \ref{Cohomology} spanning a space-time 4--manifold $M$,
\begin{equation*}
\Omega ^{p-1}(M)\overset{d_{p-1}}{\longrightarrow }\Omega ^{p}(M)\overset{%
d_{p}}{\longrightarrow }\Omega ^{p+1}(M)
\end{equation*}%
As we have seen above, cohomology classifies topological spaces by comparing
two subspaces of ~$\Omega ^{p}$:~ (i) the space of $p-$cocycles, $Z^{p}(M)=%
\func{Ker}d_{p}$, and~ (ii) the space of $p-$coboundaries, $B^{p}(M)=\func{Im%
}d_{p-1}$. Thus, for the cochain complex of any space-time 4--manifold we
have,
\begin{equation*}
d^{2}=0\quad \Rightarrow \quad B^{p}(M)\subset Z^{p}(M),
\end{equation*}%
that is, every $p-$coboundary is a $p-$cocycle. Whether the converse of this
statement is true, according to Poincar\'{e} lemma, depends on the
particular topology of a space-time 4--manifold. If every $p-$cocycle is a $%
p-$coboundary, so that $B^{p}$ and $Z^{p}$ are equal, then the cochain
complex is exact at $\Omega ^{p}(M)$. In topologically interesting regions
of a space-time manifold $M$, exactness may fail \cite{Wise}, and we measure
the failure of exactness by taking the $p$th cohomology group
\begin{equation*}
H^{p}(M)=Z^{p}(M)/B^{p}(M).
\end{equation*}


\begin{thebibliography}{99}
\bibitem{bjorken} J.D.~Bjorken and S.D.~Drell, \textit{Relativistic Quantum
Fields}, McGraw-Hill, 1965.

\bibitem{itzykson} C.~Itzykson and J.B.~Zuber, \textit{Quantum Field Theory}%
, McGraw-Hill, 1980.

\bibitem{ramond} P.~Ramond, \textit{Field Theory: A Modern Primer},
Addison-Wesley, 1990.

\bibitem{peskin} M.E.~Peskin and D.V.~Schroeder, \textit{An Introduction to
Quantum Field Theory}, Addison Wesley, 1995.

\bibitem{weinberg} S.~ Weinberg, \textit{The Quantum Theory of Fields},
Vols. 1-3, Cambridge, 1995

\bibitem{deligne} P. Deligne et al. (editors), \textit{Quantum Fields and
Strings: a Course for Mathematicians}, Am. Math. Soc. 1999.

\bibitem{zee} A.~ Zee, \textit{Quantum Field Theory in a Nutshell},
Princeton, 2003.

\bibitem{dewitt} B.S.~DeWitt, \textit{The Global Approach to Quantum Field
Theory}, Vols. 1 \& 2, Oxford, 2003.

\bibitem{nair} V.P.~Nair, \textit{Quantum Field Theory. A Modern Perspective}%
, Springer, 2005.

\bibitem{srednicki} M.A. Srednicki, \textit{Quantum Field Theory}, Cambridge
Univ. Press, 2007.

\bibitem{WittenTQFT} E. Witten, Topological quantum field theory, Com. Math.
Phys. \textbf{117}(3), 353--386, 1988.

\bibitem{Witten89} E. Witten, Quantum field theory and the Jones polynomial,
Comm. Math. Phys. \textbf{121}, 351, 1989.

\bibitem{Atiyah} M. Atiyah, Topological quantum field theories, Pub. Math.
l'IH\'{E}S \textbf{68}, 175--186, 1988.

\bibitem{TegmarkWheeler} M. Tegmark, J.A. Wheeler, Scientific American,
68--75, February, 2001.

\bibitem{DiracBook} P.A.M. Dirac, The Principles of Quantum Mechanics (4th
ed.) Clarendon Press, Oxford, 2000.

\bibitem{QuLeap} V. Ivancevic, T. Ivancevic, Quantum Leap: From Dirac and
Feynman, Across the Universe, to Human Body and Mind. World Scientific,
Singapore, 2008.

\bibitem{Frampton} P.H. Frampton, Gauge Field Theories, Frontiers in
Physics. Addison-Wesley, 1986.

\bibitem{MTW} C.W. Misner, K.S. Thorne, J.A. Wheeler, Gravitation. W.
Freeman and Company, New York, 1973.

\bibitem{BaezGauge} J.C. Baez, J.P. Muniain, Gauge Fields, Knots and
Gravity, Series on Knots and Everything -- Vol. 4. World Scientific,
Singapore, 1994.

\bibitem{GaneshSprBig} V. Ivancevic, T. Ivancevic, Geometrical Dynamics of
Complex Systems. Springer, Dordrecht, 2006.

\bibitem{GaneshADG} V. Ivancevic, T. Ivancevic, Applied Differfential
Geometry: A Modern Introduction. World Scientific, Singapore, 2007.

\bibitem{Complexity} Ivancevic, V., Ivancevic, T., Complex Nonlinearity:
Chaos, Phase Transitions, Topology Change and Path Integrals, Springer, 2008.

\bibitem{FeynQM} R.P. Feynman, Space--time approach to non--relativistic
quantum mechanics, Rev. Mod. Phys. \textbf{20}, 367--387, 1948;\newline
R.P. Feynman and A.R. Hibbs, Quantum Mechanics and Path Integrals,
McGraw--Hill, New York, 1965.

\bibitem{FeynQED} R.P. Feynman, Space--time approach to quantum
electrodynamics, Phys. Rev. \textbf{76}, 769--789, 1949; \newline
\emph{Ibid.} Mathematical formulation of the quantum theory of
electromagnetic interaction, Phys. Rev. \textbf{80}, 440--457, 1950.

\bibitem{Faddeev} L.D. Faddeev and V.N. Popov, Feynman diagrams for the
Yang--Mills field. Phys. Lett. B \textbf{25}, 29, 1967.

\bibitem{Ambjorn} J. Ambjorn, R. Loll, Y. Watabiki, W. Westra and S. Zohren,
A matrix model for 2D quantum gravity defined by Causal Dynamical
Triangulations, Phys.Lett. B \textbf{665}, 252--256, 2008.; \newline
\emph{Ibid.} Topology change in causal quantum gravity, Proc. JGRG 17,
Nagoya, Japan, December 2007; \newline
\emph{Ibid.} A string field theory based on Causal Dynamical Triangulations,
JHEP \textbf{0805}, 032, 2008.

\bibitem{Loll} R. Loll, The emergence of spacetime or quantum gravity on
your desktop, Class. Quantum Grav. \textbf{25}, 114006, 2008;\newline
R. Loll, J. Ambjorn and J. Jurkiewicz, The Universe from scratch, Contemp.
Phys. \textbf{47}, 103--117, 2006;\newline
J. Ambjorn, J. Jurkiewicz and R. Loll, Reconstructing the Universe, Phys.
Rev. D \textbf{72}, 064014, 2005.

\bibitem{VQM} B. Thaller, Visual Quantum Mechanics, Springer, 1999; \newline
\emph{Ibid.} Advanced Visual Quantum Mechanics, Springer, 2005.

\bibitem{de Rham} G. de Rham, Differentiable Manifolds. Springer, Berlin,
1984.

\bibitem{Hodge} C. Voisin, Hodge Theory and Complex Algebraic Geometry I.
Cambridge Univ. Press, Cambridge, 2002.

\bibitem{Flanders} H. Flanders, Differential Forms: with Applications to the
Physical Sciences. Acad. Press, 1963.

\bibitem{Teixeira} F.L. Teixeira and W.C. Chew, Lattice electromagnetic
theory from a topological viewpoint, J. Math. Phys. \textbf{40}, 169---187,
1999.

\bibitem{He} B. He, F.L. Teixeira, On the degrees of freedom of lattice
electrodynamics. Phys. Let. A \textbf{336}, 1--7, 2005.

\bibitem{Choquet} Y. Choquet-Bruhat, C. DeWitt-Morete, Analysis, Manifolds
and Physics (2nd ed). North-Holland, Amsterdam, 1982.

\bibitem{Gegenberg} J. Gegenberg, G. Kunstatter, The Partition Function for
Topological Field Theories, Ann. Phys. \textbf{231}, 270--289, 1994.

\bibitem{Abraham} R. Abraham, J. Marsden, T. Ratiu, Manifolds, Tensor
Analysis and Applications. Springer, New York, 1988.

\bibitem{Wise} D.K. Wise, p-form electrodynamics on discrete spacetimes.
Class. Quantum Grav. \textbf{23}, 5129--5176, 2006.

\bibitem{Labastida} J.M.F. Labastida, C. Lozano, Lectures on Topological
Quantum Field Theory. CERN-TH/97-250; US-FT-30/97; hep-th/9709192, 1997.

\bibitem{Lenz} F. Lenz, Topological concepts in gauge theories,
FAU-TP3-04/3; hep-th/0403286, 2004.

\bibitem{GEGL72} H. Georgi and S. Glashow, Unified Weak and Electromagnetic
Interactions without Neutral Currents, Phys. Rev. Lett. \textbf{28}, 1494
1972.

\bibitem{Belavin84} {{A.A. Belavin}, {A.M.} {Polyakov}}, {{A.B.} {%
Zamolodchikov}}, Infinite conformal symmetry in 2D quantum field theory, {%
Nucl. Phys.} \textbf{B241}, {333, 1984}.

\bibitem{Senechal97} {{P. Di~Francesco}, {P.}~{Mathieu}}, {{D.}~{S\'{e}n\'{e}%
chal}}, Conformal Field Theory. Springer, New York, {19}97.

\bibitem{FreedmanRMP} C. Nayak, S.H. Simon, A. Stern, M. Freedman, S.D.
Sarma, Non-Abelian Anyons and Topological Quantum Computation. Rev. Mod.
Phys. \textbf{80}, 1083 (77 pages), 2008.

\bibitem{Freedman04a} {M. }Freedman, {C. Nayak}, {K. Shtengel}, {K. Walker},
{Z. Wang}, A class of ${P},{T}$-invariant topological phases of interacting
electrons, {Ann. Phys. (N.Y.)} \textbf{310}, {428}, 2004.

\bibitem{Jones85} V.F.R. Jones, A polynomial invariant for knots via von {N}%
eumann algebras. Bull. Amer. Math. Soc. \textbf{12}, 103, 1985.
\end{thebibliography}
\end{document}